\documentclass[10pt,preprint]{aastex}
\input epsf.tex
\begin{document}
\title{Calibration of Equilibrium Tide Theory for Extrasolar Planet Systems II} 
\author{Brad M. S. Hansen\altaffilmark{1}
}
\altaffiltext{1}{Department of Physics \& Astronomy and Institute of Geophysics \& Planetary Physics, University of California Los Angeles, Los Angeles, CA 90095, hansen@astro.ucla.edu}


\lefthead{Hansen }
\righthead{Two Tides II}

\begin{abstract}

We present a new empirical calibration of equilibrium tidal theory for extrasolar planet
systems, extending a prior study by incorporating detailed physical models for the internal
structure of planets and host stars. 
The resulting 
 strength of the stellar tide produces a coupling that
is strong enough to reorient the spins of some host stars without causing catastrophic orbital evolution, thereby potentially
explaining the observed trend in alignment between stellar spin and planetary orbital angular momentum. By
isolating the sample whose spins should not have been altered in this model, we also show evidence for two different
processes that contribute to the population of planets with short orbital periods.
 We apply our results to 
estimate the remaining lifetimes for short period planets, examine the survival of planets around evolving stars,
and determine the limits for circularisation of planets with highly eccentric orbits.  Our analysis suggests that the  
survival of circularised planets is strongly affected by the amount of heat dissipated, which 
 is often large enough to lead to runaway orbital inflation and Roche lobe
overflow.

\end{abstract}


\section{Introduction}

The fact that tidal dissipation plays a role in the organisation of planetary systems has been clear since the first detections of
extrasolar planets. The absence of eccentric orbits (commonplace at larger separations) with small periastra indicates that tidal circularisation must
take place for planets with periods $< 4$~days. This likely represents a measure of the dissipation in the planet itself, but
the discovery of planets with orbital periods $<$1~day suggests that tidal dissipation in the star may also set a limit on the
long-term survival of close-in planets. This has acquired particular relevance recently, with the discovery of correlations
between the properties of the host star and the planetary orbit (Winn et al. 2010a).

A wide range of theoretical calculations of tidal interactions have been performed in order to investigate potential mechanisms
for the microphysical dissipation, and to illuminate plausible values of gross measures of dissipation, such as the `tidal Q' 
parameter (Rasio et al. 1996; Marcy et al. 1997; Terquem et al. 1998; Ogilvie \& Lin 2004, 2007; Ivanov \& Papaloizou 2004; Wu 2005; Jackson, Greenberg \& Barnes 2008; Goodman \& Lackner 2009). This has led to a wide range of estimates for the strength of tidal dissipation in
these systems.
In order to properly understand how such estimates match to theory, we need to consider the ensemble of observations within the context
of a single formalism. In Hansen (2010) (hereafter H10), we presented an analysis of the extant exoplanet database within the context
of equilibrium tide theory, as laid out in the exposition of Eggleton, Kiseleva \& Hut (1998). In that simplified model, we assumed
a single bulk dissipation constant per unit mass, $\sigma_*$, for all host stars in the range 0.7--1.4 $M_{\odot}$ and a second
similar parameter, $\sigma_p$, for all gas giant planets $> 0.3 M_J$. We were able to demonstrate that it is possible to 
 calibrate these two parameters to yield a coherent
description of a wide range of observed exoplanet systems. 

It is desirable, however, to go beyond this simple assumption. The analysis in H10 used single values of the dissipation constant
for all stars and planets respectively, and stopped short of adopting
a detailed microphysical model. Nevertheless, it is possible to calculate plausible values of $\sigma_*$ and $\sigma_p$ if one adopts a model
for the internal structure of an object, and calculates the amount of dissipation expected from the turbulent viscosity engendered in
the convection zone of a given star or planet. That is our goal in this paper -- to go beyond the assumption of uniform values of $\sigma_*$ and
$\sigma_p$, and to calculate their values as a function of mass and age, for both planets and host stars. The advantage is that we may then explore possible trends with stellar and planetary properties within the context of the model, and potentially illuminate further the role that tides play in sculpting the architecture of exoplanet systems.

In the following, \S~\ref{Model} discusses the dissipation model we adopt, as well as the construction of stellar and planetary background structures that will determine the overall level of dissipation that occurs. In \S~\ref{Stellar} we apply this model to the observed population of stellar and planetary systems that are relevant for the purposes of constraining the strength of tides. In \S~\ref{Discuss} we apply our newly calibrated tidal theory to a series of issues including survival limits for planets and trends of misalignment with stellar type, reviewing our final conclusions in \S~\ref{Conclusion}.

\section{Dissipation Model}
\label{Model}

In H10 we calibrated the Eggleton et al (1998) model (hereafter EKH98) for tidal dissipation using the
observed properties of extrasolar planet systems. The attractive feature of that particular model was
that the macroscopic evolution equations were cast in terms of a bulk dissipation constant $\sigma$,
which related the rate of energy dissipation to the rate of variation of the quadropole moment of 
the body due to the tidally induced circulation within the body. We consider this a more transparent formulation than the traditional approach of using a
tidal `quality factor', Q. Nevertheless, the approach taken in H10 is still a gross approximation,
as it left the underlying microscopic physics unspecified, and thus offered no obvious path to generalise
the formalism beyond our canonical solar-like stars and Jupiter-class planets. In this paper we wish to
try and extend our analysis to a more physically motivated model.

We assume that the underlying dissipative
mechanism is turbulent dissipation in convection zones.
EKH98 provide the formalism to do this as well. They demonstrate how the bulk dissipation
constant $\sigma$ is related to the turbulent dissipation in the star via the integral
\begin{equation}
\sigma = \frac{2}{M^2 R^4 Q_E^2} \int w \ell \gamma(r) dm.  \label{sigma_integral}
\end{equation}
In this expression, $M$ and $R$ are the mass and radius of the object and $Q_E$ (the Q of EKH98, which we have renamed
to avoid confusion with the commonly used "tidal Q") is the normalised quadropole moment. The integral is taken
over the turbulent regions of the star, with $w$ and $\ell$ the relevant velocity and mean free path. The function
$\gamma$ is a radial weighting determined by the tidally-induced velocity field in the star. It is the action of
the turbulent viscosity on this time dependent flow field that ultimately provides the tidal torque.

In this paper we will assume the turbulence is the result of convection, with $w$ and $\ell$ determined from
the standard mixing length theory. We will thus have a viscosity $\nu$ defined by
$$
\nu = \nu_0 v_c H_p f(P)
$$
where $\nu_0$ is a normalisation constant to be determined from the overall analysis but which we expect to be
of order unity, $H_p$ is the pressure
scale height, $v_c$ is the mixing length convective velocity and $f(P)$ is a function included to address the
possibility of a frequency dependance in the coupling between tidal forcing dissipation. There are two
prominent proposals in the literature. The first is that due to Zahn (1977; 1989), who proposed a form
$$
f(P) = min(1, P/2 \tau),
$$
where P is the forcing period and $\tau$ is the eddy turnover time. On the other hand, Goldreich \& collaborators
(Goldreich \& Nicholson 1977; Goldreich \& Keeley 1977) favour a form
$$
f(P) = min(1, \left( P/2 \pi \tau \right)^2).
$$
Guidance as to the correct form is not yet wholly forthcoming from analysis of the microphysics. Recently 
Penev, Barranco \& Sasselov (2009,2011) have performed numerical simulations of forced anelastic convection and
find a scaling closer to the linear model at the periods of interest to us. At shorter periods they find a 
sharper fall-off in dissipative efficiency, suggesting that both the Zahn \& Goldreich et al. models may be
valid in different regimes. On the other hand, Ogilvie \& Lesur (2012) find better agreement with the quadratic
model. We will proceed with the linear Zahn scaling as it will enable us to better compare our model with others
in the literature.

 \subsection{Stellar Models}
\label{StarModels}

To describe the structure
of main sequence stars of various mass and age, we use the EZ-Web models (Paxton 2004). 
Figure~\ref{CZsurf} shows the mass of
the surface convection zone for solar metallicity stars of various masses from 0.3$M_{\odot}$ to 1.5$M_{\odot}$, for two different
ages (0.1~Gyr and 1~Gyr). We see that the mass in the convection zone becomes comparable to brown dwarf or planet
masses once the host star mass $>1.1 M_{\odot}$. For stars greater than this mass the degree of coupling between
the convection zone and the rest of the star is of relevance, as it will determine how easy it is for the star to spin up
the surface layers to synchronism.
Also of relevance  is the question of how much of these convection zones
actually couple strongly to perturbations, if the convective overturn times are much larger than the forcing period. Figure~\ref{Turnover} shows the profile of convective eddy turnover time as a function
of radius for stars with masses $0.8 M_{\odot}$, $1.0 M_{\odot}$ and $1.3 M_{\odot}$. We see that the much of the
greater mass fraction in convection zones has eddy turnover times of many days, suggesting that they might possibly
not play as important a role as the layers closer to the surface, where the eddy times are shorter. This will serve
to equilibrate the response partially across the mass spectrum. The actual result of these competing trends is shown in Figure~\ref{Sig}, where we
show the effective stellar dissipation $\bar{\sigma}_*$ (assuming $\nu_0=1$ and the normalisation of H10) for a range of stellar masses, assuming
a stellar age of 1~Gyr and a forcing period of 1~day, using the Zahn scaling. We see that the lower-mass stars, with larger convection zones,
do possess stronger dissipation, while the main sequence A stars show much reduced dissipation. The mean value for $\bar{\sigma}_*$ derived
in H10 is realised for $M \sim 1.2 M_{\odot}$.

The size of the surface convection zone also evolves with time, especially for those stars that proceed far enough to become
subgiants and giants when they are observed.
 Figure~\ref{CZevol} shows the evolution of the surface convection zone for a 1.5~$M_{\odot}$ star as it evolves.
Although a 1.5$M_{\odot}$ star begins with a very small surface convection zone, it acquires a mass comparable
to a Jupiter mass planet within $\sim$1~Gyr. This will become relevant when we consider the planets observed around
subgiants (\S~\ref{HotSubs}), where the radial velocity technique is easier to apply than for hotter main sequence stars.
 The rapid increase in convection zone mass along the red giant branch
is evident as well.


Dissipation in the star produces two principal effects, orbital decay and stellar spin-up.
The characteristic timescale for orbital decay due to stellar dissipation is given by equation~(10) of H10,
\begin{equation}
T_{orb} = 8.2 \times 10^9 years \left( \frac{a}{0.02 AU} \right)^{8} \left( \frac{R}{R_{\odot}} \right)^{-10}
\left( \frac{M_p}{M_J} \right)^{-1} \left( \frac{\bar{\sigma}_*}{7.8 \times 10^{-8}} \right)^{-1}. \label{Torb}
\end{equation}
The characteristic time to change the spin of the star is, using the formulae in EKH98 and H10, 
\begin{eqnarray}
T_{spin} & = & \frac{\Omega_*}{\dot{\Omega_*}} = 2 k_0^2 \frac{R_*^2 \Omega_*}{M_p} \left( \frac{M_*+M_p}{G a (1 - e^2)} \right)^{1/2} T_{orb} \nonumber \\
 & = &  \frac{3.1 \times 10^9 years}{(1-e^2)^{1/2}} \left( \frac{a}{0.02 AU} \right)^{7.5} \left( \frac{R}{R_{\odot}} \right)^{-8} \left( \frac{k_0^2}{0.1}\right) \left( \frac{30 days}{P_*}\right)
\left(\frac{M_*+M_p}{M_{\odot}} \right)^{1/2} \left( \frac{M_p}{M_J} \right)^{-2} \left( \frac{\bar{\sigma}_*}{7.8 \times 10^{-8}} \right)^{-1}, \label{Tspin}
\end{eqnarray}
where $k_0$ is the stellar radius of gyration. This will become particularly important when we evaluate the change in orbital obliquity relative
to the star (\S~\ref{RM}). 

 \subsection{Planet Models}
\label{PlanetModels}

Turbulent convective dissipation is also potentially a source of viscosity in gas giant planets, as such objects
are largely convective throughout their evolution. Thus, we apply the same model to planetary structure models calculated
with the models in Hansen \& Barman (2007), although we have extended the mass range as required.

 Figure~\ref{Sigp} shows
the resulting values of $\bar{\sigma}_p$ for a series of giant planets. In the top panel, we show $\bar{\sigma}_p$ for
a $1 M_J$ planet at age 1~Gyr, using both the Zahn \& Goldreich/Nicholson prescriptions. In the bottom panel, we show
the same function for planets of younger age (0.1~Gyr -- more suitable for the epoch of circularisation) and masses of 0.3$M_J$, 1$M_J$ and 3$M_J$. A common point of 
comparison is the effective $Q'$ parameter derived for Jupiter. The
1~Gyr, $1 M_J$ models suggest $Q' \sim 5 \times 10^7$ using the Zahn presciption, and $Q' \sim 3 \times 10^{10}$ using
the Goldreich \& Nicholson prescription, assuming a forcing period of 10~hours, using equation~(11) of H10. These values are consistent with previous
estimates (e.g. Goldreich \& Nicholson 1977), in the sense that they are too large to explain the origin of the Laplace resonance amongst the Galilean
satellites. Indeed, they lie well above the recent empirical value measured by Lainey et al. (2009).

This highlights a note of potential caution in applying our approach to the planets in these systems.
 The relatively low rotation of the planet host stars is an important
motivation for why the equilibrium tide model, as outlined in H10 \& \S~\ref{StarModels} might be applicable. Indeed, we found (H10) that the level of dissipation in close
stellar binaries is well above that found in planet-hosting stars of similar mass. We attribute this discrepancy to
an enhanced level of dissipation in synchronously rotating bodies 
(resulting from a greater commensurability between the tidal forcing and internal oscillation modes of the star, e.g. Ogilvie \& Lin 2007). In the systems under discussion, we expect the planets, unlike their host stars, to be
synchronously rotating, and thus they could also potentially have higher rates of dissipation (Ogilvie \& Lin 2004; Wu 2005;
 Goodman \& Lackner 2009). This may also be the explanation for the discrepancy between the inferred Jovian numbers
and those derived from our model.

Nevertheless, our motivation in this study is not to provide a comprehensive description of potential contributors to tidal
evolution, but to evaluate the simplest possible, widely applicable, model and the degree to which it can explain the observations.
We will therefore proceed with the initial assumption that the convective dissipation model does indeed underlie the dissipation in the
planets, and evaluate where a greater level of dissipation is actually required by the observations, 
 as might be provided by a dynamical contribution to the tide. Such calibrations could represent a long-term average over the
shorter term variability inherent in such modal excitations.

Our planetary models provide not only internal models to evaluate $\bar{\sigma}_p$, but also the evolution of the
planetary radius with time, which can strongly affect the strength of tidal interactions. We do not include possible extra
sources of energy invoked to explain some anomalously large observed radii, because these anomalies are now believed to
be related to the level of irradiation received (e.g. Showman \& Guillot 2002; Fortney, Marley \& Barnes 2007; Demory \& Seager 2011)
and the bulk of the circularisation occurs at separations where irradiation is energetically irrelevant.
A more important
  process in this context is the radius inflation due to tidal dissipation within the planet. We use our cooling models to 
establish a relationship between internal 
entropy\footnote{For most of the cooling sequence this is equivalent to the
more conceptually transparent central temperature, but becomes more complicated for strongly heated planets, i.e. when significant
radius expansion occurs.} and radius. The exact nature of this relationship depends on the manner of the energy
dissipation, described
in appendix~\ref{Planet_Heat}. In that section we motivate our model to treat the 
 heating due to tidal dissipation as a simple reversal of the cooling along the
same track.

 The evolution of a particular planet depends on the competition
between cooling and tidal heating within a given scenario, but we can get an approximate criterion for the division between
survival and disruption by determining the locus of semi-major axis
and eccentricity such that the tidal heating balances the cooling for a young planet (because circularisation times
are $\sim 10^7$--$10^8$~years). 
 Figure~\ref{Disrupt} shows this criterion between
disruption and survival for a 0.5$M_J$ planet. We see that the curve of constant angular momentum (because tidal dissipation
in the planet removes energy much more efficiently than angular momentum) that grazes this
boundary produces circularised orbits at approximately the right separation to match the observed concentration around
an orbital period of three days. This lends support to the assertion of
Faber, Rasio \& Ford (2005) that the characteristic orbital periods of hot Jupiters are related to the Roche lobe
limit. Systems that lie initially above the dashed line in Figure~\ref{Disrupt} will eventually be heated to the
point of disruption, before they reach tidal circularisation.

\section{Observational Tests}
\label{Stellar}

With interior structure models for both exoplanets and stellar hosts, and a model for the tidal
dissipation, we can now calculate the tidal evolution of planet-star systems and compare them
to observations. However, before we compare to the planetary systems themselves, we note that there
is a class of stellar binary that is also applicable to the problem of constraining the strength
of tidal dissipation.

\subsection{Giant Stars}
\label{VPsec}

An exercise of the kind we perform here has already been performed in the context of compact objects orbited by giant stars
with large convective envelopes. 
Verbunt \& Phinney (1995) tested the equilibrium tide model for convective envelopes by comparing
models to the observed eccentricity-period relation for spectroscopic binaries containing evolved
stars in open clusters. They claim that the convective dissipation model of Zahn describes the
observations well.  An internally consistent model for the dissipation of tides in convective zones should
be able to incorporate this constraint as well as those from exoplanet systems. Thus, we
 wish to compare our model with the same observations as Verbunt \& Phinney (hereafter VP) to 
estimate plausible ranges for the parameters of our model and to demonstrate the consistency across
multiple stellar regimes.

For the purposes of direct comparison with the analysis of VP, we can recast equation~(3) of
H10 in an equivalent form, and calculate the integrated (over stellar evolution) change in eccentricity (in the
low eccentricity limit) to be
\begin{equation}
\Delta \ln e  = -1.7 I(t) \left[ \frac{M_2}{M} \left( \frac{M+M_2}{M} \right) \left( \frac{R_{\odot}}{a} \right)^8 \right],
\end{equation}
where $I$ is now
\begin{equation}
I(t) = 23.82 M_{\odot} R_{\odot}^2 \int_0^t \frac{M}{M_{\odot}} \left( \frac{R}{R_{\odot}} \right)^{10} \sigma_*.
\end{equation}
 The effect of dissipation in the evolving star is captured in the function $I$.

Figure~\ref{I2} shows $I$ as a function of stellar radius $R$ (a more useful quantity observationally than time), along with
the corresponding behaviour of $\sigma_*$, normalised as in H10, for a 2.3 $M_{\odot}$ star, as it evolves up the
giant branch, through core Helium burning, and back to the AGB.
VP set a criterion of $\Delta \ln e = -3$ as the value that
marks the transition between systems that should have been circularised and those that were not. The stars that are
most important for the model comparison are those designated as c and o in Table~2 of VP. We have repeated their
analysis for these systems using our model. The stellar radius inferred from the photometry for star o allows for it to be in 
multiple possible evolutionary stages, and we have adopted a status as a red clump star for this object, based on
the same rationale as VP (that stars of corresponding radius spend more time in this stage than ascending the red giant
branch for the first time). The resulting distribution of $\Delta \ln e$ for the various VP stars is shown in Figure~\ref{dle}.

We see that that c and o bracket the observed transition from circularised to eccentric orbits. The fact that this transition
brackets $\Delta \ln e = -1$ suggests that the nominal normalisation $\nu_0 \sim 1$ is a satisfactory fit to
the observations. Setting $\nu_0>1$ will move the points in Figure~\ref{dle} to the left, which preserves the
fidelity of the model as long as system~c retains $-\Delta \ln e < 1$, i.e. $\nu_0 <6.7$. If we insist that system~o
conform to the requirement $-\Delta \ln e > 3$, then require $\nu_0 >1.3$. Thus, we have a broad constraint that
$1.3< \nu_0 <6.7$. The lower limit here is based on a single circular orbit. Since stars can, in principle, form
with circular orbits, a more conservative estimate would use the next circular system, star~d, which imposes a
looser constraint $\nu_0 > 0.06$.

In closing we also note that this comparison tells us little about the appropriate scaling of dissipation with
forcing period, as the relative orbital (i.e. forcing) periods are in the range 100-200 days, and the turnover
times in the convective envelope of a giant at the tip of the red giant branch are $<90$~days. Thus, the full spectrum
of convective eddies couples to the giant envelope. Nevertheless, it sets a constraint on the normalisation of the
model that will then constrain the application to the planet  and planet host sample.

\subsection{Exoplanet Systems}

For the exoplanet systems, the mass ratio between planet and host star, in addition to the separation, will determine the relative importance
of dissipation in the planet and the star. 
We therefore analyse a series of three different planet-star samples. 
In each case, we choose a wide range of initial semi-major axis and eccentricity, and evolve the system forward using the formalism
outlined in EKH98 and H10. The dissipation constants $\bar{\sigma}_p$ and $\bar{\sigma}_*$ are now functions of time, because they
are calculated using the models for the evolution of the host star and planet. In the case of the planets, we also include the resulting
instantaneous tidal heating as a heat source, which combats the planet cooling. We assume that the planet follows the same relationship
between radius and internal entropy throughout, so that strongly heated planets reverse their evolution along the cooling track. For
sufficient heating, planets can overflow their Roche lobes.

\subsubsection{Massive Planet Hosts}
\label{BigHosts}

Our first sample of exoplanet systems is that containing massive ($M>5 M_J$) planets. 
As discussed in H10, the relative evolution timescales for stellar and planetary tides scale
as 
$$
 \frac{T_*}{T_p} \propto \left(\frac{M_*}{M_J} \right)^2 \left( \frac{R_p}{R_*} \right)^{10},
$$
 so our best constraints on the properties of the host stars are to be found from stars which host massive companions in
close orbits. The systems that provided the best constraints in that case were WASP-14b, XO-3b,
HAT-P-2b, WASP-18b and HAT-P-34b. The sample and it's properties are collected in Table~\ref{Tab1}. We note that these
systems can only provide upper limits on $\nu_0$, because the uncertain origins of these planets means that
some could have been formed on circular orbits, in which case the lack of eccentricity offers no constraint.
However, given an observed eccentricity and age, we can determine that an orbit would have circularised if
$\nu_0$ was too large.


The planetary system XO-3b (Johns-Krull et al. 2008; Winn et al. 2009a) consists of $\sim 12 M_J$ planet
in a 3.2~day orbit about a 1.2~$M_{\odot}$ star, according to the most recent analysis (Southworth 2010). We have applied our dissipation model to the evolving
surface convection zone of a 1.2~$M_{\odot}$ star, and are able to match the period and eccentricity 
of the system. Satisfactory matches are
achieved for $\nu_0<7$, assuming an age range for the system of 2--3.4~Gyr. Figure~\ref{xo3_init} shows
the range of final periods and eccentricities for an XO-3 star-planet combination, assuming $\nu_0=1.5$ in both star and planet.
 One potential evolution is shown in Figure~\ref{xo3_ex}. We see that the stellar tide plays a significant
role in the evolution of this system, as significant orbital angular momentum is transferred to the star. It is
possible to spin the star up to the observed rotation period, and we will return to this issue in \S~\ref{RM}.


 HAT-P-2b (Bakos et al. 2007b; Pal et al. 2010) is similarly well fit by the model with $\nu_0<6$, given
the recently upwardly revised stellar mass ($1.36 \pm 0.04 M_{\odot}$) and  allowing an age range of 2--3.1~Gyr. 
WASP-14b has a lower mass host star ($1.2 M_{\odot}$) and consequently larger convective zone
and more dissipation. However, the low estimated age of 0.5-1.0~Gyr (Joshi et al. 2008) means that the
constraint is very loose, allowing a wide range of eccentricity for WASP-14b.

The star WASP-18b (Hellier et al. 2009) has a mass $1.28 M_{\odot}$ (Southworth et al. 2009),
and is orbited by a planet of mass 10.3$M_J$ once every 0.94 days. Once again, the model 
fits the observations well for ages $\sim 1$--2~Gyr. We can match period and eccentricity well with $\nu_0<3$.
Figure~\ref{WASP18} shows the allowed parameter space for $\nu_0=1.5$. The large planet mass and short orbital
period means that stellar tides are particularly important in this system, and we see that significant stellar
spin-up is expected for this star. However, this constraint is somewhat weakened by the reanalysis of the radial
velocity data by Pont et al. (2011) \& Husnoo et al. (2012). They find that the previously reported non-zero eccentricity is not statistically
significant, and determine only an upper limit of $e<0.018$. For this planet, and other systems for which Husnoo et al. 
derive weaker constraints, we will note limits based on both the literature values and the weaker constraints, to illustrate
the insensitivity of our results to residual uncertainties in the observed values.

The recently announced planet HAT-P-34b has a mass of 3.3~$M_J$ and orbits a star with mass $1.4 M_{\odot}$ and age
$1.7\pm 0.5$~Gyr. The two sigma range of the eccentricity for this object requires that $\nu_0 <4$. The host star
in this system actually rotates faster than the orbital period of the star, but the planet is too distant for
the stellar spin to have a significant effect.

These five systems placed significant constraint on the model presented in H10, but are well matched within
the mass-dependant formalism presented here. In part this is because all four systems orbit relatively massive main
sequence stars, so that that small size of the convection zone (and concomitant reduced level of convective dissipation)
reduces the importance of the stellar tide in these systems. There are other massive planets, such as
 CoRoT-3b, CoRoT-14b, KOI-13.01 or WASP-30b, which could also potentially
offer useful constraints, but their eccentricities are not well enough measured to be useful at present.


The most interesting object in the massive planet sample is found in the binary star system HD41004, although it 
was excluded from the analysis of H10 because the host star mass was lower than the
calibrated range. Santos et al. (2002) originally monitored the spectrum of the A component, a K~star, for
radial velocity periodicity. A signal was detected but, on the basis of a correlation
with variability in the line bisector of the combined spectrum, it was deduced that the host
 was actually the fainter M-dwarf component of the system, HD41004~B. 
 This has since been confimed
by Zucker et al. (2003,2004), using cross-correlation methods developed for double-line spectroscopic
binaries. The remarkable aspect of this system is that the companion is itself a potential brown dwarf
($\sim 18 M_J$) in a very short period orbit (1.3~days). Santos et al. infer an age for the K-dwarf
companion of 1.6~Gyr, based on chromospheric activity. Thus, we will allow models of ages 1--2~Gyr.

This system is an excellent test of our model because the M dwarf is almost fully convective, so that the
dissipation now occurs throughout the star. Furthermore, the high mass of the companion and the short period
orbit guarantee significant tidal interactions. Despite this, the planet is well outside the location
where it is likely to be swallowed within the age of the system -- orbital periods as low as 8~hours would
survive for $>1$~Gyr with our final stellar dissipation calibration. 

However, the measured eccentricity proves problematic -- dissipation at the calibrated levels in the
convective brown dwarf companion should have reduced the eccentricity below the limit measured by
Zucker et al. (2004). Figure~\ref{HD41004} shows the allowed separation-eccentricity space for the
HD~41004~B system. We require $\nu_0<0.5$ to match the observed eccentricity, which is formally
inconsistent with our constraint from \S~\ref{VPsec}.

Unfortunately, the complexity of the system limits the degree to which we can really use this as a 
constraint. Zucker et al. (2004) note that the stellar companion is close enough that it could conceivably
provide sufficient perturbations to excite the eccentricity, and also find hints of a possible linear
trend in the data that may betray an additional perturbing companion.

Given the possibility of external perturbations, we restrict ourselves to deriving a characteristic timescale for circularisation
in this system. We repeat our above fitting procedure, but assume that the system is 1.5~Gyr at the start
(this results in a smaller companion radius and lower dissipation rate in the planet). We find that the
characteristic timescale to damp $e<0.05$ at P=1.3~days is $\sim 0.07$~Gyr, for an assumed value $\nu_0=1.5$.  


\subsubsection{Planets that define the period-eccentricity envelope}
\label{Envelope}

The massive, short period planets are the best probes of the strength of the stellar tide. The strength of
the planetary tide is best constrained by those systems that define the edge of the period-eccentricity distribution
at larger periods and eccentricities, where the stellar tide is negligible. In Table~\ref{Tab3} we collect several systems that trace this edge, selecting
 planets with masses $>0.1 M_J$ (smaller planets could be potentially rock or ice-dominated, leading to possibly different
dissipation processes). We will fit to the
$2 \sigma$ limits on the eccentricity, and therefore only select systems that have non-zero eccentricities at this level.
We  probe a range
of masses, from 0.2~$M_J$ to 3.2$M_J$ (in addition to some of the more massive systems discussed in \S~\ref{BigHosts}).

The short period end of the envelope is fixed by the systems HAT-P-16b, WASP-10b and WASP-6b. HAT-P-16b (Bucchave et al. 2010)
is not massive enough to make the sample in Table~\ref{Tab1}, but with a planetary mass $4.2 M_J$ it is still more massive
than average. The original non-zero eccentricity is confirmed by Pont et al. The age is also sufficiently well constrained
($2 \pm 0.8$ Gyr) that our models restrict $\nu_0<2.5$. The constraint is shown in Figure~\ref{HATP16}.

The constraints based on systems WASP-10b and WASP-6b are less clear.
 Christian et al. (2009) remark
on the `surprising' finite eccentricity of the WASP-10b orbit, but we find that the 2$\sigma$ limit on the eccentricity
can accomodate $\nu_0 < 2$, assuming an age $>2$~Gyr.  WASP-6b is a similar system (Gillon et al. 2009),
which means that it provides a stricter limit at 2$\sigma$ ($\nu_0<1.5$), even though the error bars are larger. However, the Husnoo
et al. (2012) re-analysis places  weaker constraints on
the eccentricity for these systems ($e<0.11$ and $e<0.08$ respectively), which would allow larger values of $\nu_0$. Nevertheless,
it is encouraging that even the marginally detected eccentricities are consistent with the constraints from the better constrained
system HAT-P-16b.

Moving further up the envelope, we encounter three systems (HD118203, HD185269 and WASP-8) clustered in close proximity.
HD118203b is a 2.2$M_J$ planet orbiting a K subgiant (Da Silva et al. 2006). The slightly evolved status affords us a reasonable
constraint on the age (4.6$\pm$0.8 Gyr). Once again, the eccentricity and separation can be matched with $\nu_0<4$.
HD185269b is also a subgiant (Johnson et al. 2006) and also allows much larger planetary dissipation, with $\nu_0<15$. This is, in part,
due to the weaker bounds on the eccentricity. However, 
WASP-8b (Queloz et al. 2010) offers a similarly weak constraint, even though the eccentricity is much better constrained.
Despite the similarities in $\nu_0$,  the planet mass is a factor of two larger and stellar mass 
significantly larger too. Figure~\ref{WASP8} shows the position of the system relative to the
$\nu_0=1.5$ envelope.

At the high eccentricity end of the envelope, we have three systems HD108147b (Pepe et al. 2002), COROT-10b (Bonomo et al. 2010) and HD17156b (Fischer et al. 2007; Barbieri et al. 2009).
HD108147b has substantial error bars on the eccentricity, so that it allows $\nu_0<10$. For COROT-10b, the
error bars are smaller, but the star has a large allowed age range (1--6 Gyr), which allows
an even larger range $\nu_0<20$. HD~17156b is also consistent with $\nu_0<10$.

\subsubsection{Synthesis}
\label{Synthesis}

Figure~\ref{Nup} summarizes the constraints on $\nu_0$. The strongest constraints come from the shorter period planets,
although the eccentricities in some of these systems may be overestimated. The most solid constraint appears to be from
the planet HAT-P-16b.  When combined with the constraints from \S~\ref{VPsec}, we find a fairly tight constraint
$1.3 < \nu_0 < 2.5$, leading us to adopt a canonical value of $\nu_0 = 1.5$.
 In a fully consistent model, this value applies for both the
stars and the planets, but we have already noted that the physical foundation for the model is better motivated in the case
of the slowly rotating host stars than it is for the planets. Therefore, we note that Figure~\ref{Nup} does potentially allow
for a factor of several increase in the strength of $\nu_0$ in planets with forcing periods longer than six days, which might indicate the action of enhanced dissipation due to a dynamical tide.

In order to compare this number with prior estimates, we note that $\nu_0=1.5$ leads to $\bar{\sigma}_* \sim 4.5 \times 10^{-7}$
for a middle-aged, solar mass star and forcing periods $\sim 1$~day. Putting this number into equation~(\ref{Torb}) yields a
characteristic orbital decay timescale of 1.4~Gyr, assuming a Jupiter-mass perturber in a 1~day period orbit. If we estimate
the same quantity using the formulae in Zahn (1977), we get 0.5~Gyr, so that the numbers are quite comparable. Another commonly
quoted estimate is that of Rasio et al. (1996), which appears to be much larger at first glance. However, they used the Goldreich-Nicholson
scaling for the coupling efficiency of turbulent convection. If we repeat their estimate, but restrict the coupling to the upper $6 \times 10^{-3} M_{\odot}$
of the convective envelope (the part that couples efficiently in the Zahn prescription), then we get an equivalent timescale of 0.6~Gyr.
Thus, our estimated viscosity agrees well with historical estimates, in the case of the planet host stars.

In order to compare our dissipation model to prior estimates of planetary dissipation, we estimate the equivalent tidal $Q'_p$ for a 1~Gyr old, Jupiter mass
model, with a forcing period of 10~hours to be $Q'_p \sim 10^8$. This lies in between the estimates of Hubbard (1974), which assumed full coupling to the
convection, and that of Goldreich \& Nicholson (1977), which assumed a quadratic efficiency coupling. Nevertheless, the physical picture is still in agreement
with Goldreich \& Nicholson, in that
the outer 3\% of the mass dominates the viscosity, and the numerical value derived is not large enough to explain the features of the Laplace resonance by
simple tidal evolution. Another useful comparison is to the results of Ivanov \& Papaloizou (2004), who evaluate the strength of the equilibrium tide in
the case of a planet on a highly eccentric orbit. They estimate the orbital evolution timescale as $\sim 3 \times 10^{10}$ years, for a $1 M_J$ planet
in a three-day orbit about a $1 M_{\odot}$ star. Our estimate for the equivalent situation is $\sim 2 \times 10^{10}$ years, in excellent agreement.

Our numbers also agree well with estimates based on numerical simulations. Penev \& Sasselov (2011) estimate stellar $Q'_*$ based on the numerical
convection simulations of Penev et al. (2009, 2011), and argue for $Q_* \sim 10^8$--$3 \times 10^9$.
 Figure~\ref{QQ} shows
the present-day Q' estimated for the stars and planets in systems with orbital periods $< 3$~days, calculated in our model assuming $\nu_0=1.5$. For most
stars, $10^7 < Q'_* < 10^8$, although it can be smaller for individual systems. The inferred advanced evolutionary
state of the host star of WASP-48b, for example, implies a larger convective zone and $Q'_*<10^7$. These numbers are consistent with the results of Penev \& Sasselov (2011),
given the suggested accuracy of their estimate.
 The estimated planetary $Q'_p$
are also smaller, with many $Q'_p<10^7$, although still well in excess of the traditional estimate based on scalings from Jupiter.

\section{Discussion}
\label{Discuss}

The result of the new calibration is that we can produce a consistent description of tidal dissipation
in convection zones, with $\nu_0 \sim 1.5$. This is consistent with the properties of binaries containing
red giants and compact objects, massive exoplanets in short orbital periods around main sequence stars
as well as the observed period-eccentricity relation of exoplanets around both main sequence stars and
subgiants. This calibration has a substantial advantage over that in H10, as it can now self-consistently
describe the full range of stellar hosts, including low mass stars and evolved stars.

\subsection{Short Period Orbits}

The most direct consequence of the stellar tide is to limit the lifetime of planets in close orbits,
even if the orbit is circular (which rules out any dissipation in the planet). In Figure~\ref{StabLim}
we show the initial orbital period that affords a planet of a given mass a lifetime of 1~Gyr (solid line)
or 0.1~Gyr (dotted line), around a 1$ M_{\odot}$ star. Comparing this to the observed planets (solid and open
circles) and Kepler planet candidates (crosses) shows that
most of the known planets are stable against being swallowed by their host stars during the observed system
ages, and that the observed distribution appears to show an edge consistent with the 1~Gyr criterion. 
There is a handful of systems that populate a tail to shorter lifetimes, with WASP-19b, WASP-18b, WASP-12b
and COROT-14b being those with the shortest remaining lifetimes. Nevertheless, we can demonstrate that the 
system parameters are consistent with the estimated ages of the systems, and that these are simply the tail
that has been caught during the final, accelerating inspiral. A similar conclusion was reached by Brown et al. (2011)
for WASP-18b and WASP-19b (see also Hellier et al. 2011).
 Figure~\ref{4Ex} shows example orbital
histories for these for planets, assuming $\nu_0=1.5$. 

The strength of the stellar tide depends on the structure of the star as well, but changes the period
limits in Figure~\ref{StabLim} by $\sim 10$\% in either direction, as long as the main sequence star
has a mass $0.8 M_{\odot} < M_* < 1.4 M_{\odot}$. For lower mass stars, the limiting periods move to
lower values, so that planets can survive at significantly shorter periods around stars with masses $<0.8 M_{\odot}$.
The planet WASP-43b is an example of this. With an orbital period less than 1~day, this planet would have
a survival time $< 10^8$ years around a solar mass star (see Figure~\ref{StabLim}), but has an estimated
survival time $\sim 2$~Gyr, given the observed properties of its host (Gillon et al. 2012). Similarly,
the planetary system around KOI-961 illustrates that planets with orbital periods as short as 0.45 days (Muirhead et al. 2012) are stable around
M dwarfs.

Also evident in Figure~\ref{StabLim} is the importance of the Roche limit. Since the inspiral time depends
on the properties of the star, the only planetary parameter of relevance is the mass, and the solid and
dotted lines in Figure~\ref{StabLim} make no assumptions about planetary structure. However, the lower
dashed line indicates the period at which a given mass planet will overflow it's Roche lobe, assuming a
uniform radius of $1 R_J$ for all planets. Furthermore, if we allow for the planets to be placed on these
orbits while still hot (either as a result of youth or heating due to tidal circularisation) the limit is
even stronger, as shown by the upper dashed line. This suggests that the observed division between
planets of mass $\sim 1 M_J$ and those of mass $<0.1 M_J$ at orbital periods less than 3 days may be due
to the fact that Roche lobe overflow sets a limit to how close young, extended gaseous planets can get
to the star. The lower mass ($<0.1 M_J$) planet population orbits below this line because they are
enriched in heavy elements and are therefore denser.

\subsubsection{Period Derivatives}

The same physics can be reflected in the current rates of period change, and so
 we have examined the current rate of orbital period
decay for all systems with orbital periods $<1.5$~days plus a few others of clear interest. The
data for these is collected in Table~\ref{Tab2}. For a given system, evolution models yield some variation 
depending on initial conditions, but the results are found to be within 30\% of the rate estimated by simply
dividing the observed period by the inspiral time given in equation~(\ref{Torb}), as long as one uses the
appropriately calibrated $\bar{\sigma}_*$. WASP-18b is the most promising system in this regard, with
an expected
rate of orbital period change of 
$\dot{P} \sim 0.5 \, {\rm ms \, yr^{-1}}$. 
The current stated precision of eclipse timing is $\sim 75$ milliseconds for WASP-18b, suggesting that it
will take $\sim 150$ years to detect inspiral unless the precision improves. Recently, a series of 21 lightcurves
obtained over 6 years was obtained for OGLE-TR-56b (Adams et al. 2011). They revised the system parameters and
place a limit of $\dot{P} = -2.9 \pm 17 \, {\rm ms \, yr^{-1}}$ for this system. Our expectation is 0.6~$ \rm ms \, yr^{-1}$
and therefore very consistent with these numbers. Unfortunately, the time it would take for this to become observable
at current precisions is $\sim 30$~years.

\subsection{Misaligned Orbits}
\label{RM}

Winn et al (2010a) have demonstrated that  the degree of orbital misalignment
of various exoplanets with host star spin appears to correlate with the effective temperature
of the host star. They note that this correlation suggests that the larger surface convective 
zones of lower mass/cooler main sequence stars produce stronger tidal coupling to the planets
and reduces the planetary obliquity. Their model claims, however, to require a degree of
decoupling between the surface convection zone and the radiative region below, and possibly
also the action of a magnetic wind to remove angular momentum. 

In order to apply our newly calibrated model to this issue, we first note that
tidal dissipation in the planet does not change the orbital plane, so we focus here only on the tidal
dissipation in the host star. Figure~\ref{Tl} shows the characteristic timescale to change the orbital semi-major
axis due to
tidal dissipation in the star (equation~\ref{Torb}), for all systems for which a misalignment angle has been measured.
In each case we use the current orbital period as the forcing period for each system. We
see that only the WASP-18b, WASP-19b, CoRoT-18b and HAT-P-23b planets are expected to have undergone significant orbital evolution
in response to the spin-up of the central star. On the other hand, the star itself often has a smaller moment
of inertia than the planetary orbit, and the stellar spin-up time $T_{spin}$ (equation~\ref{Tspin}) can differ from the orbital evolution
time by an order of magnitude or more for some systems, despite some claims made in the literature. Therefore, Figure~\ref{Tsp}
 shows the misalignment plotted against $T_{spin}$ for each star-planet system. 
There are 12 observed systems with  $T_{spin}<10^{10}$ years, and none of the retrograde
systems are among them. The largest misalignment of this group is that of XO-3b. These characteristic times are calculated assuming a slow rotation for the
star, regardless of the observed rotation rate, as the goal is to identify systems for which the tidal interactions are expected to adjust the stellar spin.
In the case of XO-3b, the host star is observed to be rotating almost synchonously (Johns-Krull et al. 2008), as one would expect from this estimate.

A consequence of tidal spin-up is that the stellar spin may be increased significantly. Figure~\ref{LV2} shows the
measured $V \sin i$ versus $\lambda$ for each of the observed systems. We also show three sample theoretical trajectories, calculated using equation~4 of H10.
There is some ambiguity in performing the comparison between theory and observation because only the sky-projected
angle and velocities are measured (see Winn \& Fabrycky 2010 for an exposition of the difference between $\Psi$ and $\lambda$).
We show a sample trajectory for WASP-18b, assuming that the system was initially rotating retrograde. The short value
of $T_{spin}$ for this system indicates that stellar tides could have completely reversed the stellar spin.
We also sample evolutions of COROT-1b and XO-3b, two moderately misaligned systems with $T_{spin} \sim 10^{10}$ years or less. For stars
with a significant initial spin perpendicular to the orbital plane, it is impossible to completely align the orbit without significant 
decrease of the semi-major axis, 
 as the tides only affect the stellar spin parallel to the orbit. Furthermore, the $V \sin i$ measured for the star
contains the entire velocity associated with the projection of the stellar spin vector on the sky, which can contain a
component of the spin perpendicular to the orbit as well. In such cases, the reversal of the parallel spin does not pass 
through $V \sin i=0$. Despite these uncertainties, Figure~\ref{LV2} demonstrates the character of the spin-misalignment evolution
in this model. The degree of spin-up required is not necessarily enough to dramatically alter the stellar parameters, as we
can see from Figure~\ref{LV2} that $V \sin i > 8 km/s$ can be found for retrograde systems, for which there is little spin-up.
Pont (2009) and Husnoo et al. (2012) have claimed evidence for excess rotation in several planet host stars associated with tidal interactions. For seven of
these systems (CoRoT-2b, CoRoT-18b, HAT-P-2b, HAT-P-20b, WASP-19b, WASP-43b and XO-3b) our spin-up times are consistent with some alteration of stellar spin, although
three others  (CoRoT-3b, HD189733b and $\tau$~Boo~b) appear to be too weakly coupled to have evolved significantly.

Previous attempts at modelling these effects (Barker \& Ogilvie 2009; Winn et al. 2010a) have emphasized the importance of magnetic stellar winds
in spinning down the host stars. However, this is necessary primarily because the strength of the tidal interaction assumed in those studies
was greater than is found here, so that a spin-down mechanism was needed in order to reduce the expected stellar spin-up to the observed level.
To demonstrate this, we have repeated our spin-up calculations using the spin-down law from Barker \& Ogilvie
\begin{equation}
\frac{d \Omega_{\perp, \parallel}}{dt} = -1.5 \times 10^{-14} \gamma \,  \Omega^2 \Omega_{\perp,\parallel} \, {\rm years}
\end{equation}
(where $\gamma=1$ for G,K,M stars, and $\gamma = 0.1$ for F stars), and found no change in our conclusions. This is because the effects of the magnetic wind are
weak when the spins are slow. We can derive a magnetic breaking timescale equivalent to equation~(\ref{Tspin}),
\begin{equation}
T_{mb} = 1.3 \times 10^9 years \, \frac{1}{\gamma} \left( \frac{P_{rot}}{30 days} \right)^2
\end{equation}
which is plotted as a band in Figure~\ref{Tsp}, and shows that the bulk of the aligned systems should experience an approximate equilibrium between
tidal spin-up and magnetic breaking. This braking will also not affect the trajectories in Figure~\ref{LV2}, but only the rate of motion
along the trajectories.

Thus, with our calibration of the tidal dissipation, the correlation noted by Winn et al. (2010a) can be consistently interpreted as
demonstrating the effect of tides primarily on the stellar spin, without significant orbital evolution in most cases. Only four
systems in this sample are likely to have undergone much orbital evolution, but as many as 12 are likely to have experienced
some modifications of their spins.
 The fact that 
 hotter stellar temperatures appear to favour misaligned orbits is consistent with the reduced amount of dissipation in the thin surface convection
zones, making it more difficult to change the spin of the star. Winn et al also proposed a model in which the tides couple only to the stellar
surface convection zone, out of concern that the moment of inertia of the entire star would prove to be too large a sink of angular momentum,
causing an unacceptably rapid rate of inspiral. With our calibration this concern proves to be unfounded. Indeed, if we use the moment of inertia
of just the convection zones to calculate equation~(\ref{Tspin}), shown in Figure~\ref{Tsp2}, many of the misaligned systems exhibit unacceptably short synchronisation times -- an efficient transfer of angular
momentum from the dissipative regions to the radiative interior is actually an essential requirement for consistency in this model.

Triaud (2011) has recently claimed a correlation between the degree of misalignment and the age of observed systems, but only 5 of the 22 systems in that
sample have $T_{spin} < 10^{10}$~years, and 4 of those are in the young group. As such, if the misalignment is being removed over time, it is not due to
the action of stellar tides. Lai (2011) has addressed the question of misalignment versus orbital decay by postulating that planets on retrograde orbits can couple
more easily to stellar internal modes than those on prograde ones, allowing for more rapid evolution in obliquity than semi-major axis. Although we have shown
that the difference in the stellar and orbital moments of inertia is sufficient to allow for a consistent evolution, the sample in Figure~\ref{Tsp} does show
hints of a slope, in the sense that most of the retrograde systems have $T_{spin} > 10^{12}$ years, whereas one might expect the edge of the distribution to
lie closer to $10^{10}$ years. If real, this leaves some room for stronger dissipation in retrograde systems.

\subsection{Hot Stars and Subgiants}
\label{HotSubs}

Section~\ref{RM} assumed that most host stars initially had relatively slow rotation, but some of the
known transitting planets orbit hot stars, whose spin rates are significantly higher than their
cooler counterparts. This can lead to qualitatively different behaviour, because
for rapidly spinning stars, the transfer of angular momentum can be outwards, from star to
planetary orbit. The transitting planet WASP-33b serves as a
demonstration of this.  The host star is a main sequence A star, with rotational velocity $\sim 90 km/s$
(Collier Cameron et al. 2010), so that radial velocity precision is severely compromised through line
dilution (because of the high effective temperature) and broadening (high velocity).

We model the evolution of this system in tandem with the rotational
evolution of the host star, including radial expansion and evolution of the moment of inertia.
We assume angular momentum conservation as a solid body during the expansion (because the winds from hot stars are weak). The transfer
of angular momentum from the rapidly rotating star to the planet causes the planet to move outward
in semi-major axis. As the star evolves and swells, the spin decreases and eventually the tidal coupling
causes the planet to spiral inwards. The location of the orbital turnaround point depends on 
the planet mass. For planets $\sim 1 M_J$, the maximum orbital period is $\sim 1.4$ days, but it gets
out to 2.7~days for $3 M_J$ planets and could get as large as 5 days if the planet were $\sim 10 M_J$ (although
Collier Cameron et al. place a limit of 4.1$M_J$ on this particular object).
The planet is eventually swallowed during the ascent of the giant branch, hastened by the fact that
the transfer of angular momentum from the planet orbit to the star is dwarfed by the increase in the
moment of inertia as the star evolves, so that the late-time inspiral of the planet has a negligible
spin-up effect on the star.

The spin-down/cooling of hot stars during their evolution to the giant branch has been utilised by several
groups to probe a higher mass range for stellar hosts (Sato et al. 2003; Hatzes et al. 2005; Johnson et al. 2006, 2007;
Lovis \& Mayor 2007; Liu et al 2008; Sato et al 2008; de Medeiros et al. 2009; Johnson et al. 2011). This has led to claims of
higher mass planets around higher mass hosts and a paucity of short orbital period planets. With our
calibrated model we can now evaluate the tidal stability limit for stars of different evolutionary
stages. The results are shown in Figure~\ref{Ra} for host stars of 1.2 and 1.5$M_{\odot}$. Two cases
are shown for the latter mass -- one assuming an initially slowly rotating star (30 day rotation period)
and another a rapidly rotating star (1 day rotation period). We see that tidal instability can explain the
paucity of companions to subgiants interior to 0.1~AU (Sato et al. 2008; Hansen 2010; Kunitomo et al. 2011). This effect is stronger for lower
mass stars, in the sense that the planets with semi-major axis $<0.1 AU$ are swallowed at smaller stellar
radius than for higher mass stars, and may therefore lend support to the notion that
the observed subgiant population is contaminated with lower mass stars (Lloyd 2011). However, there
is little difference in the implications for planets from 0.1--0.6~AU, so that tidal forces do not appear
to be the explanation for the lack of planets in this regime. The region out to 0.6~AU can be cleared
by tidal engulfment by the time the red giant tip is reached, in agreement with Kunitomo et al. 2011, but
the paucity of planets in this middle region around ascending stars ($R_* \sim 10 R_{\odot}$) is not
obviously explained by this mechanism. 

\subsection{Scattering, Tidal Inflation and the Roche Limit}
\label{Overflow}

Tides are often implicated in the origins of hot Jupiter short orbital periods.  Recent measurements
of orbital misalignment (Winn et al. 2010a and references therein) have generated a lot of interest in
 models in which the planets are initially on orbits with large semi-major axis but with high
eccentricities. If the periastron is small enough, the tidal forces at periastron will eventually
circularise the orbit and capture the planet into a short period orbit (Rasio \& Ford 1996; Weidenschilling \& Marzari 1996;
Wu \& Murray 2003; Takeda \& Rasio 2005; Ford \& Rasio 2006; Fabrycky \& Tremaine 2007; Nagasawa, Ida \& Bessho 2008; Chatterjee et al. 2008).

In order to assess this issue within the context of our calibration, we adopt a canonical
hot Jupiter system as a $1 M_{\odot}$ star,  with a 1$M_J$ planet in orbit around it, which has
evolved for 3~Gyr.
The left panel of Figure~\ref{xoeo}
shows the range of initial semi-major axis and eccentricity that yield final, circular orbits (
defined as $e<0.05$). At large semi-major axis, we see a horizontal band that represents planets
too far from the star to have their orbits appreciably altered by tides. Interior to this, we see
a horn extending up to larger eccentricity, representing those planets whose orbits have been circularised
by tides. We also show a dotted curve indicating the edge of the parameter space that leads to surviving planets, even
if they are not completely circularised. To the left of this line lies the region in which planets 
overflow their Roche lobes before completing their evolution. 
Orbits that start at higher eccentricities, while remaining to the right of the dotted line, do survive, but
are not completely circularised. The edges of this wedge are essentially defined by a fixed periastron
$q = 0.025$~AU on the left, and a fixed specific angular momentum (equivalent to a final circularised radius
of 0.038~AU) on the right. 

As we have discussed, the expected pseudo-synchronisation of the planets in these systems raises the possibility 
of exciting internal modes which may enhance the tidal dissipation. To examine the consequences
of such an enhancement, the right hand side of Figure~\ref{xoeo} shows a model in which we keep $\nu_0=1.5$ for
the star, but increase $\nu_0=15$ for the planet. We see that the character of the solutions remain the same
although the upper end of the eccentricity distribution moves to slightly higher values. The edges of the wedge
are now defined by a periastron of 0.035~AU and a final circularised radius of 0.054~AU. 

This tail of high eccentricity initial conditions defines a characteristic range of final circularisation periods,
and therefore represents a specific realisation of the 
 hypothesis of Faber et al. (2005), that the pile-up at three days represents the circularisation of eccentric
orbits, limited at short periods by Roche lobe overflow. However, within the context of our model, this does not
extend to origins at large semi-major axis, as planets starting at $a>0.1$~AU are heated to the point of runaway
inflation and disruption. Figure~\ref{Edot} illustrates this point by showing the tidal luminosity in a 1~$M_J$
planet as a function of eccentricity, assuming either a fixed periastron (dotted line) or angular momentum
conservation (solid line). We see that the reason the inner edge in Figure~\ref{xoeo} is defined by a line of
constant periastron is that this luminosity asymptotes at large eccentricity, so that it will naturally map
onto the critical value of tidal dissipation needed to overwhelm cooling in a particular planet. The fact that the luminosity along
the circularisation trajectory continues to increase at large eccentricity means that there is a critical value
below which circularisation is stable against tidal inflation. One other important element in this calculation is
the quasi-linear dependance of the dissipation rate on orbital period in this model (see Figure~\ref{Sigp}). If we fix the dissipation to the value at
forcing period of the final circular orbit, the luminosity along the constant angular momentum curve is given by the dashed line. Thus,
the tidal inflation is driven, in part, by the increased coupling to the internal viscosity at longer forcing periods. Figure~\ref{xo2}
demonstrates this, by showing an equivalent plot to those in Figure~\ref{xoeo}, but now with the dissipation fixed to the value
calculated at a forcing period of 1~day (and with $\nu_0=1.5$). We see that high eccentricity planets are now circularised without
disruption. Thus, it is the increased efficiency of coupling at longer orbital periods that drives the planets to disruption.
In \S~\ref{Synthesis} we show that our calibration is also consistent with the estimates for the equilibrium tide calculated
in the parabolic approximation by 
Ivanov \& Papaloizou (2004), which suggests that our model remains viable even for very eccentric orbits.

To illustrate the planet behaviour in more detail, we derive an approximate formula based on fitting to evolutionary models. For
planets of age $\sim 10^8$ years old, we fit the normalised dissipation, as function of mass,
for this sequence of models as $\bar{\sigma} \sim 1.9 \times 10^{-7} (M_p/M_J)^{0.63} (P/1 day)$. With this model, we can estimate
the level of dissipation in each circularising planet. Our models suggest circularisation is enhanced somewhat by planetary radius
inflation due to tidal heating, so we adopt planet radii $\sim 1.3 R_J$ and estimate the characteristic circularisation locus
by requiring a circularisation time of $\sim 5 \times 10^9$~years. This yields the criterion
\begin{equation}
a_f = 0.0393 AU \left( \frac{M_p}{M_J} \right)^{-0.057} \left( \frac{M_*}{M_{\odot}} \right)^{0.023}. \label{Afinal}
\end{equation}
Figure~\ref{Ma} shows that this passes through the peak of the planet distribution, providing an explanation for the
`three-day pileup'. Figure~\ref{Ma} also shows a dashed curve which represents the locus of planets that circularise
within $10^8$ years from a starting point determined by setting $R=1.6 R_J$ and the total dissipation from tidal heating
to $\sim 10^{-6} L_{\odot}$. This represents the edge of runaway heating and disruption, as shown in Figure~\ref{xoeo}
for example. We see that this traces the inner edge of the planetary distribution quite well. It lies close to the
criterion of Ford \& Rasio (2006), for circularisation to twice the Roche limit. The two are related, although our
criterion is based on a limit to how much heating a planet can absorb without disruption.
Finally we show the 1~Gyr 
survival line versus stellar tides (see Figure~\ref{StabLim}) as well. These three criteria seem to encompass the bulk of the observed distribution.
The distribution of circularised orbits does appear to extend beyond the solid line in Figure~\ref{Ma}. As noted before, our calibration
does leave room for additional dissipation at high eccentricities (possibly due to a dynamical tide contribution). Such enhanced levels of
dissipation will circularise orbits from further out, if they don't disrupt. Alternatively, not all planets on circular orbits need have
been tidally captured, if additional migration mechanisms are at work.

Socrates et al. (2011) suggest an observational test of the model in which planets on initially highly eccentric orbits are circularised at short
orbital periods. With a sufficiently large observational sample (such as that obtained by the Kepler spacecraft), an assumption of steady state
evolution predicts the distribution of planets on high eccentricities. Our results indicate that the evolutionary tracks that lead to the hot
Jupiters should be depleted for $e>0.6$. 


\section{Conclusions}
\label{Conclusion}

The preceding discussion demonstrates that one can obtain a self-consistent description of tidal dissipation
in exoplanet systems using the simple model of turbulent dissipation in convection zones. The level of dissipation
expected in the host stars is consistent with expectations from forcing periods $\sim$~1~day to $\sim 100$~days
(if one includes the red giant binaries). This consistency is achieved by assuming the Zahn approximation for
the coupling efficiency, which is also supported by some numerical simulations (Penev, Barranco \& Sasselov 2009; 2011)

The resulting calibration also provides a consistent description of various observations regarding the nature of
exoplanet orbits. We find that the stellar tides are weak enough that only a few known systems have orbits which
are likely to have been significantly affected, but that there are several more whose stellar spins are likely to have been altered
enough to reduce the relative obliquity. Recent observations regarding the properties of planetary systems with high
obliquities are consistent with this model in that all systems which possess short enough spin-up times are found to have
low obliquity. It is also encouraging that this model does not require a decoupling between the surface convection zone
and the interior of the star. A further implication is that the distinction between misaligned and aligned orbits is not
solely due to the action of tides, because there is a distinct population of planets whose orbits are aligned despite 
having long alignment times.
Figure~\ref{nosync} shows the resulting period-eccentricity relations when we separate prograde and retrograde orbits,
excluding those systems for which obliquities could have been tidally altered.
 With the sole exception of HD~17156b, the
systems whose orbits and spins are primordially prograde have almost circular orbits, in contrast to the systems which are
either primordially retrograde or whose primordial obliquities have been potentially compromised. This suggests that
retrograde orbits might also be preferentially eccentric, and that there are at least two different pathways to forming
short period planetary systems. A similar conclusion was reached by Schlaufman (2010), although we draw different divisions
between the two samples. Furthermore, this separation indicates that eccentricity pumping by rapidly rotating pre-main sequence
host stars (Dobbs-Dixon, Lin \& Mardling 2004) is unlikely to explain the observations, as this would produce aligned, prograde eccentric planetary orbits.

We have also modelled the increase in tidal dissipation in stars as they evolve towards the giant branch, and find that
planetary systems with initial semi-major axes $<0.1$~AU will be swallowed during the subgiant stage and that the tidal
swallowing distance can reach $> 1$~AU during later evolutionary stages. The distribution of
observed systems suggests that many hosts may have masses closer to $1.2 M_{\odot}$ than $1.5 M_{\odot}$, in that the frequency
of such systems drops off sharply beyond the theoretical limit of the former case.

Convective turbulence may also play a role in providing the tidal dissipation in the planets. Our calibration is consistent
with the same normalisation being applicable to both host stars and giant planets. However, our constraints are such
that we can accommodate a factor of
several higher normalisation in the planets, leaving open the possibility of a dynamical tide contribution. 
The amount of tidal heating that accompanies circularisation proves to be
a limiting factor, as planets with initially high eccentricity orbits are often significantly inflated, to the point of undergoing
Roche lobe overflow. Our results suggest that the pile-up at $\sim $3~day orbital periods can arise from the circularisation of
initially eccentric planets, but only those
with initially $e <0.5$. This restriction is based on the assumption that Roche lobe overflow is catastrophic, and may be
relaxed if this is not the case. Furthermore, an important part of the catastrophic heating is the increased efficiency of coupling
between the turbulent dissipation and forcing period for long period eccentric orbits. If this is reduced below what is expected
from our model, then planets can survive the circularisation from further out. Nevertheless, the data do seem to support the notion
of Faber et al. (2005), that the pile-up of hot Jupiter orbital periods is the result of tidal capture from eccentric orbits, albeit
modified by some restrictions placed by tidal heating and Roche lobe overflow that restrict the initial semi-major axis to $<0.1$~AU.

The fact that the orbital distribution is consistent with a similar level of dissipation in both planets and stars is
somewhat surprising. The bulk of the exoplanet host stars are slowly rotating, but we expect most planets to have spins
synchronised to the orbital period, which increases the likelihood of coupling internal dynamical modes to the forcing
frequencies and increasing the level of dissipation. Some enhancement in dissipation in the planets is still possible, as
we have shown that such an order of magnitude increase in dissipation moves the circularisation period out from 2.7~days to 4.6~days,
which approximately covers the observed pile-up of orbital periods. However, substantial increases in dissipation will likely
exacerbate the tendency of planets on high eccentricity orbits to heat up and expand until they overflow their Roche lobes.
This is an issue that clearly deserves greater attention in the future.

\acknowledgements 
This research has made use of NASA's Astrophysics Data System Bibliographic Services, the Exoplanet Orbit Database
and the Exoplanet Data Explorer at exoplanets.org, and  René Heller's Holt-Rossiter-McLaughlin Encyclopaedia at www.aip.de.
The author would like to thank Norm Murray and Yanqin Wu for illuminating discussions.

\newpage

\appendix

\section{The response of a planet to heating}
\label{Planet_Heat}

The dissipation of tidal energy in the planet can far exceed the internal
luminosity for some parts of parameter space. This level of energy input can
significantly affect the structure of a planet. In particular, since radius is
intimately linked with the internal entropy, we expect heating to be accompanied 
by radius inflation. 

Within the context of the convective dissipation model adopted in \S~\ref{Model}, we can
calculate which part of the planet contributes most to the dissipation. Figure~\ref{SigInt}
shows the accumulation of the integral~(\ref{sigma_integral}) as a function of pressure
for a 1~$M_J$ planet, of age $10^8$~years, experiencing a tidal perturbation with frequency
 10 days, assuming the Zahn efficiency coupling. These parameters are chosen to represent the conditions for an eccentric Jupiter-mass
planet which will end up at a few days orbital period once circularised. We see that the integral
is dominated by the outer layers of the star, because that is where the convective eddy turnover
times are comparable to the forcing period and hence the coupling is strongest.

Figure~\ref{3Lums} shows the effect of enforcing a heating of different prescribed levels in
the same planet as represented in Figure~\ref{SigInt}, in which the heating is restricted
to the part of the planet with $P < 10^{10} dyn.cm^{-2}$. We see that above luminosities $\sim
5 \times 10^{-6} L_{\odot}$, this leads to an unstable situation, in which the amount of dissipated
heat is too much for the outer layers to transport without considerable radial expansion. If we assume
the same total level of dissipation, but spread equally over the planet mass, the radius expansion is
 not as large, asymptoting to a finite radius within a Kelvin-Helmholtz time.

A fully self-consistent model would integrate each evolutionary curve while simultaneously heating
the planet according to the appropriate dissipation profile. However, this is not computationally
 feasible, given the large parameter space we wish to survey. Instead, we adopt the approximation
of uniform heating of the planet, which allows us to approximate planet evolution under tidal heating
to a simple reversal of the cooling curve. The results shown here indicate that this is the conservative
assumption, since a realistic dissipation profile is more likely to result in runaway inflation and
tidal disruption (see \S~\ref{Overflow} for why this is then the most conservative option). 

 Figure~\ref{LTC} shows the our model for a 1 Jupiter mass planet -- taken to significantly hotter levels
 than most planet models, even so-called 'hot start' models. 
If the planet is heated too much it
can undergo Roche lobe overflow (e.g. Gu, Bodenheimer \& Lin 2003).
If the radius expands to the point where the planet begins to overflow it's Roche
lobe, we assume the result is catastrophic disruption of the planet. Some authors (e.g. Gu et al. 2003) have
attempted to describe the varieties of possible subsequent evolution, but we have opted to assume complete disruption of such planets
to keep the calibrated model simple. This assumption has some physical basis in that
 the mass-radius relation of
heated planets results in expansion in the event of mass loss, and the effects of a fixed amount of
tidal dissipation are greater for lower mass planets. Furthermore, the heating can lead to an unstable
feedback, in that the heating rate itself is $\propto R_p^{10}$, so that radius expansion generates an
increase in dissipation and an unstable feedback.

\clearpage

\begin{deluxetable}{lccccccccc}
\tablecolumns{10}
\tablewidth{0pc}
\tablecaption{Data for Massive Planet -- Star systems
\label{Tab1}}
\tablehead{
\colhead{Planet} & \colhead{$M_p$}   & \colhead{ $M_* $} & \colhead{R$_*$}  & \colhead{a} & \colhead{P} &  e  & \colhead{Age } & \colhead{$\bar{\sigma}_*$} & References \\
 & \colhead{$(M_J)$} & \colhead{($M_{\odot}$)} & \colhead{($R_{\odot}$)} & \colhead{(AU)} & \colhead{(Days)} & & \colhead{(Gyr)}& \colhead{($10^{-7}$)} & 
 }
\startdata
XO-3  & 11.8 & 1.4  & 1.4 & 0.0477 & 3.7 & $0.288 \pm 0.004$ & $2.7 \pm 0.7$ & 0.04 & 1 \\
HAT-P-2b & 9.1 & 1.4 & 1.6 & 0.0688 & 5.6 & $0.517 \pm 0.003$ & $2.6 \pm 0.5$ &  0.06 & 2\\
WASP-14b & 7.3 & 1.2 & 1.3 & 0.0368 & 2.2 & $0.088 \pm 0.003$ & $0.75 \pm 0.25$ & 0.2 & 3 \\
WASP-18b & 10.3 & 1.3 & 1.2 & 0.0203 & 0.94 & $0.009\pm 0.003$/$<0.018$ & $< 2$ & $>0.2$ & 4,6\\
HAT-P-34b & 3.3 & 1.4 & 1.5 & 0.0677 & 5.5 & $0.441 \pm 0.032$ & $1.7 \pm 0.5$ & 0.1 & 5 
\enddata
\tablecomments{$\bar{\sigma}_*$ is the present day value and is quoted assuming $\nu_0=1$ and can be scaled accordingly. The parameters
are drawn from the quoted references, 
(1) Johns-Krull et al. (2008); (2) Pal et al. (2010); (3) Joshi et al. (2008); (4) Hellier et al. (2009); (5) Bakos et al. (2012);
(6) Husnoo et al. (2012)}
\end{deluxetable}

\begin{deluxetable}{lccccccccc}
\tablecolumns{10}
\tablewidth{0pc}
\tablecaption{Data for Planet--Star Systems that define the period--eccentricity envelope
\label{Tab3}}
\tablehead{
\colhead{Planet} & \colhead{$M_p $}   & \colhead{ $M_*$} & \colhead{R$_*$}  & \colhead{a} & \colhead{P} & e  & 
\colhead{Age} & \colhead{$\bar{\sigma}_*$} & References  \\
 & \colhead{$(M_J)$} & \colhead{$(M_{\odot})$} & \colhead{($R_{\odot}$)} & (AU) & \colhead{(Days)} & & \colhead{(Gyr)} & \colhead{($10^{-7}$)} & 
 }
\startdata
WASP-10b & 3.1 & 0.8 & 0.7 & 0.0378 & 3.09 & $0.06 \pm 0.01$/$<0.11$ & $>2$ & 17 & 1,10\\
HAT-P-16 & 4.2 & 1.2 & 1.2 & 0.0413 & 2.78 & $0.035 \pm 0.003$ & $2 \pm 0.8$ & 0.6 & 2\\
WASP-6b & 0.5 & 0.9 & 0.9 & 0.0421 & 3.36 & $0.05 \pm 0.02$/$<0.075$ & $>3$ & 11 & 3,10\\
HD118203b & 2.14 & 1.23 & 2.15 & 0.070 & 6.13 &  $0.31 \pm 0.014$ & $4.6\pm 0.8$ & 1.2 & 4\\
HD185269 & 0.95 & 1.3 & 2.03 & 0.077 & 6.84 & $0.30 \pm 0.04$ & 4.2 & 0.6 & 5\\
WASP-8b & 2.3 & 1.0 & 0.9 & 0.0801 & 8.16 & $0.31\pm 0.003$ & $4\pm 1$ & 9 & 6\\
HD108147b & 0.26 & 1.17  & 1.3 & 0.101 & 10.90 & $0.5 \pm 0.12$ & 2.2 & 1.1 & 7\\
COROT-10b & 2.8 & 0.9 & 0.7 & 0.1060 & 13.24 & $0.53 \pm 0.04$ & (2.0) & 26 & 8\\
HD17156b & 3.3 & 1.28 & 1.51 & 0.163 & 21.22 & $0.682 \pm 0.044$ & $2.8\pm 1$ & 0.3 & 9
\enddata
\tablecomments{ (1) Christian et al. (2009); (2) Buchave et al. (2010); (3) Gillon et al. (2009); (4) Da Silva et al. (2006);
(5) Johnson et al. (2006); (6) Queloz et al. (2010); (7) Pepe et al. (2002); (8) Bonomo et al. (2010); (9) Barbieri et al. (2009);
(10) Husnoo et al. (2012)
}
\end{deluxetable}

\begin{deluxetable}{lcccccccccc}
\tablecolumns{11}
\tablewidth{0pc}
\tablecaption{Data for Short Period Planet -- Star systems: P$<$1.5~days and $M_p>0.1 M_J$
\label{Tab2}}\tablehead{
\colhead{Planet} & \colhead{$M_p (M_J)$}  & \colhead{ $M_* (M_{\odot})$} & \colhead{R$_*$/R$_{\odot}$}  & \colhead{a (AU)} & \colhead{ P 
(days)} &
\colhead{Age (Gyr)} & \colhead{$\bar{\sigma}_*/10^{-7}$} & $T_{in}$ (Gyr) & \colhead{ $\dot{P}$ (ms/yr)} & \colhead{Reference} \\
 }
\startdata
WASP-19b & 1.2 & 1.0 & 1.0 & 0.0166 & 0.80 & $>1$ & $<5.5$ & $>0.2$ & $<0.5$ & 1 \\
WASP-43b & 2.0 & 0.7 & 0.7 & 0.0153 & 0.81 & $>1.0$ & $<12$ & $>1.9$ & $<0.04$ & 2 \\
WASP-18b & 10.3 & 1.3 & 1.2 & 0.0203 & 0.94 & $<2$ & $>0.2$ & $<1.2$ & $>0.1$ & 3 \\
WASP-12b & 1.4 & 1.4 & 1.6 & 0.0229 & 1.09 & $2 \pm 1$ & 0.2 & 0.6 & 0.24& 4 \\
OGLE-TR-56 & 1.3 & 1.2 & 1.3 & 0.0239 & 1.21 & $3.1\pm 1.1$ & 0.4 & 1.6 &  0.1 & 5\\
WASP-33b & $<4.1$ & 1.5 & 1.5 & 0.0256 & 1.22 & $<0.5$ & $>10^{-5}$ & $> 1600$ & $<10^{-4}$ & 6 \\
TrES-3 & 1.9 & 0.9 & 0.8 & 0.0228 & 1.31 & $<3.7$ & $>6.9$ & $<17$ & $>0.01$ & 7 \\
WASP-4b & 1.2 & 0.9 & 0.91 & 0.0231 & 1.34 & $<10$ & $>6.2$ & $<8.7$ & $>0.02$ & 8 \\
Qatar-1b & 1.1 & 0.9 & 0.8 & 0.0234 & 1.42 & $>6$ & $<6.9$ & $>26$ &  $<0.007$ & 9 \\
OGLE-TR-113b & 1.2 & 0.8 & 0.8 & 0.0228 & 1.43 & $>10$ & $<11$ & $>9.3$ &  $<0.02$ & 5 \\
\hline 
CoRoT-1b & 1.0 & 1.0 & 1.1 & 0.0254 & 1.51 & $8\pm 4$ & 6.8 & 1.0 & 0.2 & 10 \\
HD41004B b& 18.1 & 0.4 & 0.4 & 0.0177 & 1.33 & $1.5 \pm 0.5$ & 210 & 5.7 & 0.03 & 11 \\
COROT-14b & 7.7 & 1.1 & 1.2 & 0.0269 & 1.51 & $<8$ & $>1.0$ & $<1.5$ & $>0.13$ & 12 \\
WASP-48b & 1.0 & 1.2 & 1.8 & 0.0345 & 2.1 & $8 \pm 2$ & 13 & 1.4 & 0.19 & 13 \\
KOI-13.01 & 9.2 & 2.1 & 2.6 & 0.0363 & 1.76 & $<1$ &  $4 \times 10^{-6}$ & $1.8 \times 10^3$ & $10^{-4}$ & 14 
\enddata
\tablecomments{ We quote stellar masses and radii to one decimal place because the variation between different estimates is at this level.
Some planets, like HAT-P-7b and OGLE2-TR-L9b are not shown because their host star dissipations are too small for measureable effects.
$\bar{\sigma}_*$ is the present day value and is quoted assuming $\nu_0=1$ and can be scaled accordingly. The parameters
used above are drawn mostly from the homogeneous analysis presented in Southworth (2008, 2009, 2010, 2011). Additional references are
given for papers from which other information is drawn (often ages) or which significantly update the information quoted by Southworth, or
for those systems that are not transitting.(1) Hebb et al. (2010); (2) Gillon et al. (2012) (3) Hellier et al. (2009);
 (4) Hebb et al. (2009); (5) Torres, Winn \& Holman (2008); (6) Collier-Cameron et al. (2010); (7) Sozzeti et al. (2009);
 (8) Wilson et al. (2008); (9) Alsubai et al. (2011); (10) Barge et al. (2008); (11) Zucker et al. (2004); 
 (12) Tingley et al. (2011); (13) Enoch et al. (2011); (14) Shporer et al. (2011)}
\end{deluxetable}

\begin{deluxetable}{lcccccccccc}
\tablecolumns{11}
\tablewidth{0pc}
\tablecaption{Data for Planet--Star Systems with Alignment Constraints
\label{Tab4}}
\tablehead{
\colhead{Planet} & \colhead{$M_p$}   & \colhead{ $M_*$} & \colhead{R$_*$}  & \colhead{a} &  e  &
\colhead{Age } & \colhead{$\bar{\sigma}_*$} & $\left| \lambda \right| $ & Ref (Star) & Ref ($\lambda$)  \\
 & \colhead{($M_J$)} & \colhead{($M_{\odot}$)} & \colhead{($R_{\odot}$)} & (AU) & & (Gyr) & $10^{-7}$ & ($^{\circ}$) & & \\
 }
\startdata
WASP-33b & $<4.1$ & 1.5 & 1.5 & 0.0256 & 0 & $<0.5$ & $>10^{-5}$ & $252 \pm 1$ & 1 & 1\\
HAT-P-14b & 2.2 & 1.4 & 1.5 & 0.0606 & $0.11 \pm 0.01$ & $1.3 \pm 0.4$ & 0.03 & $189 \pm 1$ & 2 & 51\\
HAT-P-7b & 1.8 & 1.5 & 2.0 & 0.0381 & 0 & $2.2 \pm 1$ & $>10^{-5}$ & $183\pm 9$ & 3 & 52 \\
HAT-P-6b & 1.1 & 1.3 & 1.5 & 0.0524 & 0 & $2.2 \pm 0.6$ & 0.2 & $166 \pm 10$ & 4 & 53\\
WASP-2b & 0.9 & 0.9 & 0.8 & 0.0309 & 0 & $<13$ & $>7$ & $155 \pm 13$ & 5, 6 & 54\\
WASP-17b & 0.5 & 1.3 & 1.6 & 0.0515 & $0.03 \pm 0.02$ & $<4$ & 0.3 & $149 \pm 5$& 7 & 54\\
WASP-15b & 0.6 & 1.2 & 1.4 & 0.0499 & 0 & $4.7 \pm 2$ & 0.8 & $140 \pm 5$ & 8 & 54\\
WASP-8b & 2.3 & 1.0 & 0.9 & 0.0801 & $0.310 \pm 0.003$ & $3 \pm 0.6$ & 9.3 & $123 \pm 4$ & 9 & 9\\
HAT-P-11b & 0.08 & 0.8 & 0.7 & 0.0526 & $0.20 \pm 0.05$ & $<12$ & 22 & $103 \pm 20$ & 10 & 55\\
WASP-7b & 0.9 & 1.2 & 1.2 & 0.060 & 0 & (2.0) & 0.2 & $86 \pm 6$ & 11 & 56\\
WASP-1b & 0.9 & 1.2 & 1.5 & 0.0390 & 0 & $3 \pm 0.6$ & 0.6 & $79 \pm 4$ & 5, 6 & 57\\
CoRoT-1b & 1.0 & 1.0 & 1.1 & 0.0254 & 0 & $8\pm 4$ & 2.0 & $77 \pm 1$ & 12 & 58 \\
HAT-P-30b & 0.7 & 1.2 & 1.2 & 0.0419 & $0.04 \pm 0.02$ & $1.2 \pm 0.6$& 0.9 & $74 \pm 9$ & 13 & 13 \\
XO-4b & 1.6 & 1.3 & 1.5 & 0.0547 & 0 & $2.1 \pm 0.6$ & 0.2 & $47 \pm 7$ & 14 & 59\\
HD80606b & 4.1 & 1.0 & 1.0 & 0.4564 & 0.93 & $4.7 \pm 3$ & 9.3 & $42 \pm 8$ & 15 & 60\\
XO-3b & 11.8 & 1.2/1.4 & 1.4 & 0.0477 & $0.288 \pm 0.004$ & $2.7 \pm 0.7$ & 0.6/0.04 & $37 \pm 4$ & 16,17,18 & 61 \\
WASP-14b & 7.3 & 1.2 & 1.3 & 0.0368 & $0.088 \pm 0.003$ & $0.75 \pm 0.25$ & 0.2 & $33 \pm 7$ & 19 & 62\\
CoRoT-3b & 22.0 & 1.4 & 1.6 & 0.0578 & 0 & $2.2 \pm 0.6$ & 0.04 & $32 \pm 16$ & 20 & 63\\
TrES-1 & 0.8 & 0.9 & 0.8 & 0.0395 & 0 & $3.5 \pm 2.5$& 10.7 & $30 \pm 21 $ & 21, 6 & 64 \\
Kepler-8b & 0.6 & 1.2 & 1.5 & 0.0485 & 0 & $3.8 \pm 1.5$ & 0.7 & $26 \pm 10$ & 22 & 22\\
KOI-13.01 & 9.7 & 2.1 & 2.6 & 0.0363 & 0 & $< 1$ & $4 \times 10^{-6}$ & $23 \pm 4$ & 23 & 65\\
WASP-38b & 2.7 & 1.2 & 1.4 & 0.0755 & $0.032 \pm 0.005$ & (2.0) & 1.0& $20 \pm 38$ & 24 & 57\\
WASP-22b & 0.6 & 1.1 & 1.2 & 0.0470 & 0 & $3 \pm 1$ & 2.2 & $22 \pm 16$ & 25 & 66\\
HAT-P-9b & 0.8 & 1.3 & 1.3 & 0.0530 & 0 & $1.8 \pm 1.6$ & 0.2 & $16 \pm 8$ & 26 & 67\\
HAT-P-23b & 2.1 & 1.1 & 1.2 & 0.0232 & $0.11 \pm 0.04$ & $4 \pm 1$& 0.9 & $15 \pm 22$ & 27 & 67 \\
HD149026b & 0.4 & 1.3 & 1.5 & 0.0431 & 0 & $2 \pm 0.8$ & 0.2 & $12 \pm 15$ & 28 & 68 \\
WASP-25 & 0.6 & 1.0 & 0.9 & 0.0474 & 0 & $<4$ & 4 & $15 \pm 7$ & 29 & 69 \\
WASP-5b & 1.6 & 1.0 & 1.0 & 0.0271 & $0.038 \pm 0.02$ & $3.1 \pm 1.4$ & 3.7 & $11 \pm 9$ & 30 & 54\\
CoRoT-18b & 3.5 & 1.0 & 1.0 & 0.0295 & 0 & $>4$ & 3.6 & $10\pm 20$ & 31 & 31\\
HAT-P-16b & 4.2 & 1.2 & 1.2 & 0.0413 & $0.036\pm 0.004$ & 2 & 0.6 & $10 \pm 16$ & 32 & 67\\
WASP-6b & 0.5 & 0.9 & 0.9 & 0.0421 & $0.05 \pm 0.02$ & $>3$ & 11 & $9 \pm 16$ & 33 & 33\\
TrES-2 & 1.2 & 1.0 & 1.0 & 0.0357 & 0 & $5.3 \pm 2.4$ & 3.9 & $9 \pm 12$ & 34, 6 & 70\\
HD17156b & 3.3 & 1.3 & 1.5 & 0.1637 & 0.68 & 2.8 & 0.3 & $9 \pm 9$& 35 & 71\\
HAT-P-8b & 1.5 & 1.3 & 1.6 & 0.0487 & 0 & $3.4 \pm 1$ & 0.35 & $9 \pm 8$ & 36 & 57, 67\\
CoRoT-2b & 3.6 & 1.0 & 0.9 & 0.0285 & $0.014 \pm 0.007$ & 2 & 4.2 & $7 \pm 5$ & 37 & 72\\
TrES-4 & 0.9 & 1.4 & 1.8 & 0.0508 & 0 & $4.7 \pm 2$ & 0.07 & $6 \pm 5$ & 38 & 73\\
WASP-19b & 1.2 & 1.0 & 1.0 & 0.0166 & 0 & $>1$ & $<5.5$ & $5 \pm 5$ & 39 & 74\\
HAT-P-4b & 0.7 & 1.3 & 1.6 & 0.0447 & 0 & $5.2 \pm 1.6$  0.3& & $5 \pm 12$ & 40 & 51 \\ 
WASP-3b & 2.0 & 1.3 & 1.4 & 0.0319 & 0 & $2.1 \pm 1.4$ & 0.2 & $5 \pm 6$ & 41 & 75\\
WASP-24b & 1.1 & 1.2 & 1.3 & 0.0365 & 0 & $3.8 \pm 1.2$ & 0.6 & $5 \pm 4$ & 42 & 57\\
WASP-16b & 0.8 & 1.0 & 1.0 & 0.0418 & 0 & $<8 $ & 4 & $4 \pm 12$ & 43 & 69 \\
HAT-P-1b & 0.5 & 1.1 & 1.1 & 0.0554 & 0 & $3 \pm 2.2$ & 2.5 & $4 \pm 2$ & 44 & 76\\
WASP-18b & 10.3 & 1.3 & 1.2 & 0.0203 & $0.009 \pm 0.003$ & $<2$ & $>0.2$ & $4 \pm 5$ & 45 & 54 \\
HD209458b & 0.7 & 1.1 & 1.1 & 0.0475 & 0 & $3.1 \pm 0.7$& 2.1 & $4 \pm 1$ & 6 & 77\\
WASP-31b & 0.5 & 1.2 & 1.3 & 0.0466 & 0 & $4 \pm 1$ & 1 & $3 \pm 3$ & 46 & 69 \\
HAT-P-13b & 0.9 & 1.2 & 1.6 & 0.0427 & $0.021 \pm 0.009$ & $5.9 \pm 1.6$ & 0.1 & $1 \pm 9$ & 47 & 78\\
HAT-P-2b & 9.1 & 1.3 & 1.6 & 0.0688 & $0.517 \pm 0.003$ & $2.3 \pm 1.1$ & 0.3 &  $1 \pm 13$  & 48 & 79\\
HD189733b & 1.2 & 0.8 & 0.8 & 0.0314 & $0.004 \pm 0.002$ & $6.8 \pm 4.4$ & 14 &$1 \pm 0.3$ & 49, 6 & 63\\
WASP-4b & 1.2 & 0.9 & 0.91 & 0.0231 & 0 & $<10$ & $>6.2$ & $0 \pm 13$ & 50 & 54
\enddata
\tablecomments{ 
(1) Collier Cameron et al. (2010); (2) Torres et al. (2010); (3) Pal et al. (2008); 
(4) Noyes et al. (2008); (5) Collier-Cameron et al. (2007); (6) Torres, Winn \& Holman (2008); (7) Anderson et al. (2010);
(8) West et al. (2009); (9) Queloz et al. (2010); (10) Bakos et al. (2010); (11) Hellier et al. (2009); 
(12) Barge et al. (2008); (13) Johnson et al. (2011); (14) McCullough et al. (2008); (15) Naef et al. (2001); 
(16) Johns-Krull et al. (2008); (17) Winn et al. (2008b); (18) Southworth (2010); (19) Joshi et al. (2008); (20) Deleuil et al. (2008);
(21) Alonso et al. (2004); (22) Jenkins et al. (2010); (23) Shporer et al. (2011); 
 (24) Barros et al. (2011); (25) Maxted et al. (2010); (26) Shporer et al. (2009);
(27) Bakos et al. (2011); (28) Sato, B. et al. (2005); (29) Enoch et al. (2011); 
 (30) Anderson et al. (2008); (31) Hebrard et al. (2011b); (32) Buchave et al. (2010);
(33) Gillon et al. (2009); (34) O'Donovan et al. (2006); (35) Barbieri et al. (2009); (36) Latham et al. (2009); (37) Alonso et al. (2008);
(38) Mandushev et al. (2007); (39) Hebb et al. (2010); (40) Kovacs et al. (2007); (41) Pollaco et al (2010); (42) Street et al. (2010);
(43) Lister et al. (2009); 
(44) Bakos et al. (2007a); (45)  Hellier et al. (2009), Southworth et al. (2009); (46) Anderson et al. (2011b);
 (47) Bakos et al. (2009); (48) Bakos et al. (2007b); (49) Bouchy et al. (2005);  (50) Wilson et al. (2008);
(51) Winn et al. (2011); (52) Winn et al. (2009b); (53) Hebrard et al. (2011a); (54) Triaud,  et al. (2010);
(55) Hirano et al. (2011a); Winn et al. (2010c); Sanchis-Ojeda \& Winn (2011); (56) Albrecht et al. (2012);
(57) Simpson et al. (2011); (58) Pont et al. (2010); (59) Narita et al. (2010a); (60) Hebrard et al. (2010);
(61) Winn et al. (2009a); Hirano et al. (2011b); (62) Johnson et al. (2009); (63) Triaud et al. (2009);
(64) Narita et al. (2007); (65) Barnes, Linscott \& Shporer (2011) ApJ, 197, 10;
 (66) Anderson et al. (2011a) ; (67) Moutou et al. (2011); (68) Wolf et al. 2007; (69) Brown et al. (2012); 
 (70) Winn et al. (2008a);
(71) Cochran et al. (2008); Narita et al. (2009); (72) Bouchy et al. (2008); (73) Narita et al. (2010b);
(74) Hellier et al. (2011); (75) Miller et al. (2010); (76) Johnson et al. (2008); (77) Winn et al. (2005); 
(78) Winn et al. (2010b); (79) Winn et al. (2007)
}
\end{deluxetable}

\clearpage

\plotone{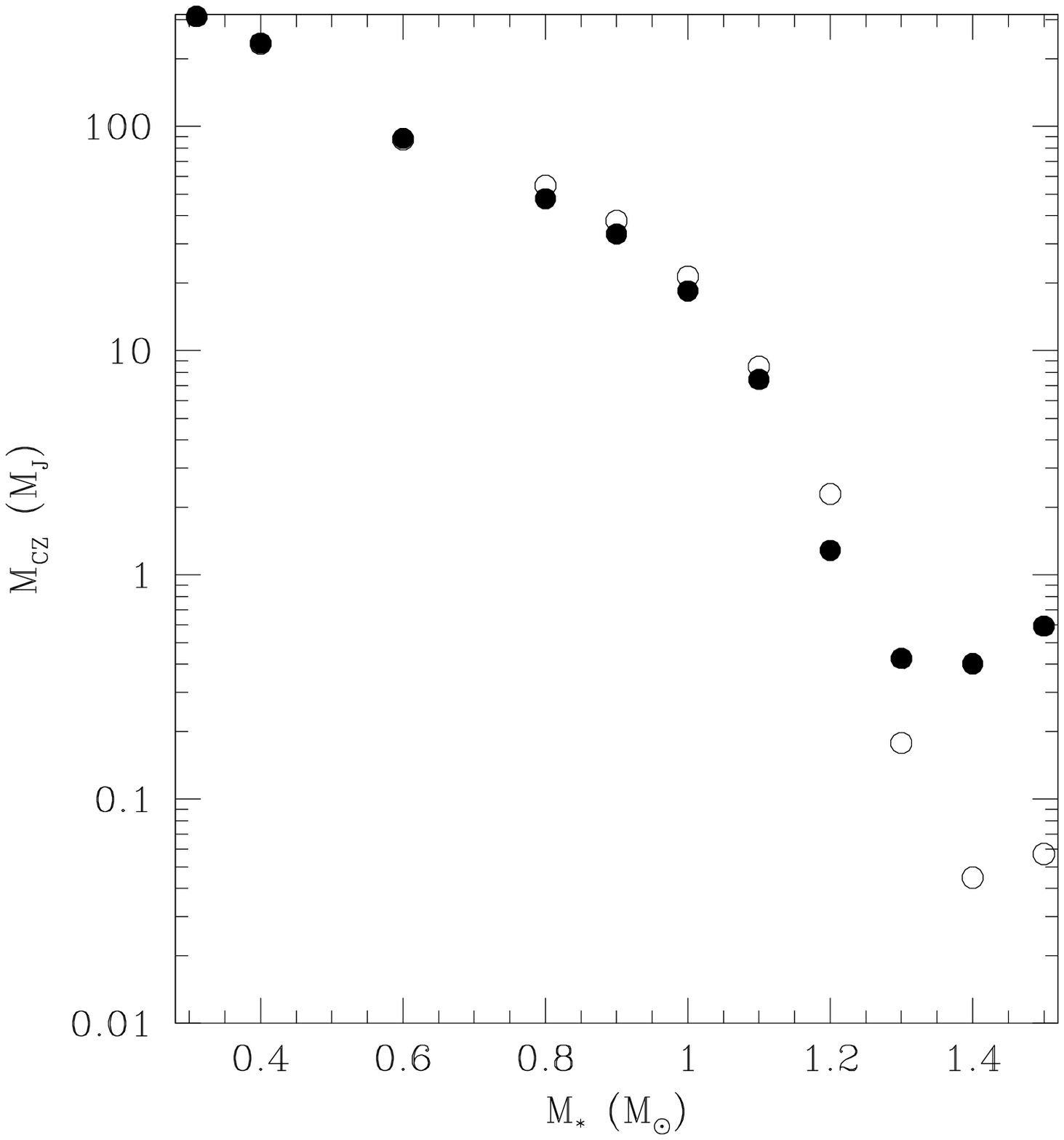}

\figcaption[CZ.ps]{The filled circles show the mass of the surface convection zone for
solar metallicity main sequence stars of different mass, at an age of 1~Gyr. The open circles show the
same quantity at an age of 0.1~Gyr. Within the context of our model, this is the part of the star
that determines the amount of dissipation.
\label{CZsurf}}

\clearpage

\plotone{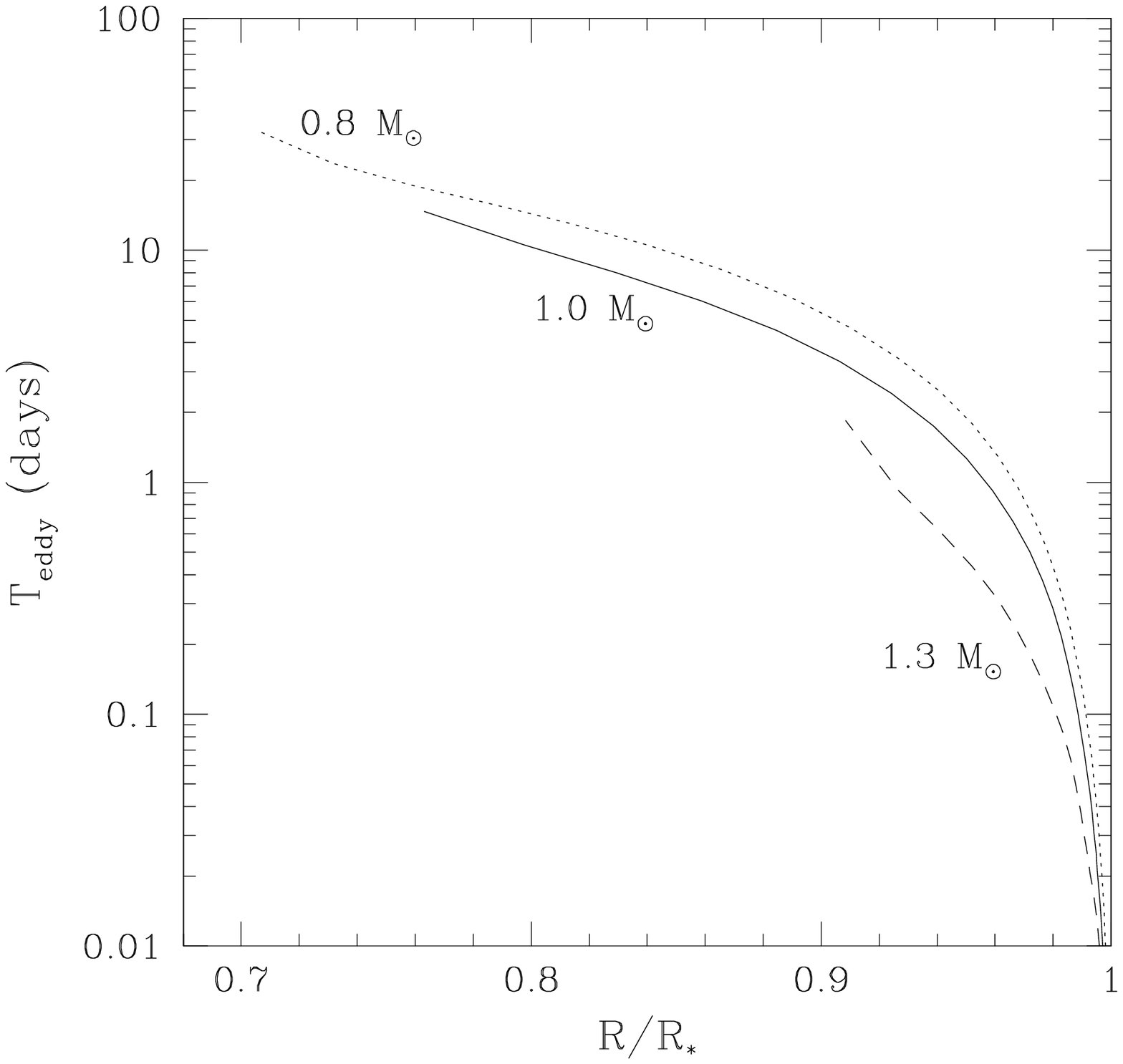}

\figcaption[Ted.ps]{The solid line shows the largest local eddy turnover time as a function
of scaled radius in the surface convection zone of a $1 M_{\odot}$ star. The dashed line shows
the same for a star of mass 1.3~$M_{\odot}$ and the dotted line for a star of mass $0.8 M_{\odot}$.
All stars have an age of 1~Gyr. Planetary tidal perturbations, with forcing periods of several days,
will likely couple well to the entire convection zone for the more massive stars, but only part of
the zone for lower mass stars.
\label{Turnover}}

\clearpage

\plotone{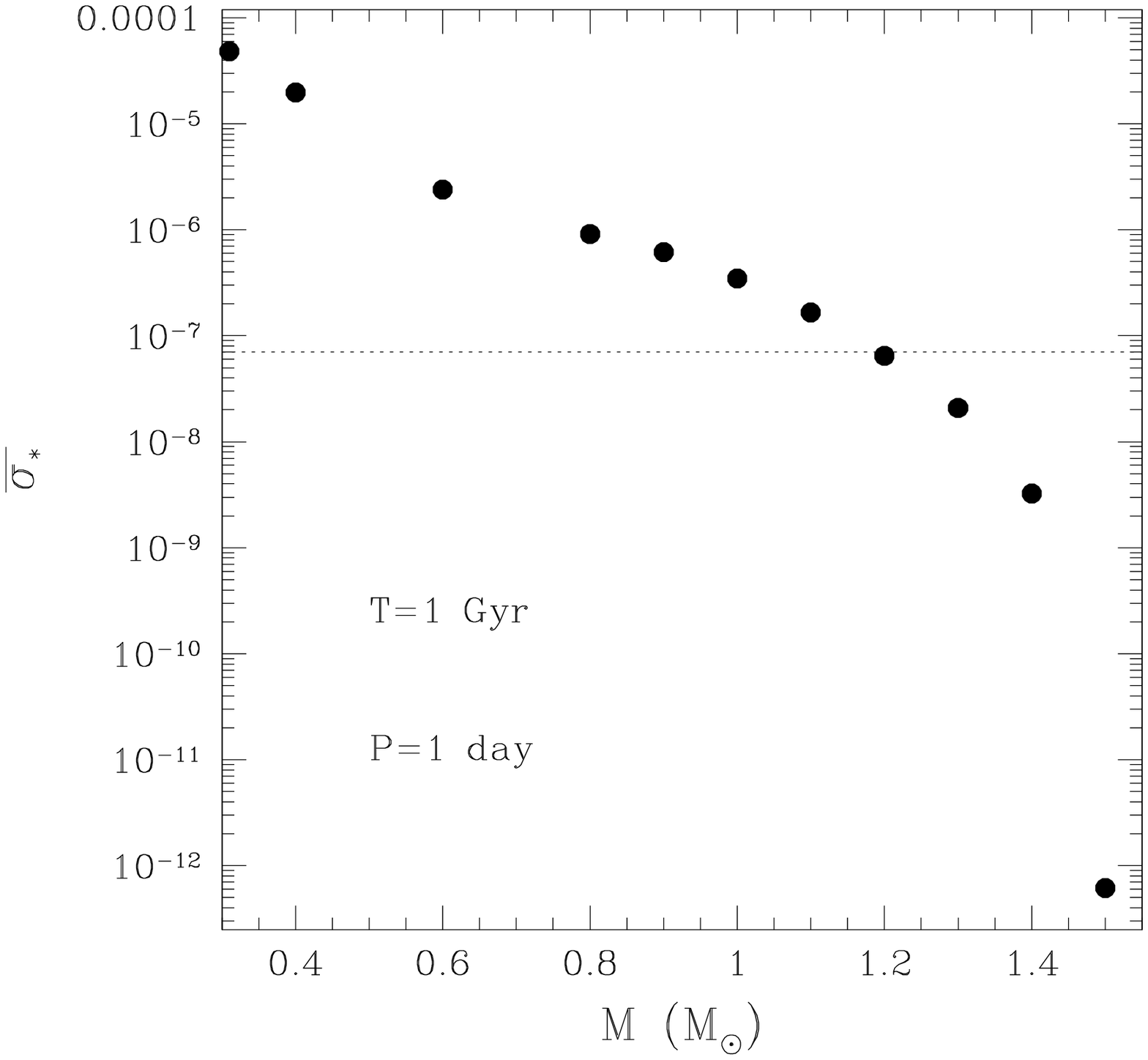}
\figcaption[Sig.ps]{The solid points show the normalised bulk dissipation per unit mass, as defined in H10, as a function of stellar mass, for an age of 1~Gyr and a forcing period
of 1 day. The dissipation is assumed to take place only in the convection zone, and with an efficiency given by Zahn's prescription. The horizontal dotted line is the calibration
from H10 over all main sequence stars. We see that the most massive stars are somewhat less dissipative and lower mass stars are more dissipative. 
\label{Sig}}

\clearpage

\plotone{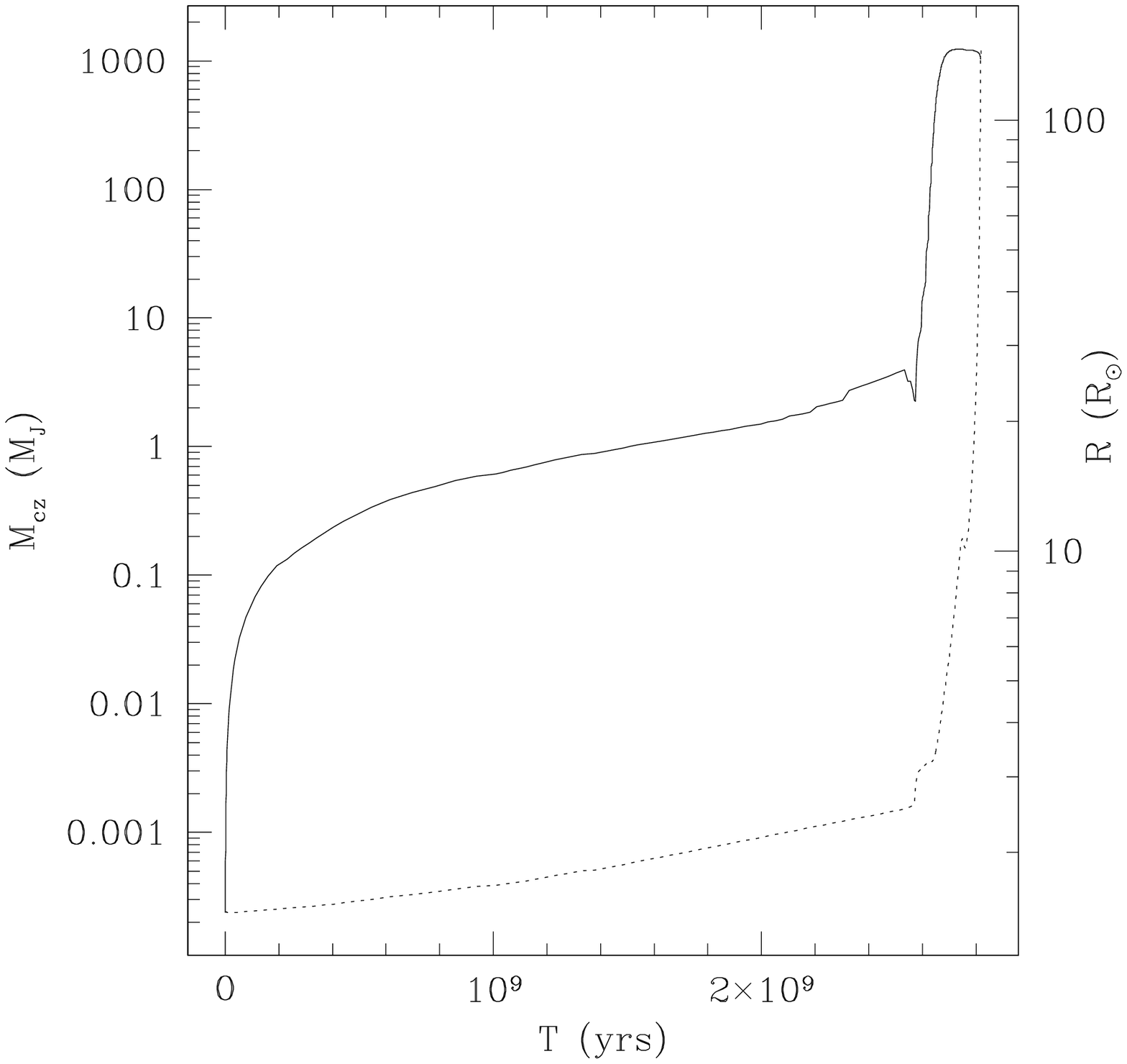}

\figcaption[Cevol.ps]{The solid line shows the evolution of the surface convection zone mass, for a
solar metallicity star of mass 1.5$M_{\odot}$ (scale shown on left axis). The dotted line shows the 
evolution of the stellar radius (scale shown on right axis). Both the increase in the size of the
surface convection zone, and the increase in stellar radius will act to accelerate orbital decay.
\label{CZevol}}

\clearpage

\plotone{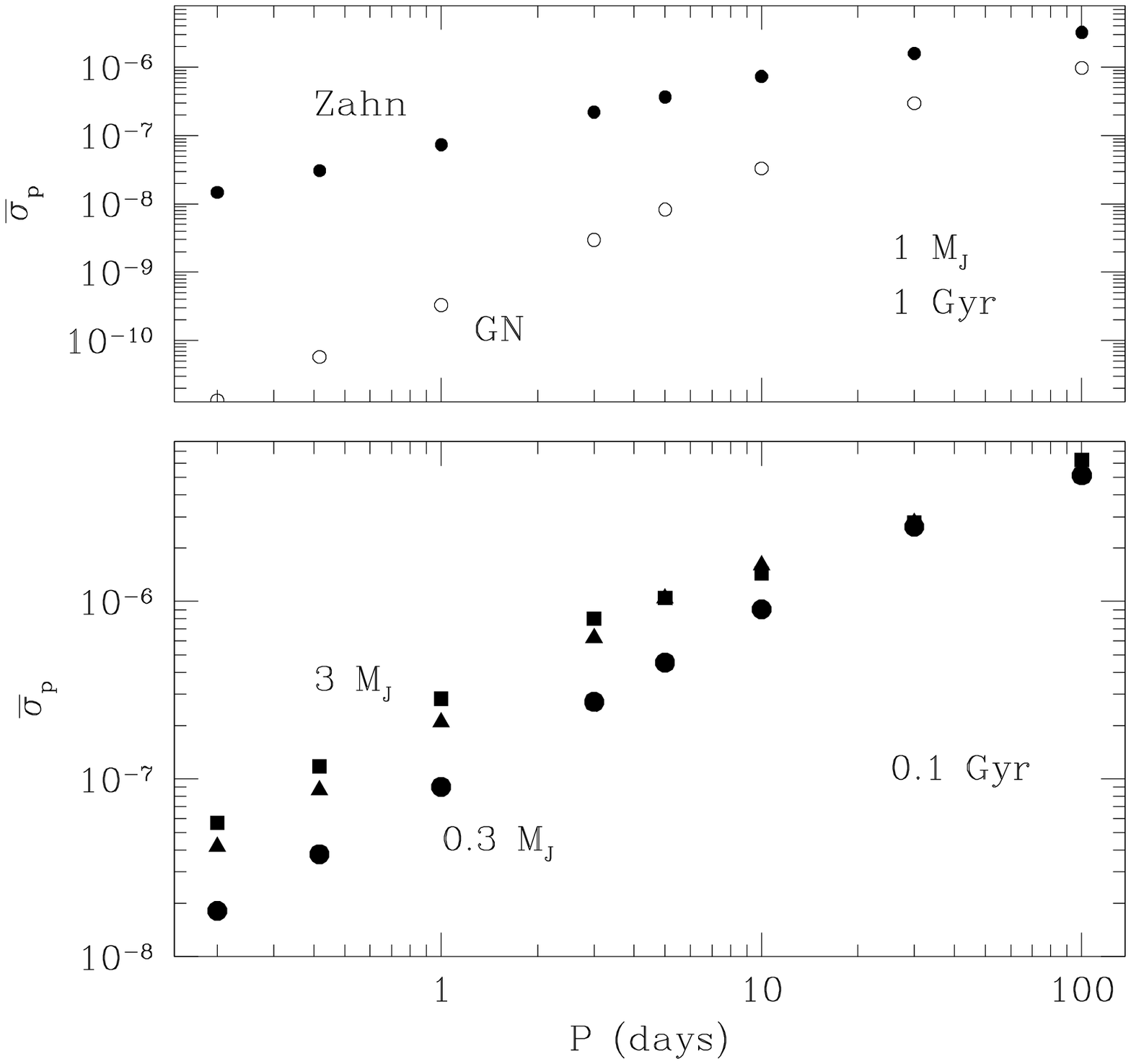}

\figcaption[Sigp.ps]{The upper panel shows the expected period dependance of the normalised bulk
planetary dissipation $\bar{\sigma}_p$, for a 1~Gyr old Jupiter-mass planet. The filled circles
indicate the results assuming the linear, Zahn efficiency. The open circles show the corresponding
results Goldreich \& Nicholson scaling, for comparison with that estimate based on a model
of Jupiter. The lower panel shows the corresponding scalings for three planets ($0.3 M_J$, $1 M_J$
and $3 M_J$ -- bottom to top) at a younger age of 0.1~Gyr. This is more appropriate for the extrasolar planet case,
as the tidal circularisation times turn out to be of this order.
\label{Sigp}}

\clearpage
\plotone{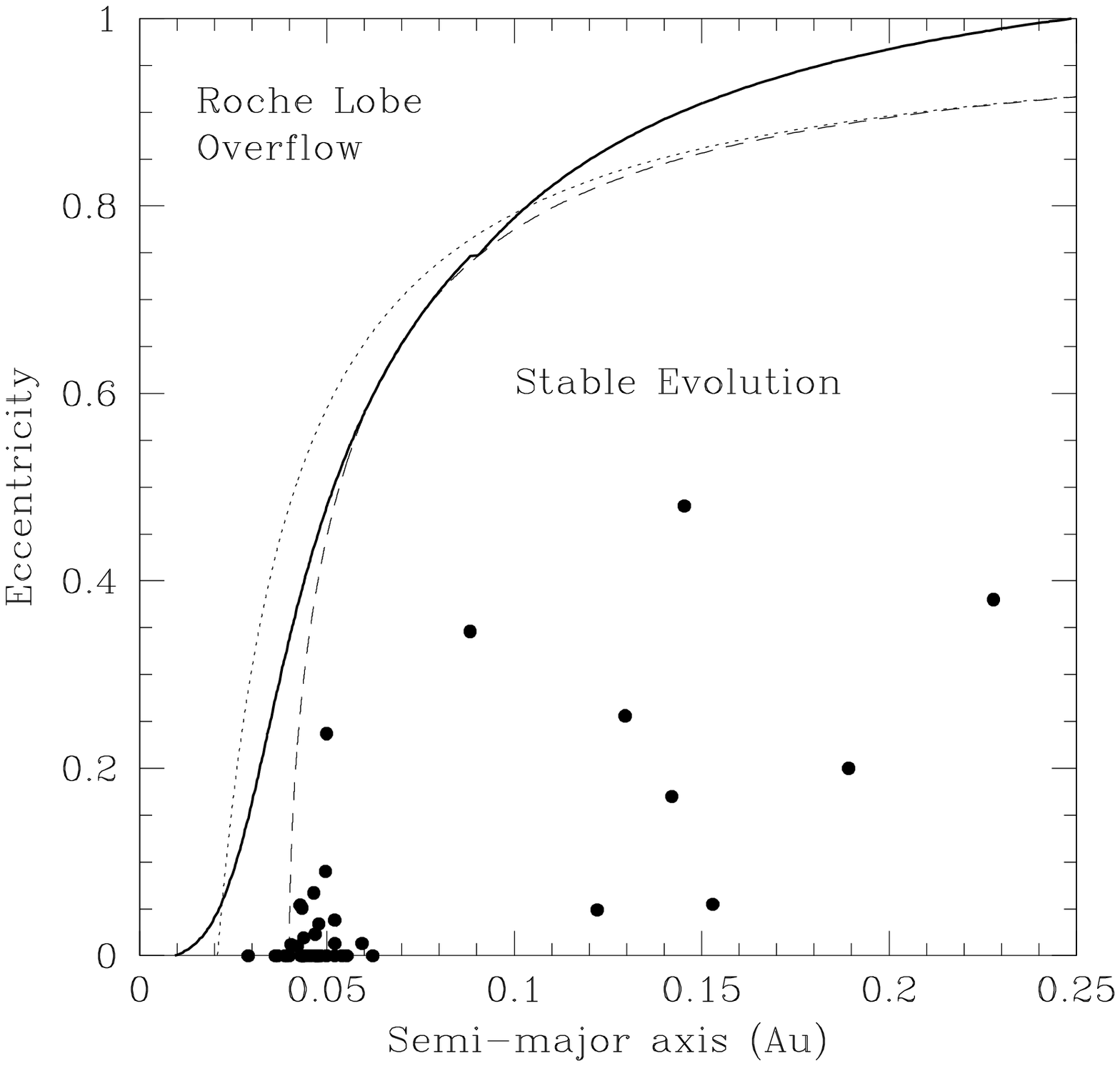}

\figcaption[Disrupt.ps]{ The solid line indicates the value of semi-major axis and eccentricity
that generate a self-consistent tidal dissipation luminosity that maintains the planet at a fixed
radius of 1.6~$R_J$ (characteristic of a 0.5$M_J$ planet at young ages) in the absence of
thermal evolution. The dashed line indicates
a curve of constant specific angular momentum that just grazes this curve. The dotted line indicates
the periastron curve corresponding to the same initial orbit. The solid points show known planets
with mass $0.5 \pm 0.2 M_J$.
\label{Disrupt}}

\clearpage
\plotone{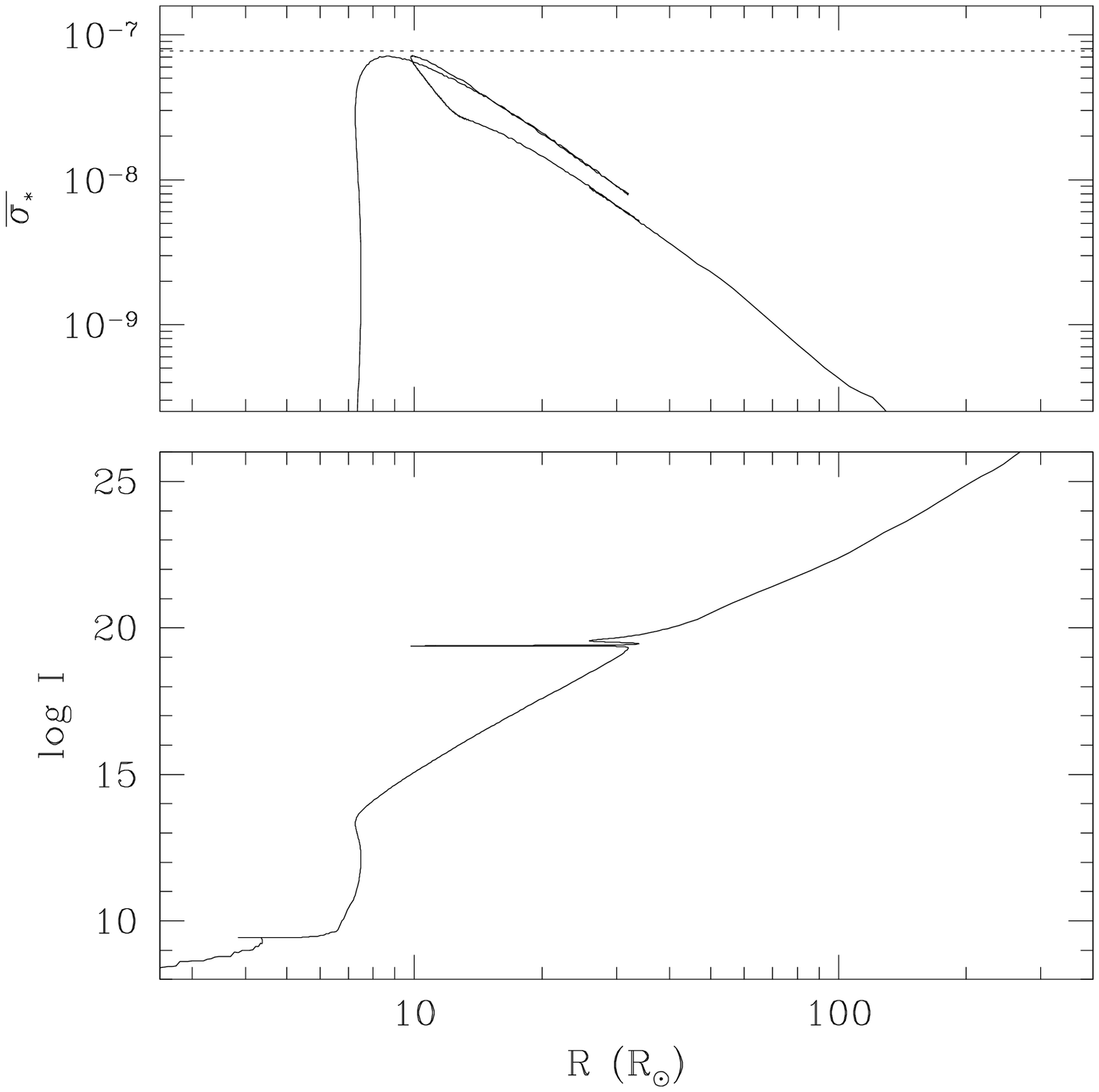}

\figcaption[I2.ps]{The lower panel shows the evolution of the integral $I$ with stellar radius. The
horizontal section at $\log I \sim 20$ corresponds to the core Helium burning stage. The upper panel
shows the equivalent $\bar{\sigma}_*$ for the star, at each point in the evolution. The horizontal dotted
line represents the approximate value derived in H10 for main sequence stars. Although the effective 
dissipation per unit mass decreases as the star ascends the giant branch, the increase in stellar radius
nevertheless dominates the increase in I.
\label{I2}}

\clearpage

\plotone{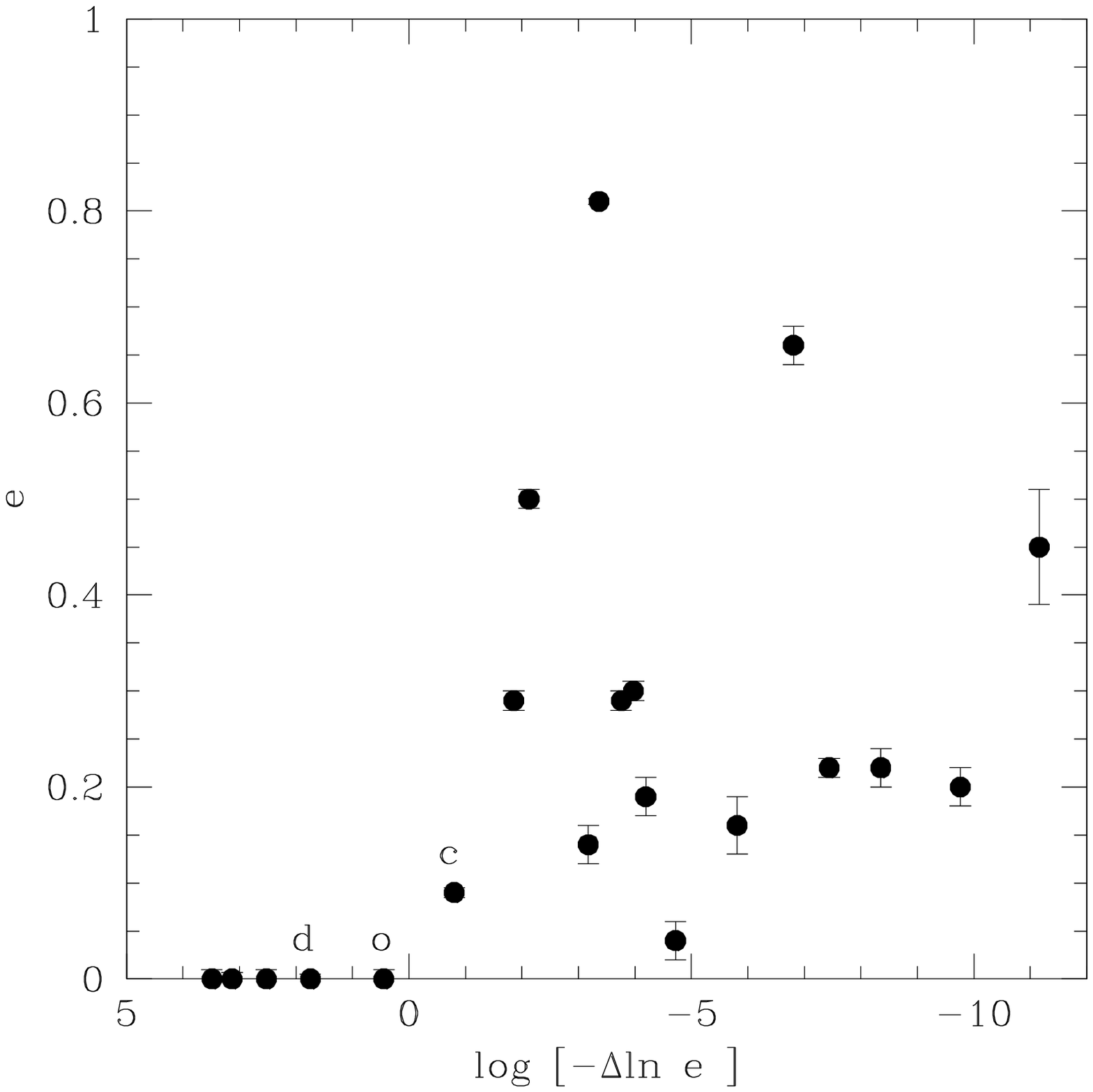}

\figcaption[dle.ps]{ The expected change in eccentricity $ - \Delta \ln e$ due to tidal interactions
is shown, as a function of eccentricity, for $\nu_0=1$. We have labelled specific systems (using the designations of
VP) that bracket the circularisation boundary. Note that this sample is a subsample of that shown in
VP, restricted to primaries with masses $2.2 \pm 0.2 M_{\odot}$.
\label{dle}}

\clearpage

\plotone{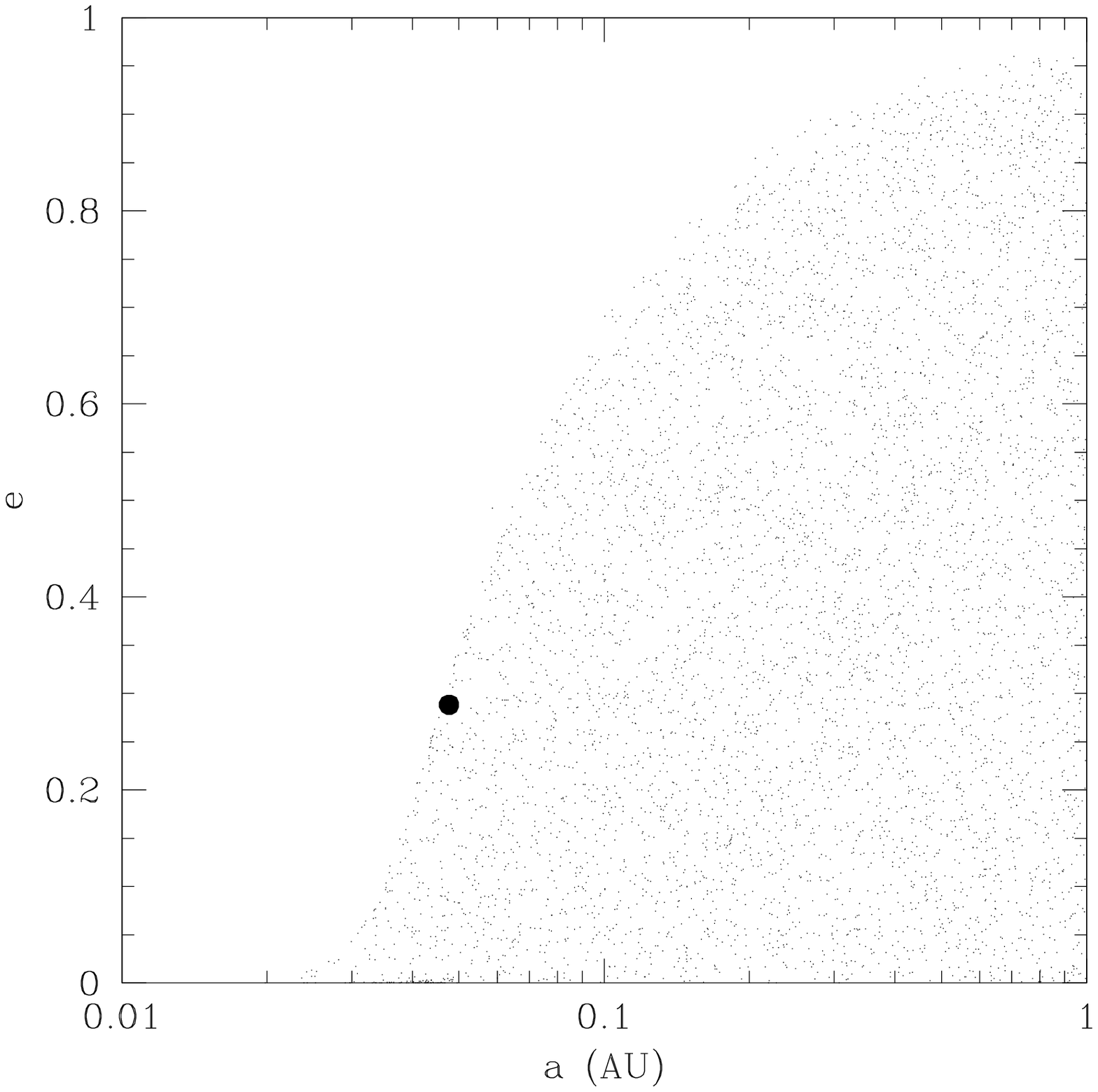}

\figcaption[xo3_init.ps]{The solid point indicates the present semi-major axis and eccentricity of the XO-3b planet.
Models were generated with a full range of semi-major axis and eccentricity, using the stellar and planetary masses
of the system. The dotted points show the parameters of systems that survive for a range of ages of 2--4~Gyr. The
edge of the surviving distribution is defined by two processes -- the circularisation of eccentric orbits and planetary
disruption driven by tidal heating and Roche lobe overflow.
\label{xo3_init}}

\clearpage

\plotone{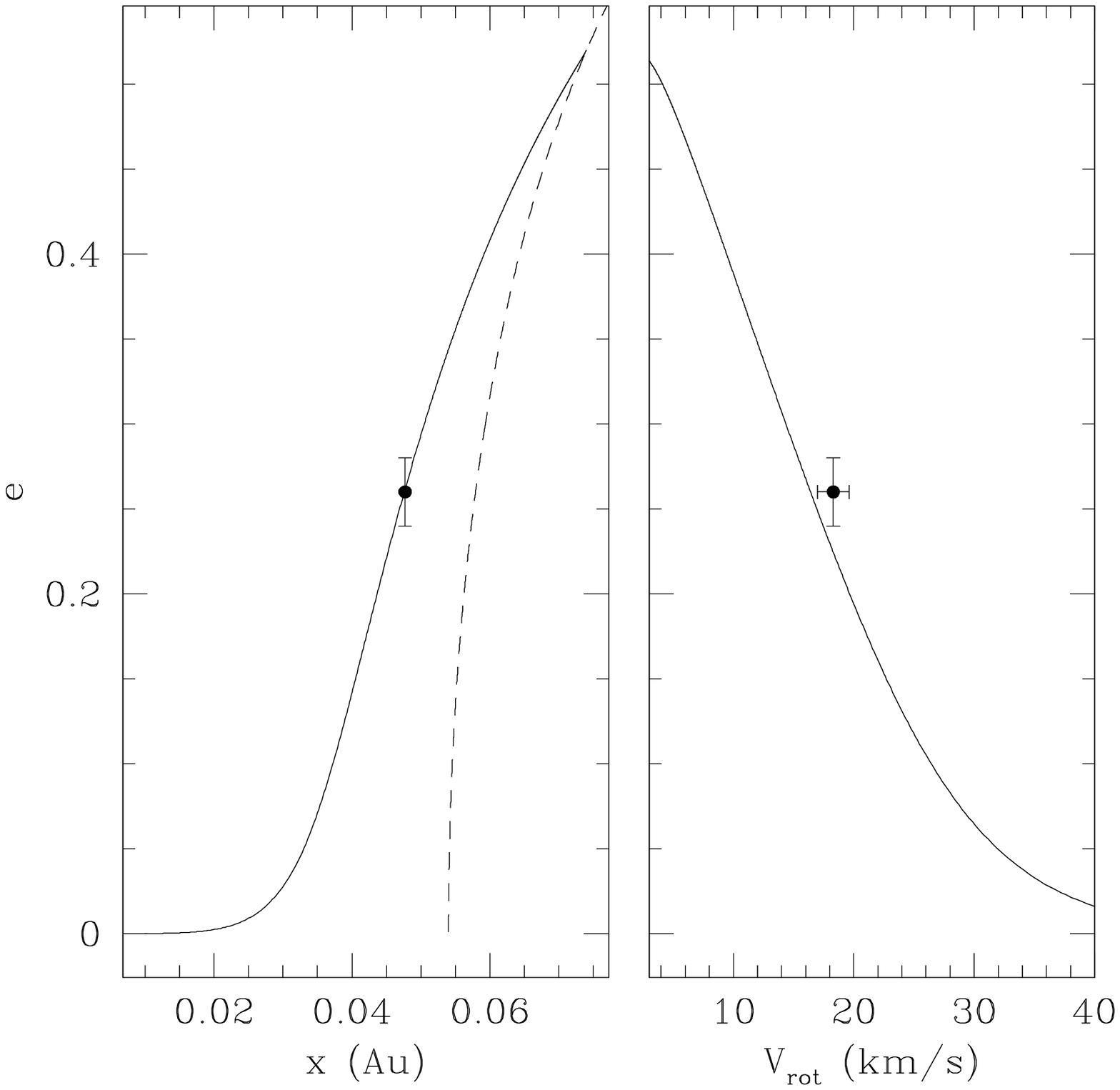}

\figcaption[xo3_ex.ps]{Left hand panel shows the evolution of semi-major axis versus eccentricity, while the right
hand panel shows the latter plotted against stellar rotational velocity. The original semi-major axis and eccentricity
are 0.074~AU and 0.52 respectively, while the star was initially spinning with a canonical spin period of 30~days. The
dashed line in the left hand panel shows the evolution if constant orbital angular momentum is maintained (as would happen
if there was no stellar contribution to the tidal interactions).
\label{xo3_ex}}

\clearpage

\plotone{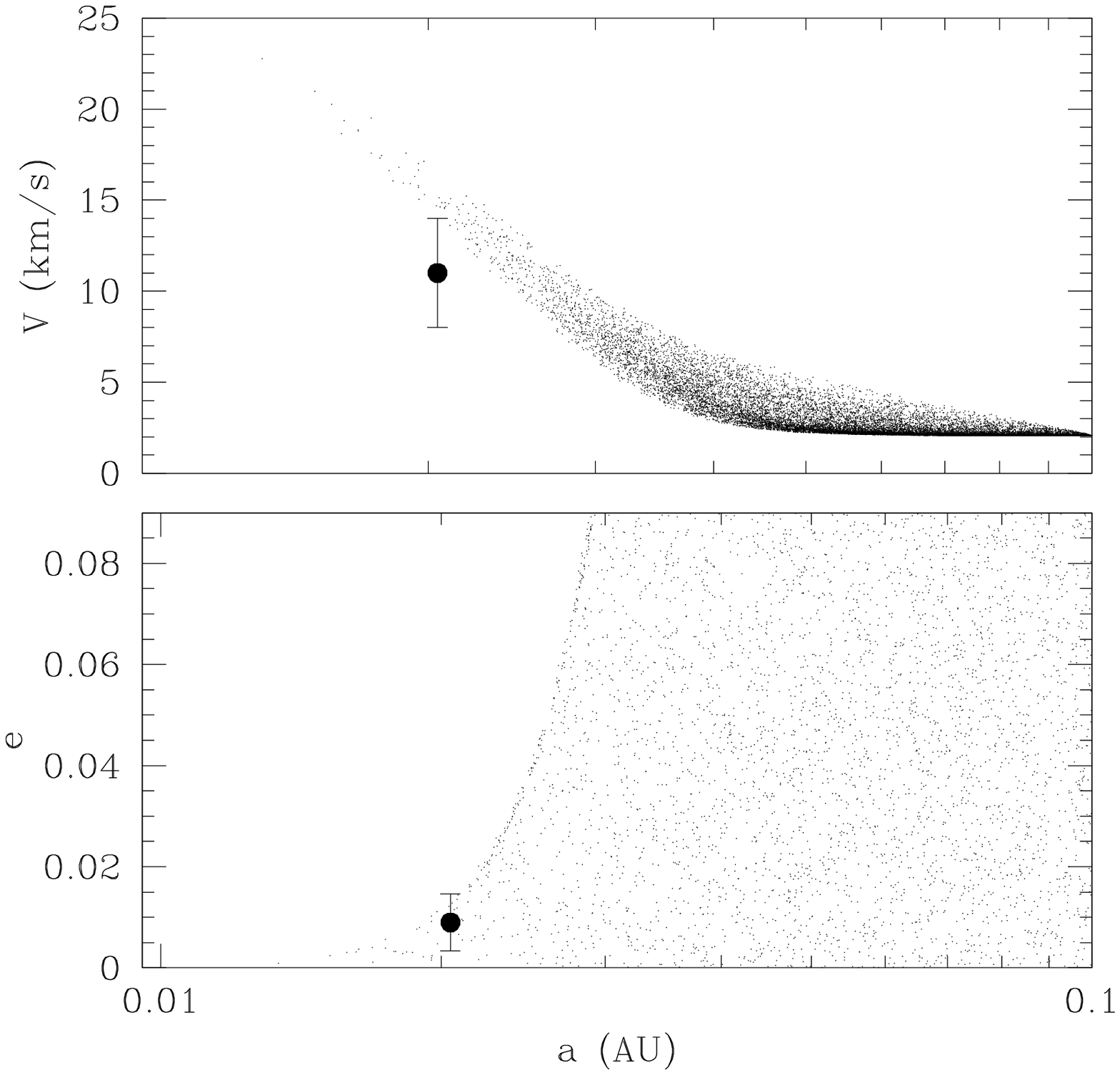}

\figcaption[WASP18.ps]{The organisation of this figure is much like that of Figure~\ref{xo3_init}, but using the
parameters of the WASP18b system. In the upper panel we see that significant stellar spin-up is expected due to the
transfer of angular momentum to the star as the planet spirals in. In the lower panel, the zoom into low eccentricities
allows us to see the wedge extending to $\sim 0.04$~AU, that results from the circularisation of highly eccentric orbits
due to tides in the planet, at distances too large to be strongly affected by the stellar tide.
\label{WASP18}}

\clearpage

\plotone{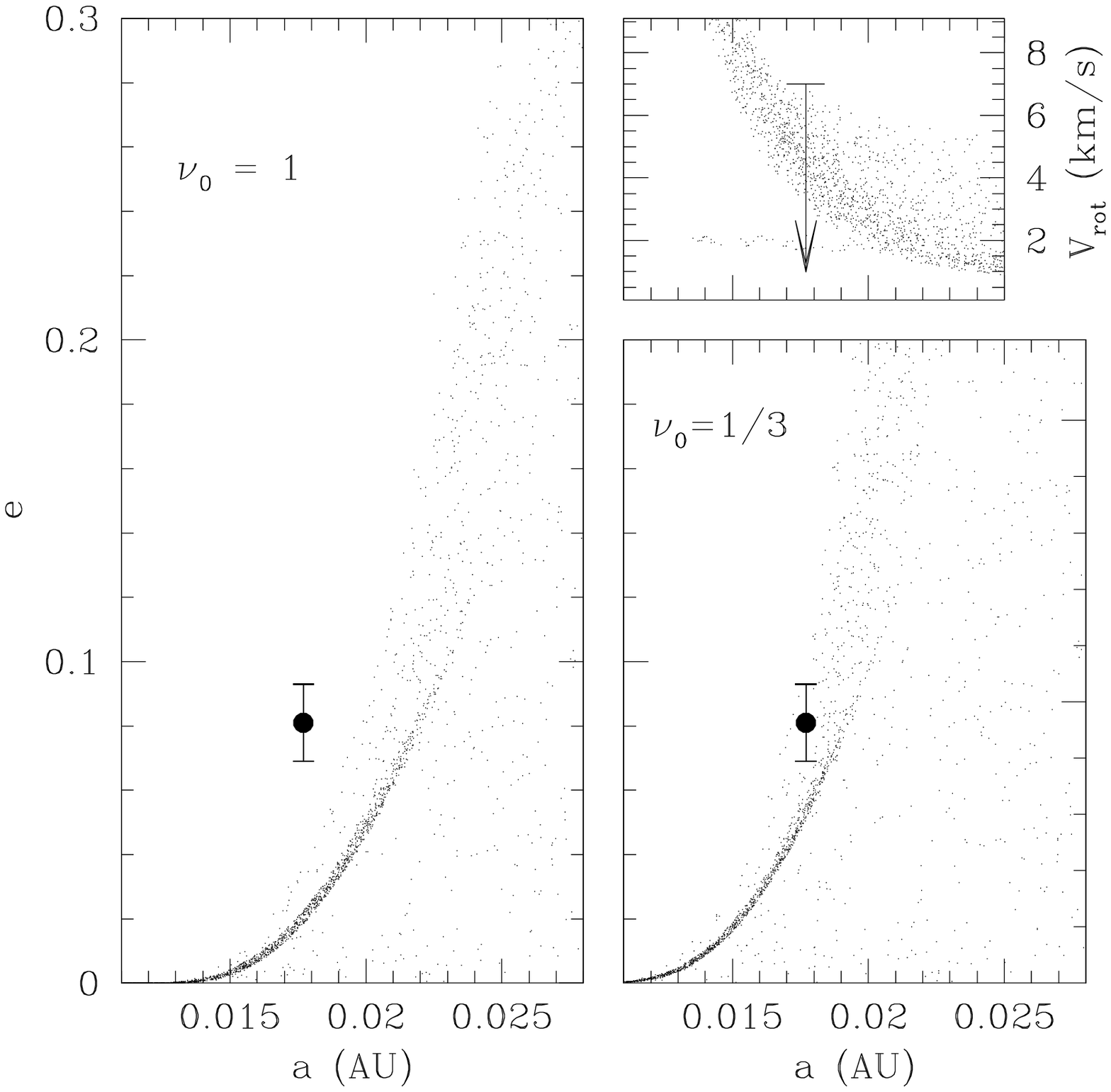}

\figcaption[HD41004.ps]{The left hand panel shows the range of realised semi-major axes and eccentricities
for models of the HD41004B system, assuming ages from 1--2~Gyr, and a normalisation of $\nu_0=1$. We see
that the results are inconsistent with the observational values (shown as a solid point). On the right, in
the lower panel, we
show a similar comparison but with $\nu_0=1/3$. In the upper right panel, we show the expected degree of
spin-up for the host star. This is constrained by the broadening of the templates in the analysis of
Zucker et al. (2004), represented by the upper limit.
\label{HD41004}}

\clearpage

\plotone{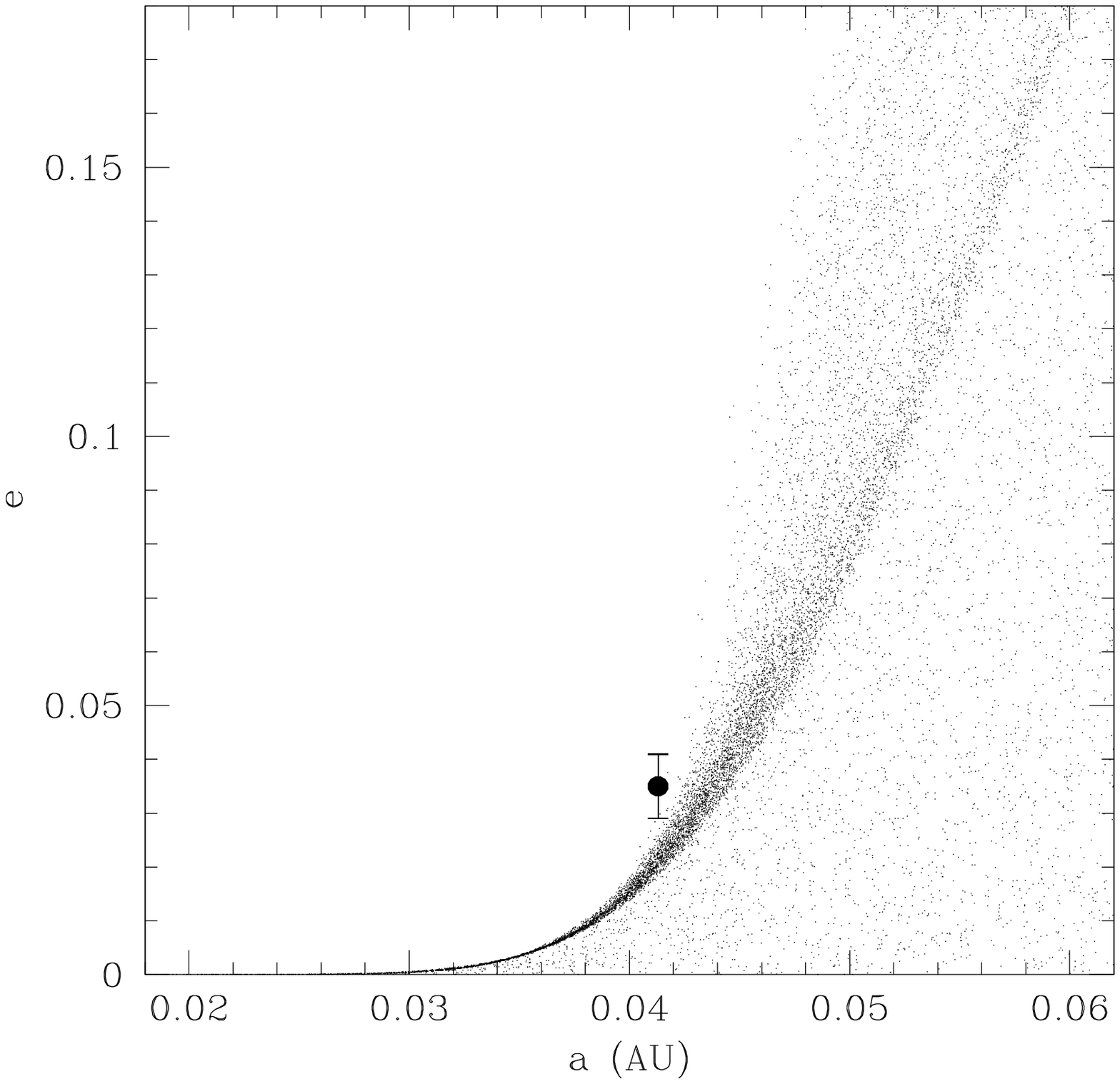}

\figcaption[HATP16.ps]{The solid point shows the HAT-P-16b parameters, and the small points indicate possible
system values for different initial conditions and an age between 1.2 and 2.8~Gyr. The leftmost edge of the allowed
region is defined by the youngest systems, although the age spread is not important for the HAT-P-16b parameters.
\label{HATP16}}

\clearpage

\plotone{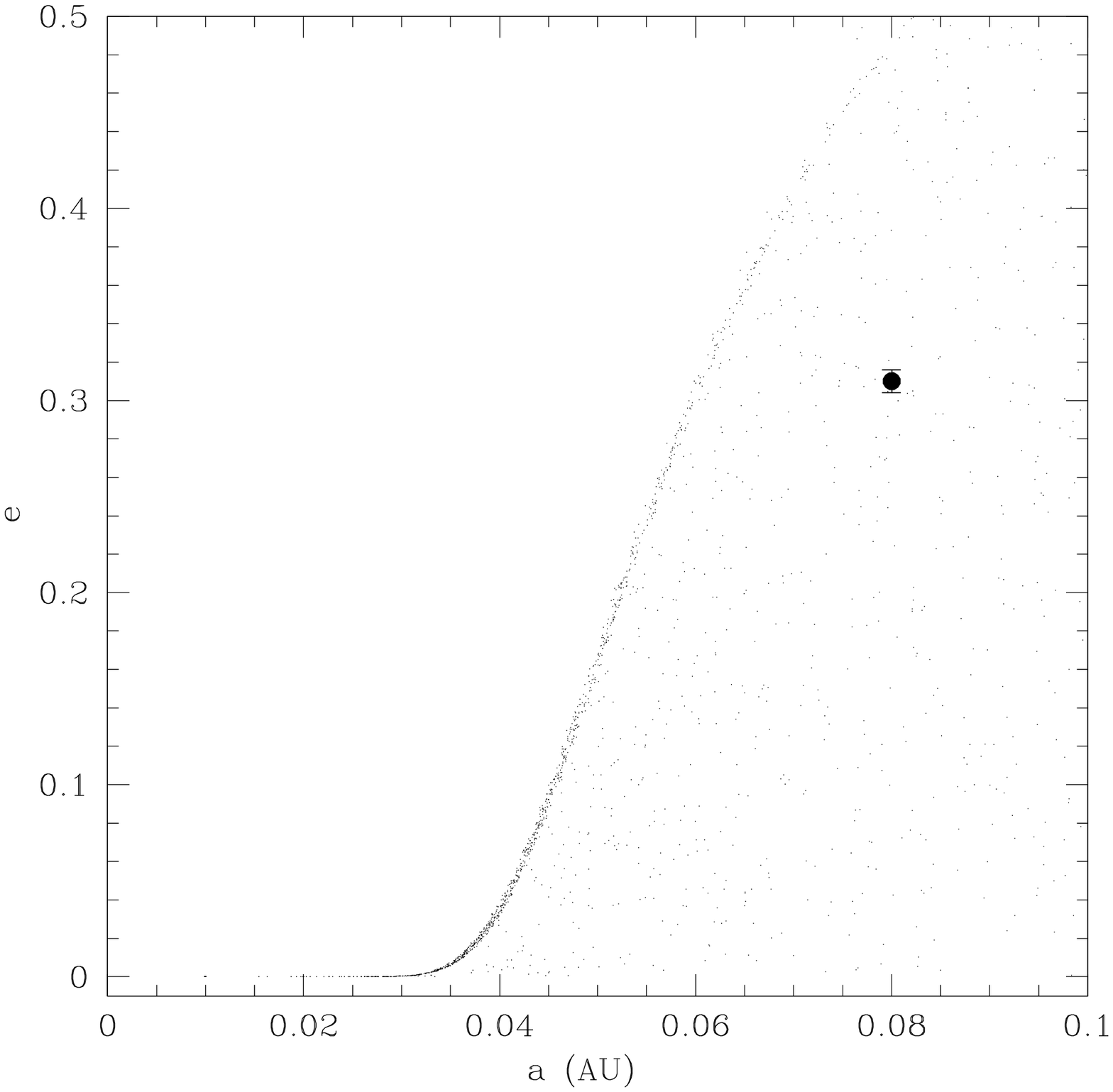}

\figcaption[WASP8.ps]{The solid point shows the WASP-8b planetary parameters. The dotted points are
model points for $\nu_0=1.5$ and assuming an age of $4\pm 1$~Gyr.
\label{WASP8}}

\clearpage

\plotone{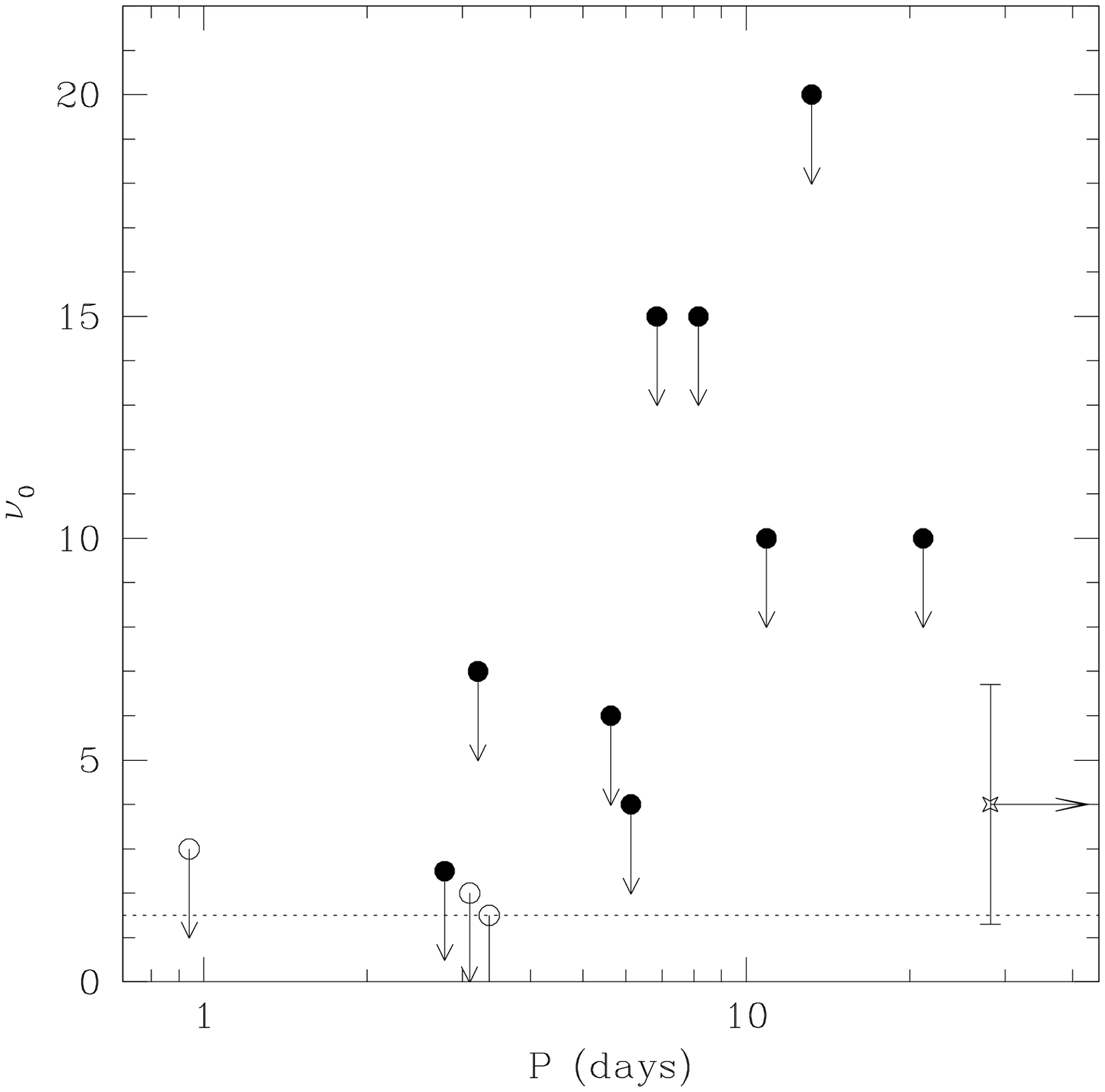}
\figcaption[Nup2.ps]{The upper limits represent the constraints on $\nu_0$ from the individual systems discussed
in \S~\ref{Envelope}. The solid points indicate those with robust eccentricity detections, while the open circles
indicate systems with initially reported non-zero eccentricities that have been called into question by the reanalysis
of Pont et al. (2010). The star symbol 
 on the right represents the constraint from the analysis of the red giants
in \S~\ref{VPsec}. The horizontal dotted line indicates our adopted value of $\nu_0=1.5$.
\label{Nup}}

\clearpage

\plotone{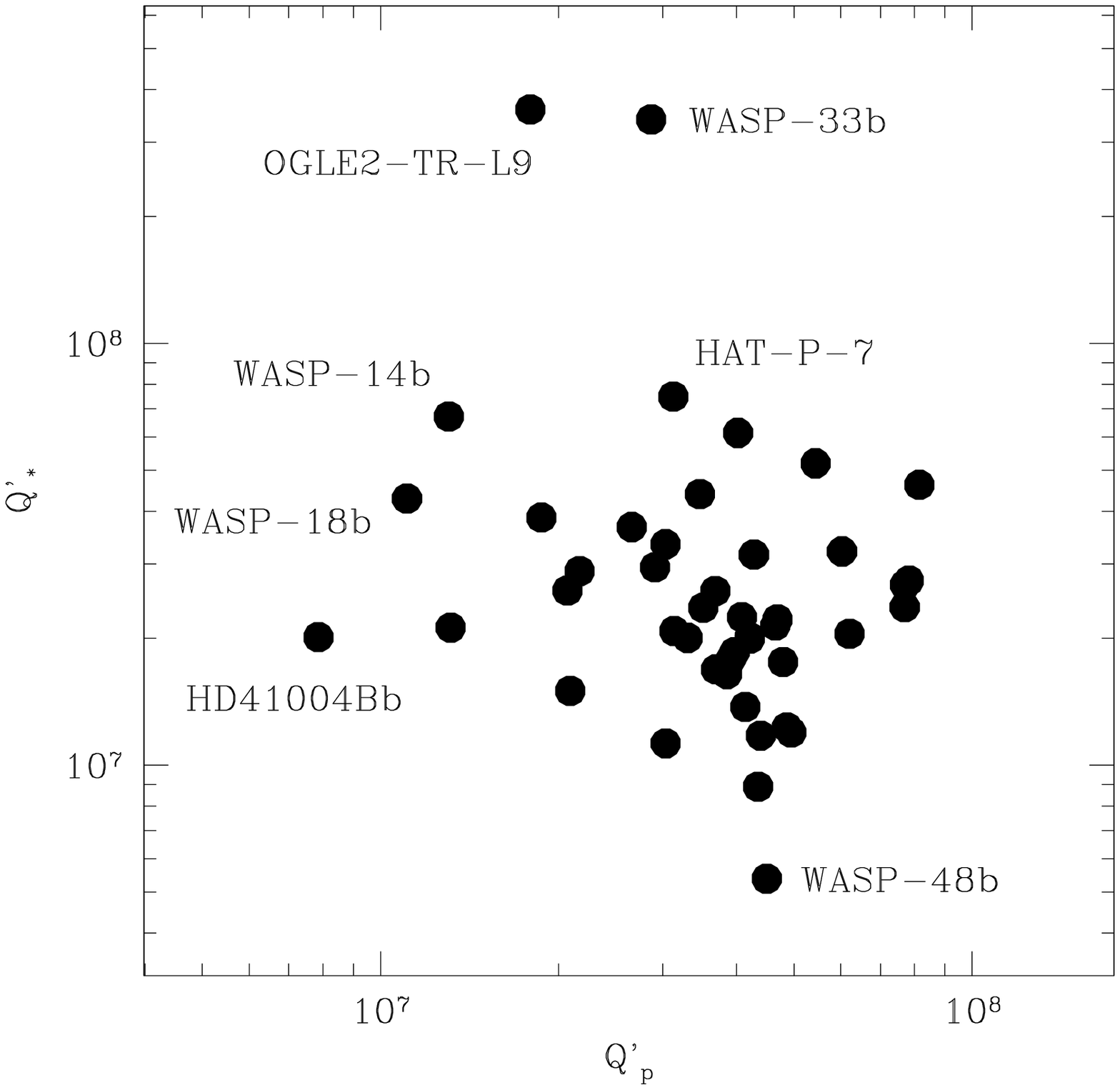}
\figcaption[QQ.ps]{Each point shows the estimated $Q'_*$ and $Q'_p$ for a planet-star system, using the observed
masses, radii and orbital period. Shown are estimates for all planets with orbital periods less than three days
(except for a handful of planets with masses $M_p<0.1 M_J$, which may not fit within the same gas giant paradigm).
Some of the systems are labelled, to illustrate the trends. The highest $Q'_*$ are found for host stars with
masses $\sim 1.5 M_{\odot}$, which have small surface convection zones and thus less dissipation. On the other hand,
slightly evolved host stars can lead to larger dissipation and smaller $Q'_*$, such as in the case of WASP-48b.
\label{QQ}}

\clearpage
\plotone{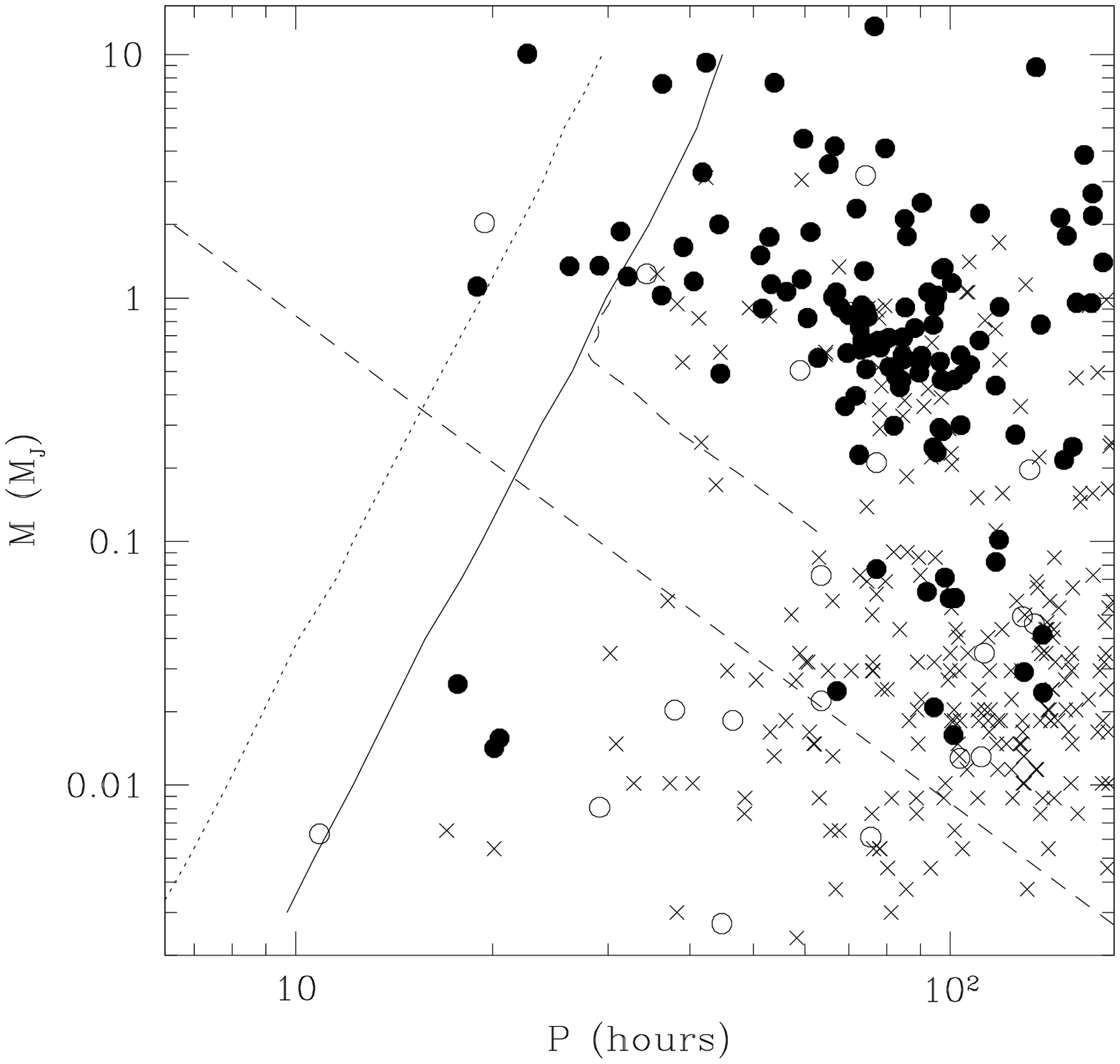}
\figcaption[Stab2.ps]{Theoretical tidal stability limit for planets and host stars of different masses. The solid curve
indicates a lifetime of 1~Gyr, around a $1 M_{\odot}$ star, while the dotted line indicates a lifetime of 0.1~Gyr.
 The dashed
line indicates the Roche lobe overflow limit for a gas giant of radius 1$R_J$. Solid points are RV-confirmed planets
orbiting stars with $M>0.8 M_{\odot}$ and open circles are confirmed planets orbiting stars with $M<0.8 M_{\odot}$. The 
short period open circle is WASP-43b, which has a lifetime $\sim 2$~Gyr despite lying to the left of the dotted line,
because the host star is lower mass, which moves the corresponding curve inwards (the decrease in stellar radius dominates
over the increased size of the convection zone).
The crosses indicate candidate Kepler objects, where we have estimated masses based on $M = 1 M_{\oplus} (R/R_{\oplus})^{2.3}$.
\label{StabLim}}

\clearpage

\plotone{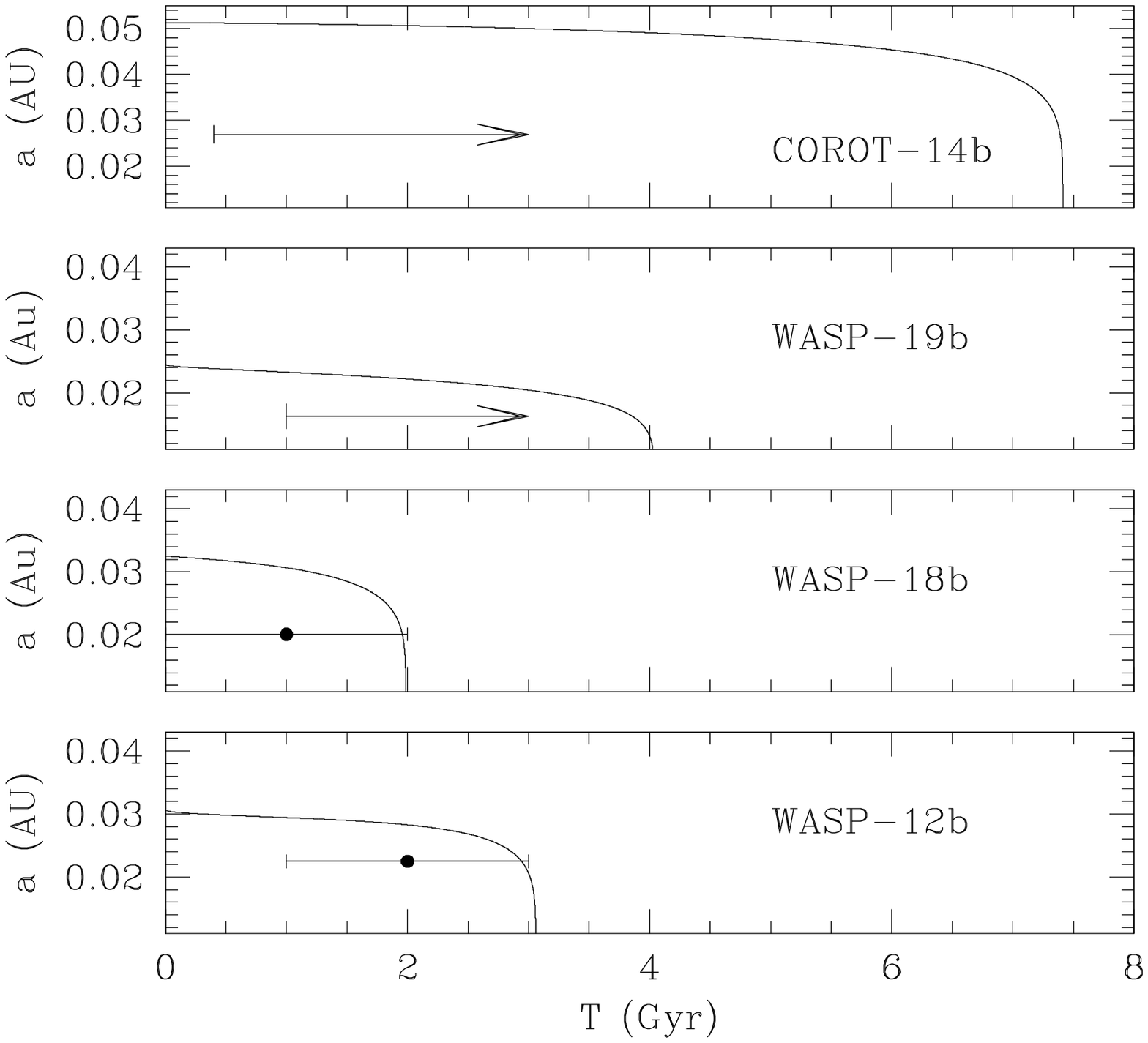}
\figcaption[4Ex.ps]{The four panels show potential evolutionary trajectories for the four systems which
have the shortest characteristic inspiral times. In each case we see that it is possible to locate initial
conditions such which realise the current configuration at the estimated age of the system, as shown.
\label{4Ex}}

\clearpage

\plotone{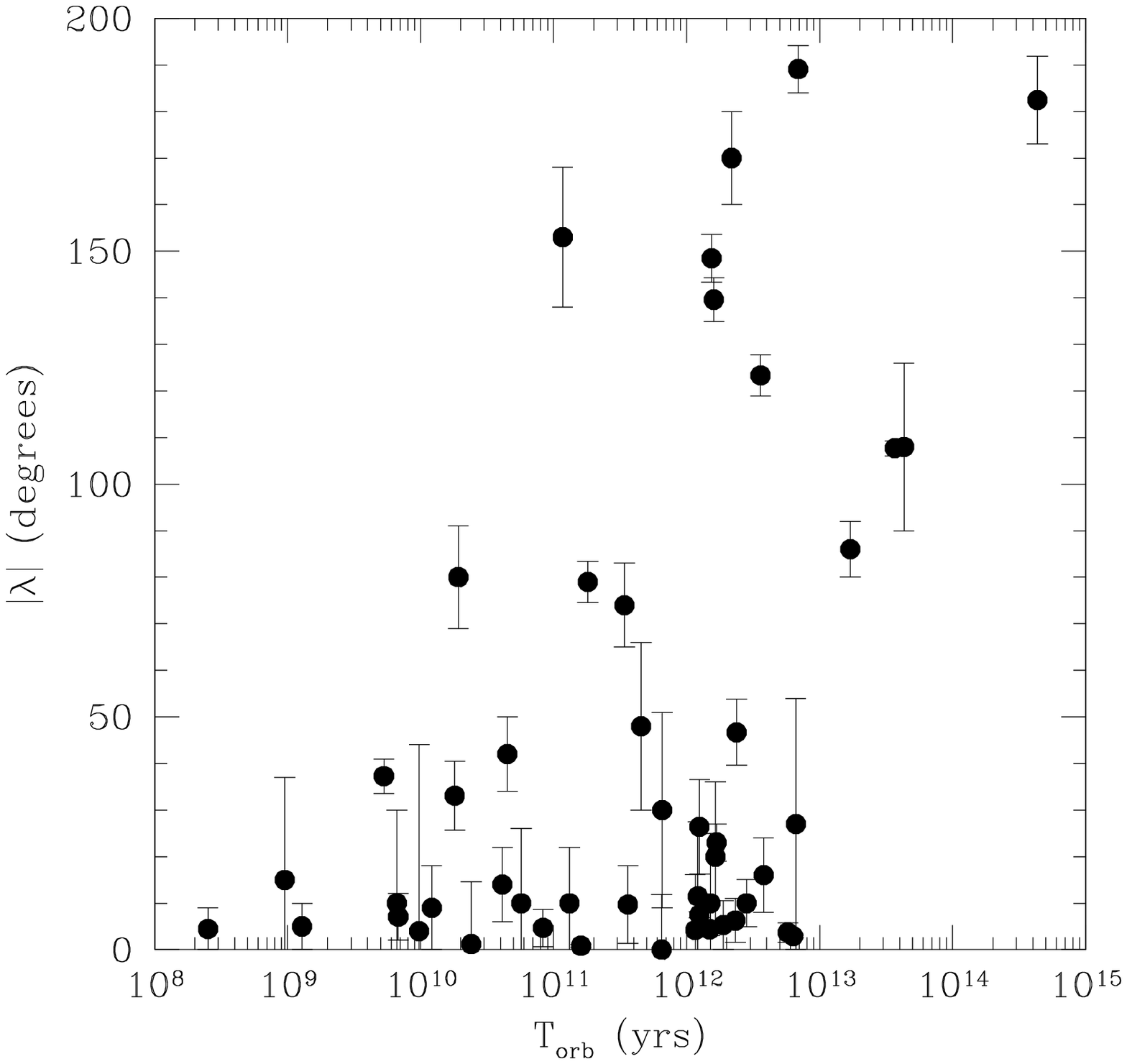}

\figcaption[Tl.ps]{The points show the measured orbit/stellar spin misalignment, as a function
of characteristic orbital evolution timescale due to stellar tides, given in equation~(\ref{Torb}). The fact that nearly
all the known systems have $T_{orb}>10^{10}$~years indicates that the relative obliquity is not strongly
affected by orbital evolution. The alignment of the few that do have short enough timescales is consistent with
our model.
\label{Tl}}

\clearpage

\plotone{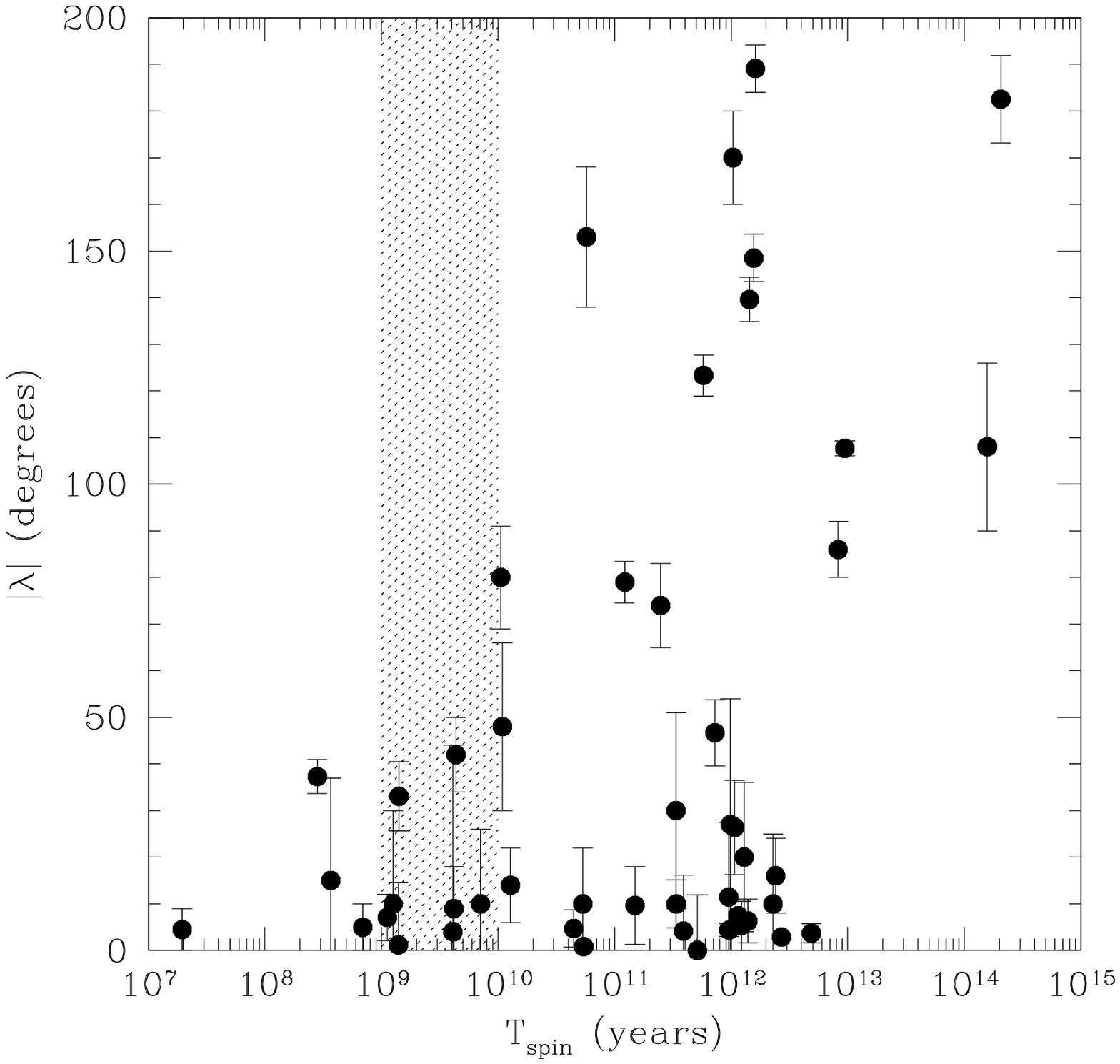}
\figcaption[Tsp.ps]{The points show the measured orbit/stellar spin obliquity, as a function of
stellar spin-up time, given in equation~(\ref{Tspin}). The degree of alignment seems to correlate
well with $T_{spin}$, suggesting that alignment results from the torqueing of the stellar spin.
We note that these estimates of $T_{spin}$ use the full stellar moment of inertia, so that we do
not require any decoupling between surface convection zones and stellar interior. The shaded region
shows the expected timescale for spin-down from magnetic stellar winds, allowing for the different
wind properties of F, G \& K stars. The one point with significant misalignment and short spin-up time
is XO-3b, in which the host star is indeed rotating almost synchronously.
\label{Tsp}}

\clearpage

\plotone{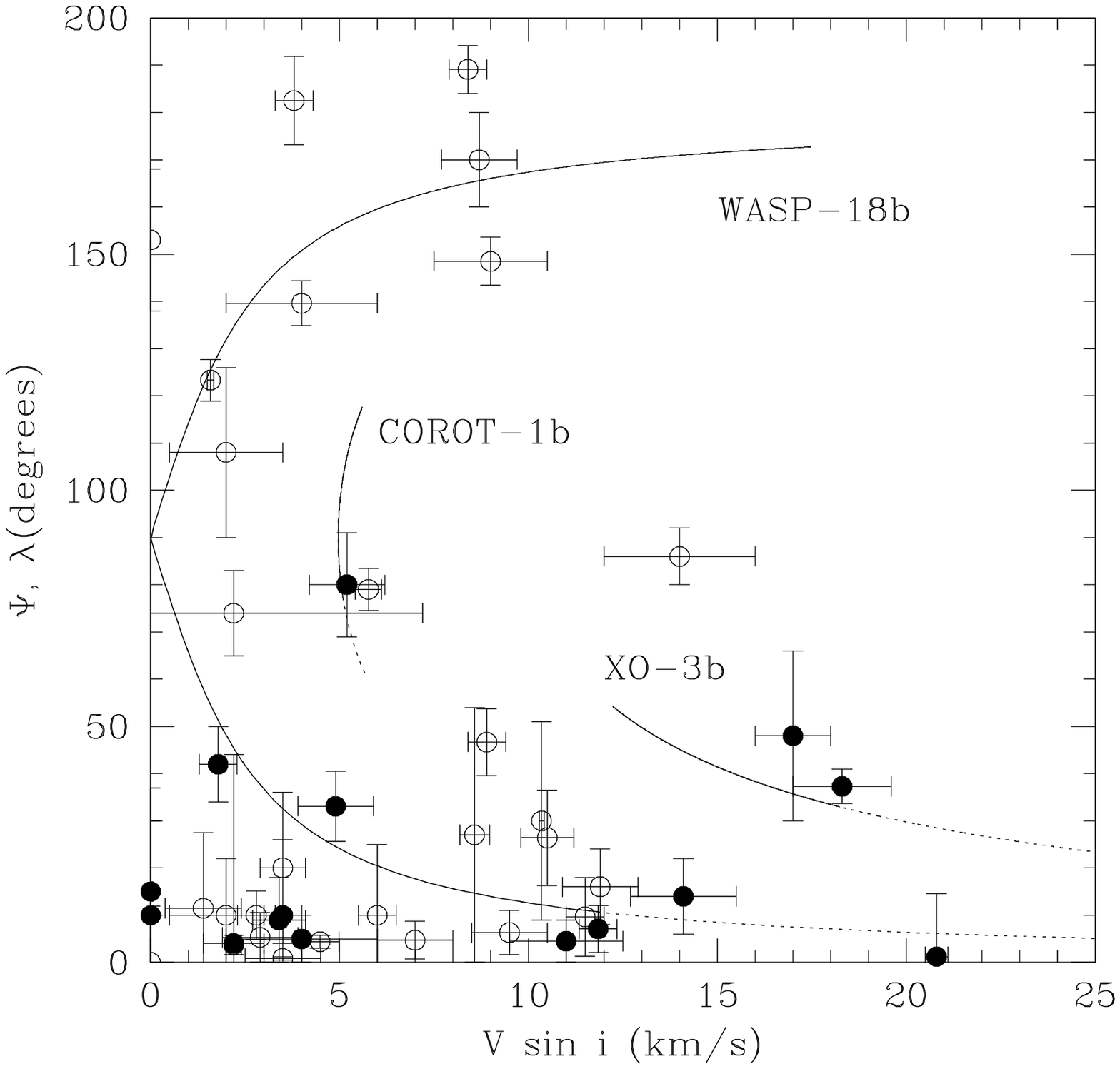}
\figcaption[LV2.ps]{The solid points indicate planets with $T_{orb}<2 \times 10^{10}$ years, while open points
indicate $T_{orb}>2 \times 10^{10}$~years. The curves indicate sample trajectories for the parameters of
WASP-18b, XO-3b and COROT-1b (solid for the stated age of the systems and dotted into the future). The
strength of tidal interactions in WASP-18b are sufficiently large that they could have completely
reversed the spin of the star within the lifetime. For COROT-1b and XO-3b, observations require a significant
stellar rotation component perpendicular to the planetary orbital plane, so that complete
alignment is impossible.
\label{LV2}}

\clearpage

\plotone{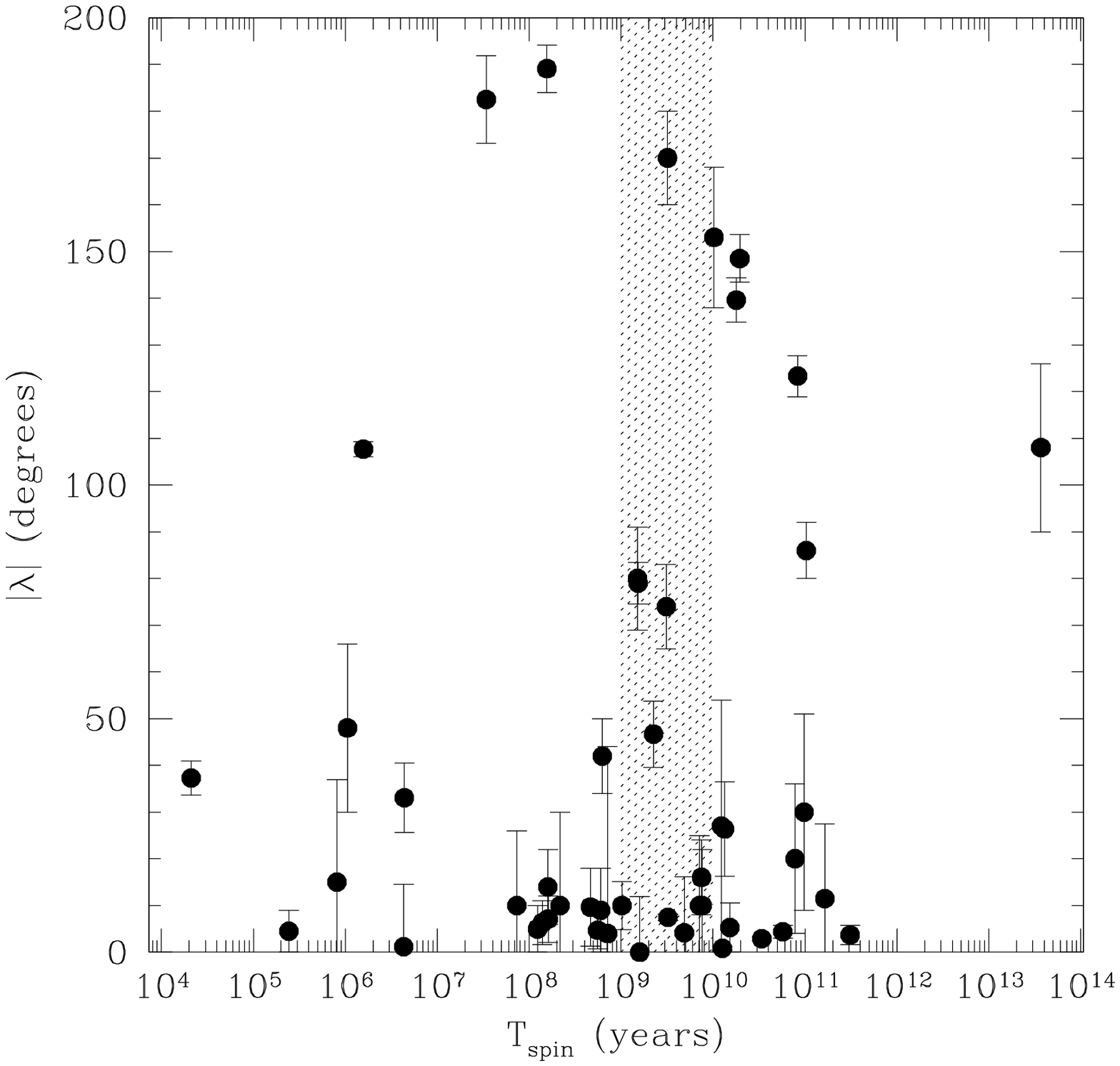}
\figcaption[Tsp2.ps]{The points show the spin-up times for the planet host stars assuming only
the moment of inertia of the surface convection zones, as suggested by Winn et al. 2010. We see
that the spin-up times of several retrograde systems are now well below expected stellar
lifetimes. The dotted line shows the equivalent magnetic braking spin-up time using the adjusted
spin-down law in Winn et al. 
\label{Tsp2}}

\clearpage

\plotone{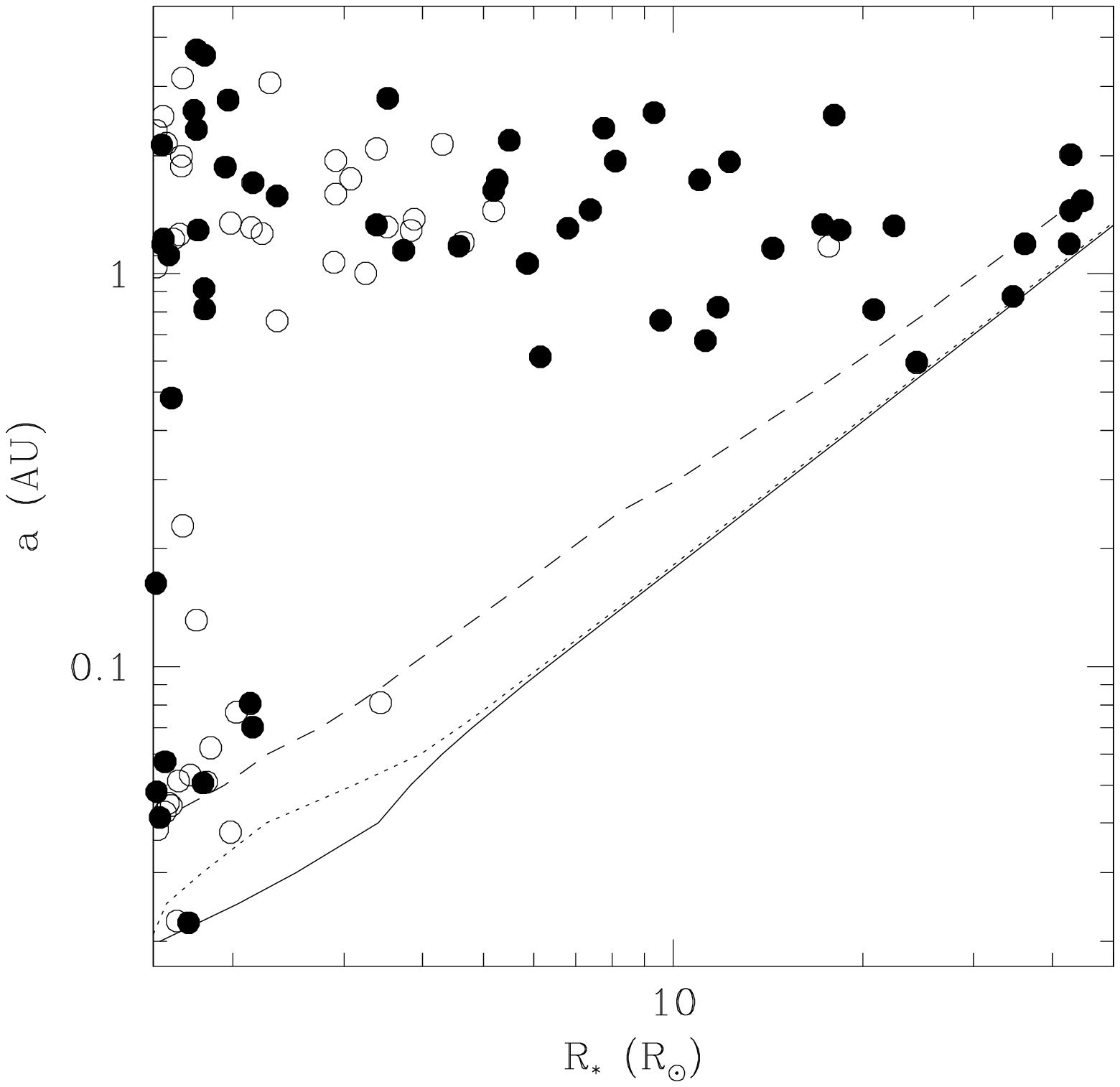}
\figcaption[Ra.ps]{The solid points indicate known planets with $M> 2 M_J$ and open circles indicate $M< 2 M_J$.
The dotted line indicates the survival radius, against tidal swallowing, of a planet in orbit around a slowly rotating 1.5$M_{\odot}$
star. The solid line indicates the corresponding criterion for the 1.5$M_{\odot}$ star that rotates with a 1~day period on the
main sequence and conserves angular momentum as it expands. The dashed line is the survival radius for a slowly rotating 1.2$M_{\odot}$
star. The observations definitely seem to imply an edge associated more with the 1.2 $M_{\odot}$ than the 1.5$M_{\odot}$ model.
\label{Ra}}

\clearpage

\plotone{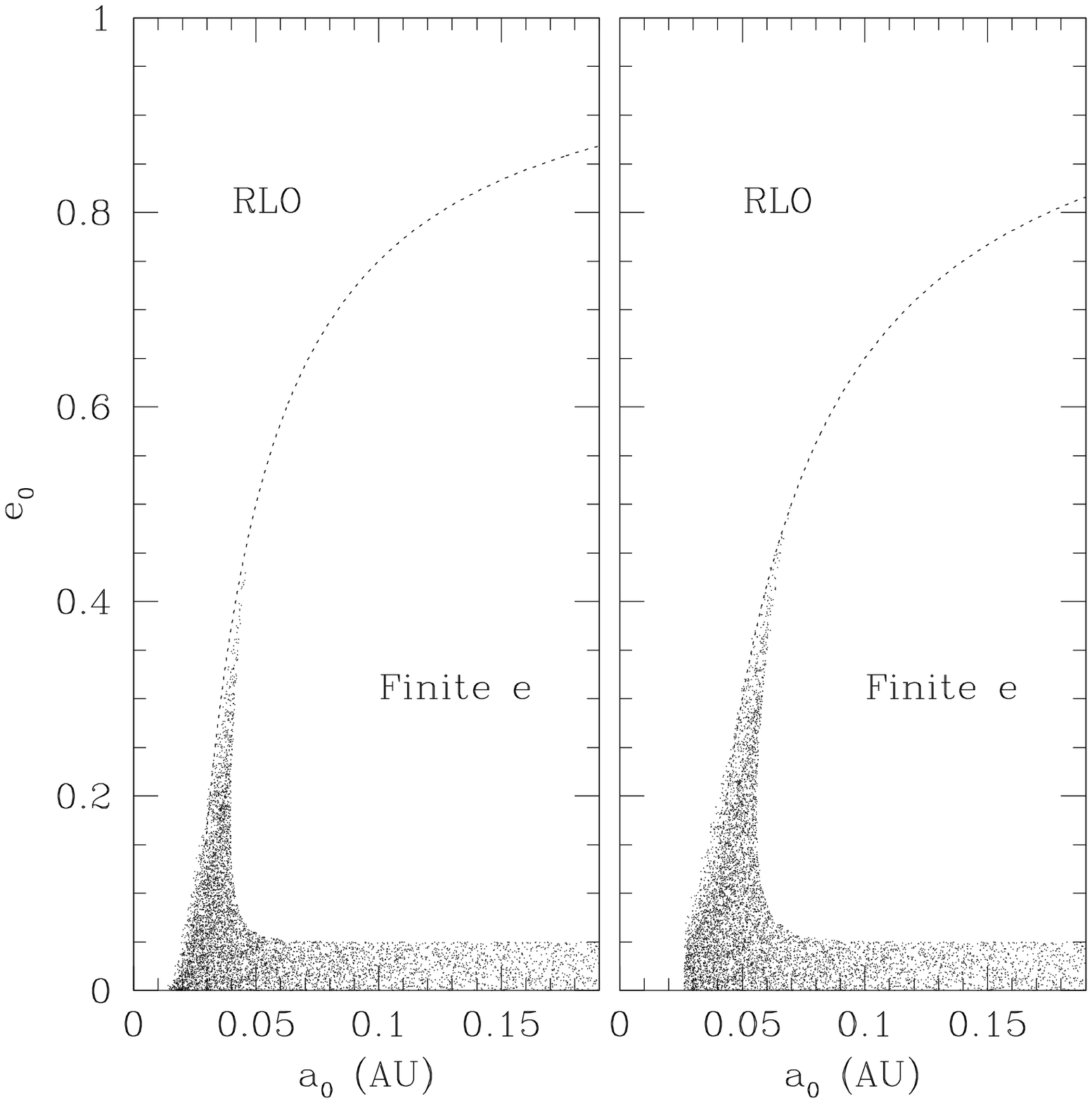}
\figcaption[xoeo.ps]{Both panels show those initial orbital parameters that produce a circular orbit within
5~Gyr, assuming a 0.5$M_J$ planet orbiting a 1$M_{\odot}$ star. The left hand panel assumes $\nu_0=1.5$ for
both planet and star, while the right hand panel uses $\nu_0=15$ for the planet. The dashed line delineates
the initial orbits that allow a planet to survive Roche lobe overflow.
\label{xoeo}}

\clearpage

\plotone{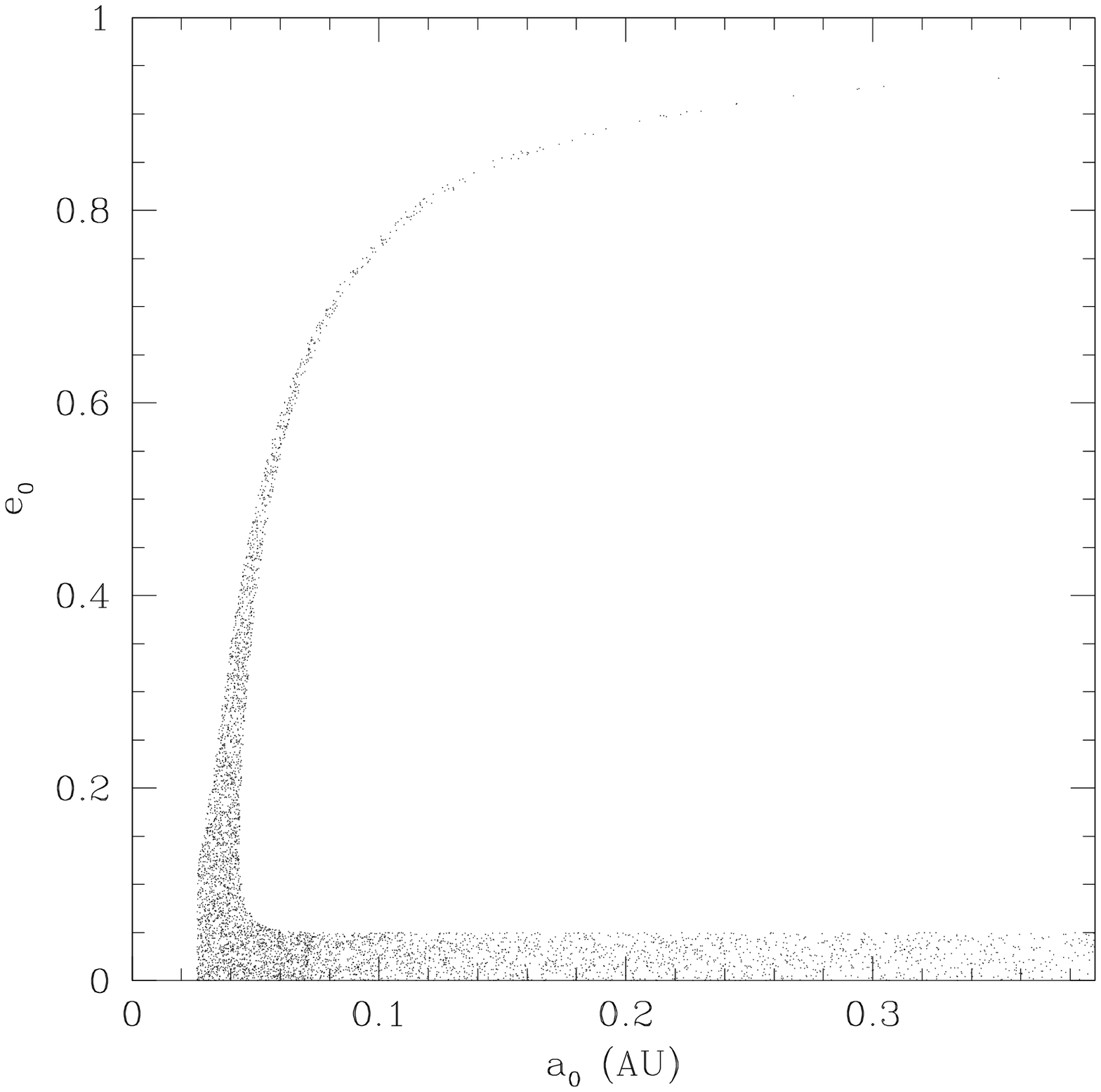}
\figcaption[xo2.ps]{The distribution of initial semi-major axis and eccentricity for a 0.5~$M_J$ planet orbiting a host
star of mass $1 M_{\odot}$, assuming a final (at age 5~Gyr) circular ($e<0.05$) orbit. 
 The difference between this and Figure~\ref{xoeo} is that
the dissipation in the planet was kept fixed at the normalised value for a forcing period of 1~day. This allows survival
for orbits from much farther out, as is postulated in several current origin scenarios.
\label{xo2}}

\clearpage

\plotone{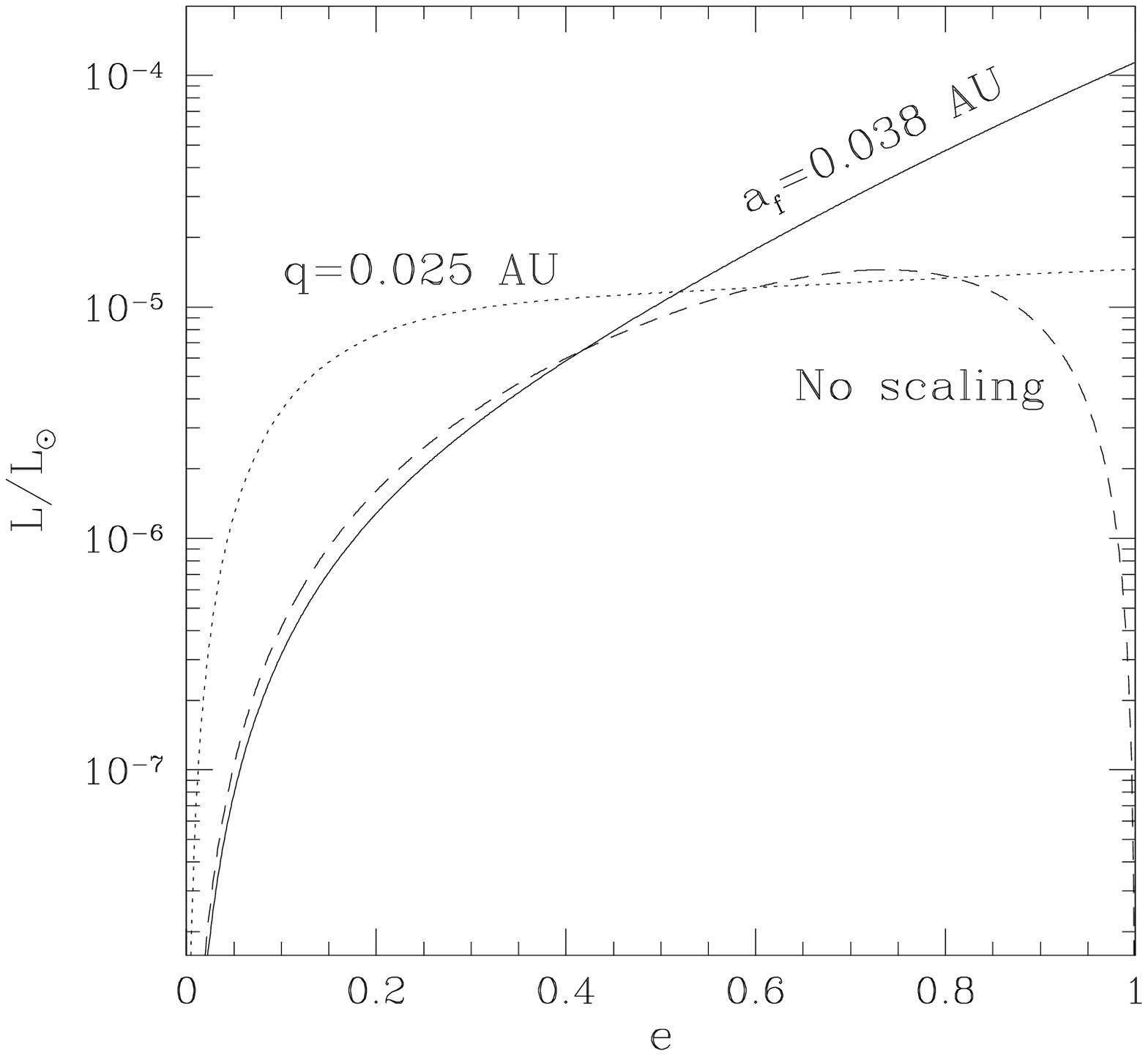}
\figcaption[Edot.ps]{Tidal heating for a $0.5 M_J$ planet around a 1$M_{\odot}$ star. All models assume a planet radius
of $1.6 R_J$. The dotted line represents constant periastron, while the solid line represents a line of constant angular
momentum. The dashed line is also a line of constant angular momentum, but one in which the dissipation constant is held
fixed, rather than scaled with forcing period according to the Zahn prescription.
\label{Edot}}
 


\clearpage
\plotone{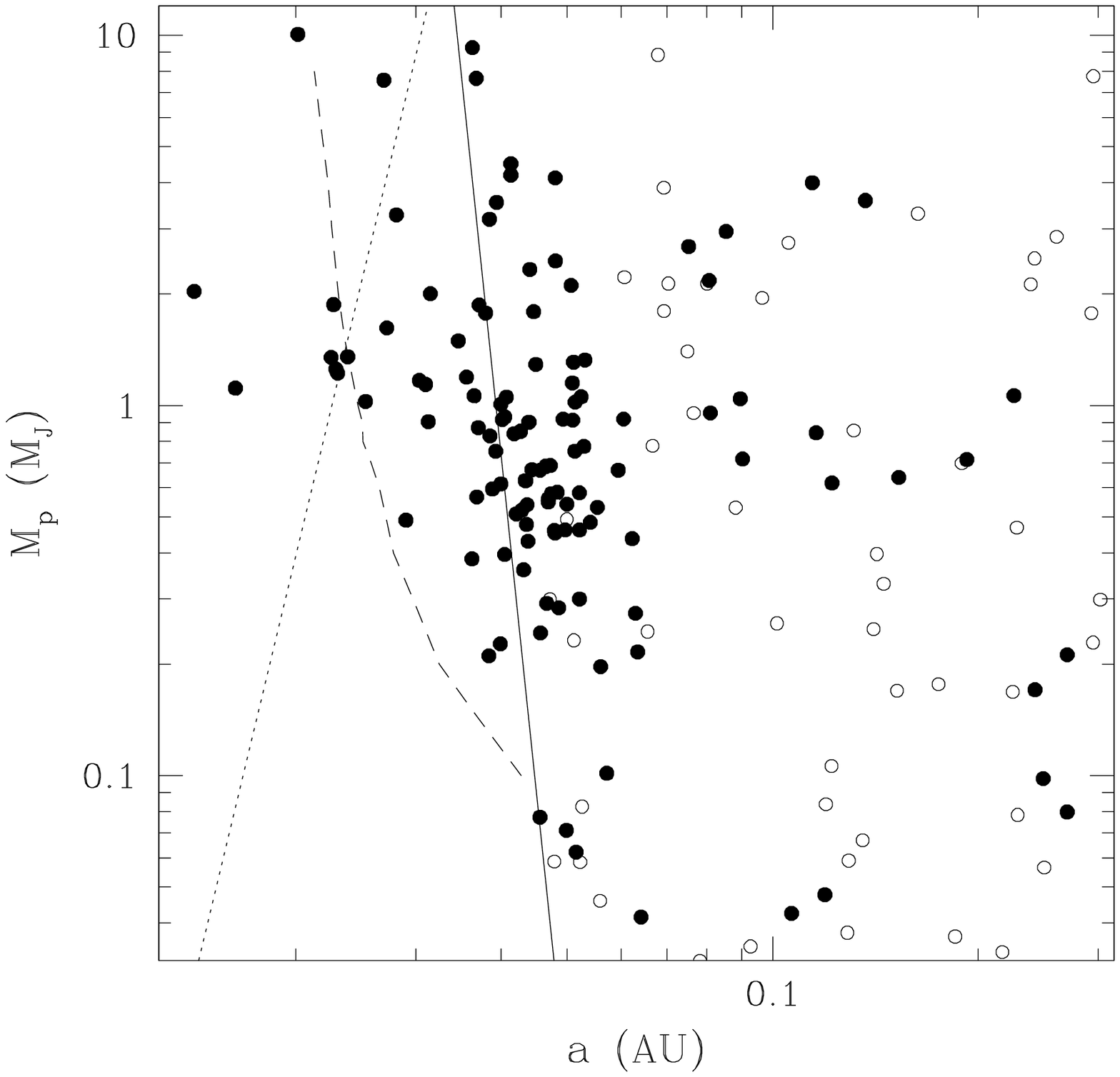}
\figcaption[Ma.ps]{The solid line shows the equation~(\ref{Afinal}) for a star of $1 M_{\odot}$.
 The dashed line represents the final locus of orbits that approach the edge of disruption from tidal heating during 
the circularisation process, but survive. 
 The dotted
line is the 1~Gyr survival limit against stellar tides (for a $1 M_{\odot}$ star) shown in Figure~\ref{StabLim}. 
Solid circles show known planets whose eccentricities $e < 0.1$ and stellar hosts $M_* > 0.7 M_{\odot}$. Open circles have
 $e > 0.1$ and $M_* > 0.7 M_{\odot}$. 
\label{Ma}}


\clearpage
\plotone{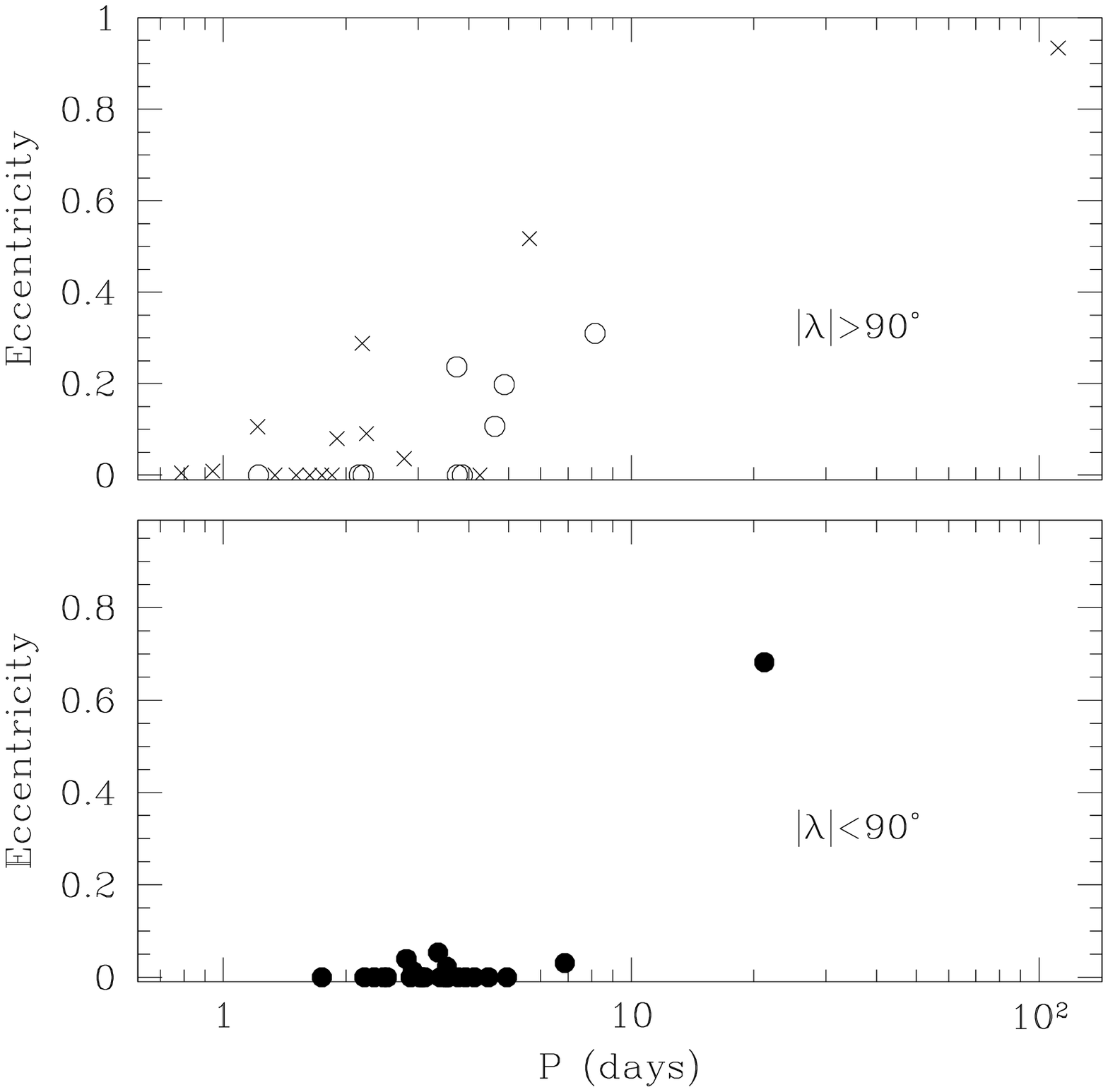}
\figcaption[Pe_nosync.ps]{The lower panel shows the period eccentricity relation for those systems whose measured
obliquities are $<90^{\circ}$ and which have $T_{spin}>2 \times 10^{10}$~years. The upper panel shows, as open circles,
those systems with  $>90^{\circ}$ and which have $T_{spin}>2 \times 10^{10}$~years. The crosses indicate all
systems (regardless of $\lambda$) with $T_{spin}<2 \times 10^{10}$~years. In this case, the original obliquity may have been affected due to the
tidal transfer of angular momentum to the star. We see that very few eccentric systems are found amongst the primordially
aligned systems.
\label{nosync}}

\clearpage
\plotone{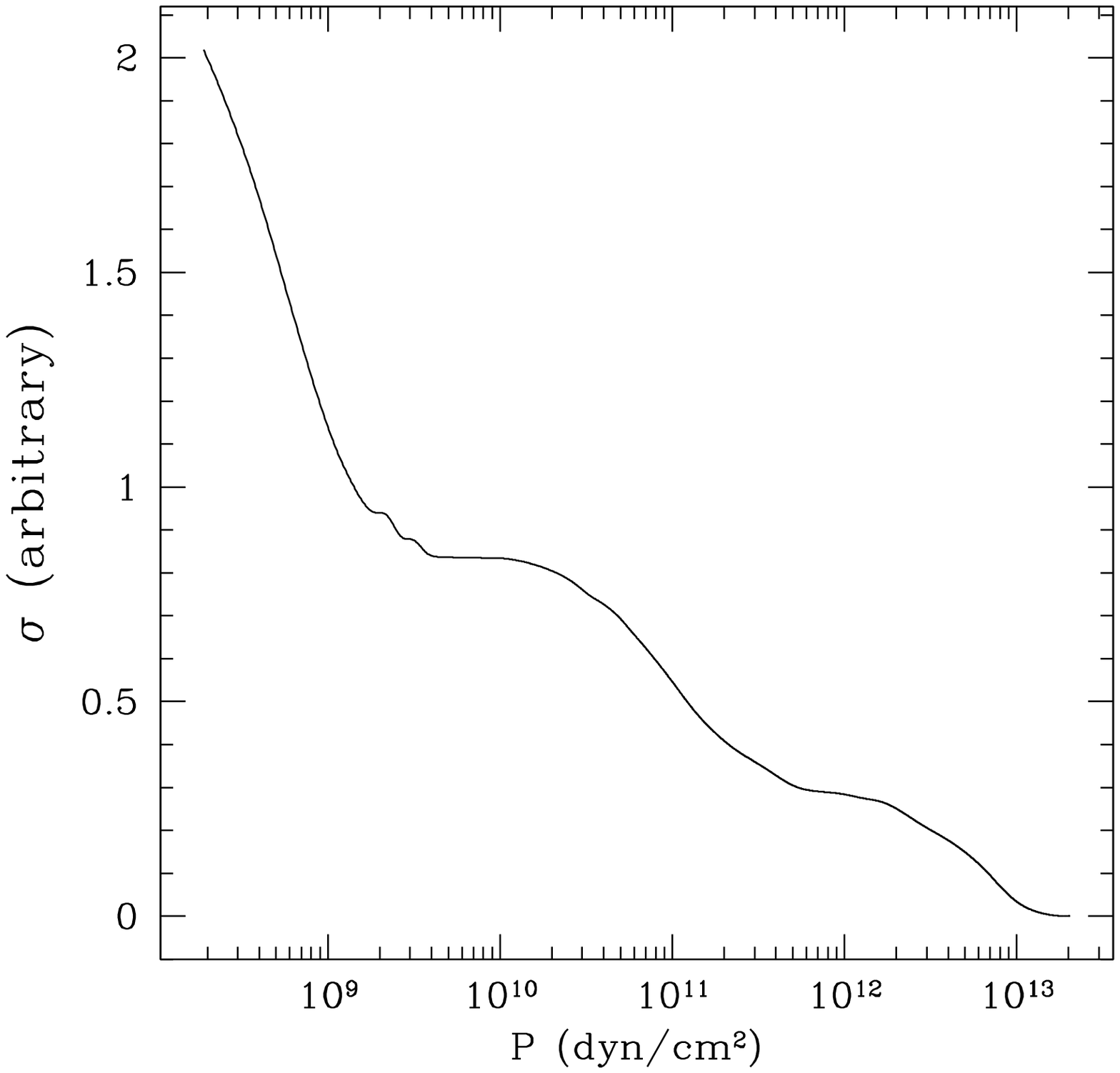}
\figcaption{The integration of equation~(\ref{sigma_integral}) is shown, plotted here against pressure instead
of mass. We see that the bulk of the contribution to the dissipation comes from lower pressures, with more than
50\% of the final total contributed by regions of the planet with $P < 10^{10} dyn/cm^2$.
\label{SigInt}}

\clearpage
\plotone{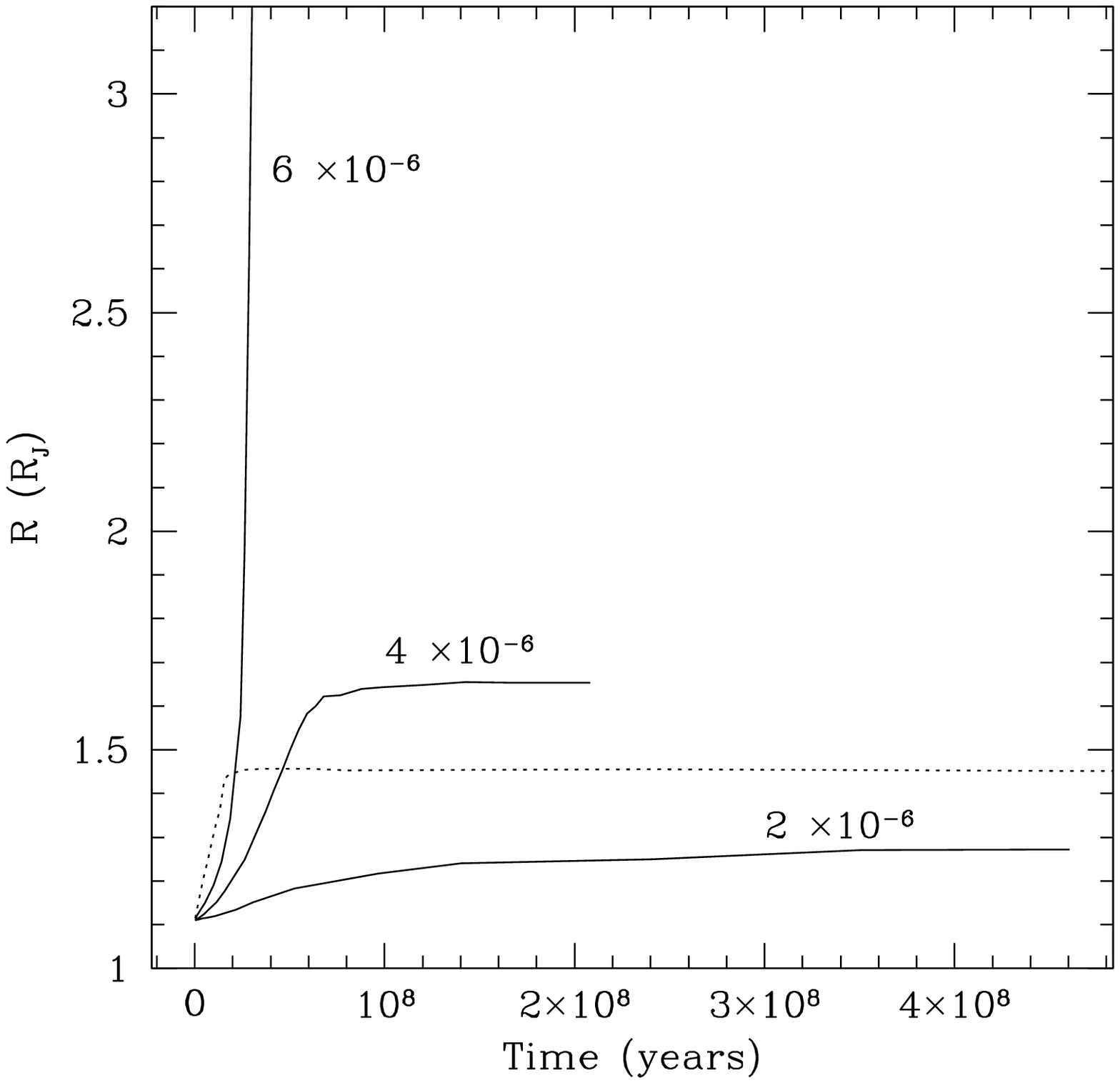}
\figcaption{The three solid curves show the evolution of a $10^8$~year old, $1 M_J$ planet, subjected to
different amounts of heating (each curve is labelled with the total input, in units of $L_{\odot}$. In
all three cases, the heating is restricted to regions of the planet with $P < 10^{10} dyn/cm^2$. The
dotted line shows the evolution of the same planet, subjected to $6 \times 10^{-6} L_{\odot}$ of heating,
but with the heat distributed uniformly by mass throughout the planet.
\label{3Lums}}

\clearpage
\plotone{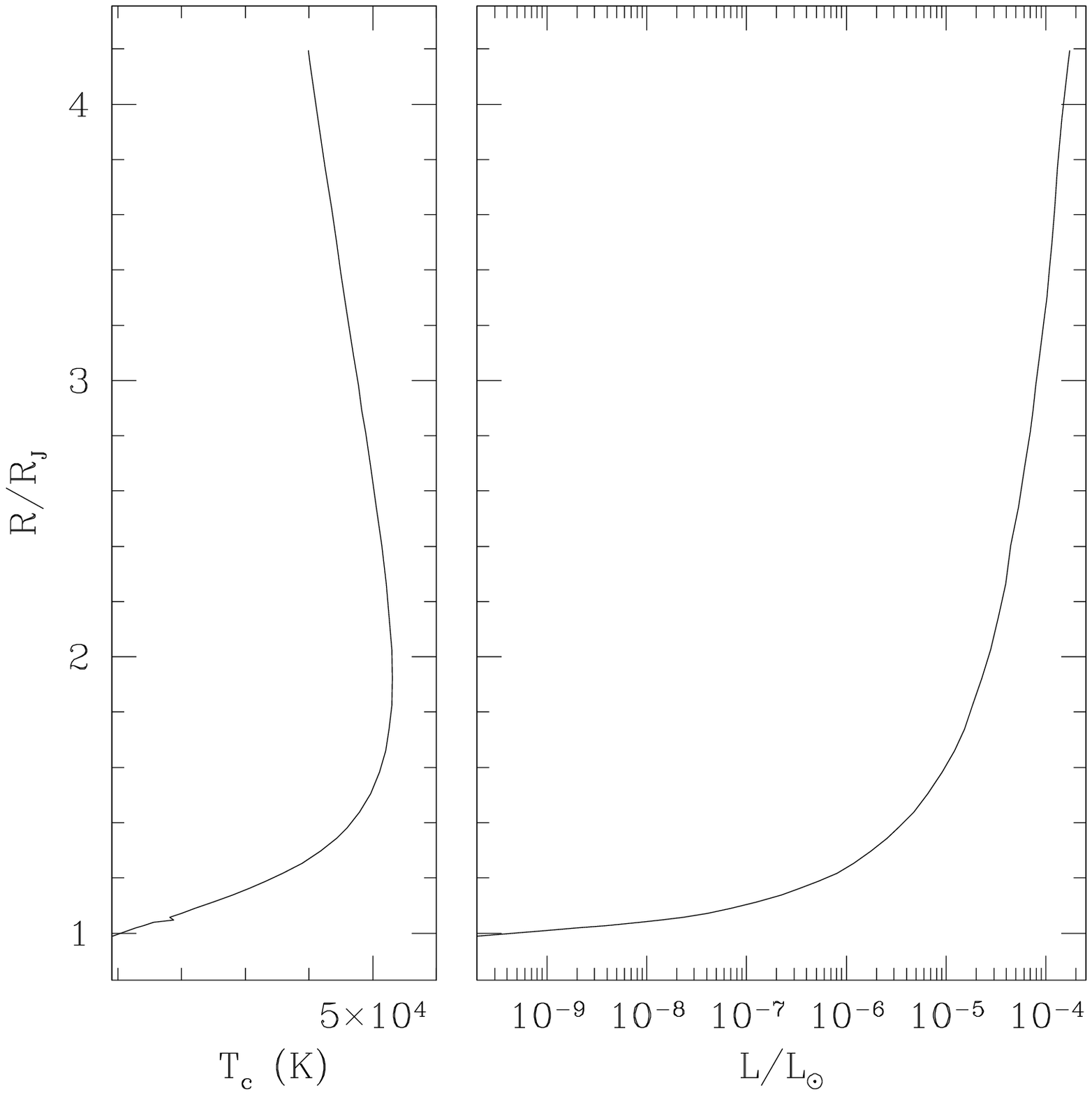}

\figcaption[RTL.ps]{The left-hand panel shows the relationship between central temperature
and radius for a $1 M_J$ planet. We see that the relation is not monotonic, because models
with large radii and large internal entropies actually have lower temperatures, due to the
storage of energy in gravitational potential rather than thermal energy. The right hand side
shows that the relationship between radius and luminosity is monotonic, however.
\label{LTC}}

\end{document}